\documentclass[final,5p,times,twocolumn,sort,compress]{elsarticle}

\usepackage[mathlines,switch]{lineno}
\newif\iflinenos
\linenosfalse
\iflinenos\linenumbers\fi

\usepackage{amsmath}
\usepackage{amsfonts}
\usepackage{amssymb}
\usepackage{bbold}
\usepackage{epsfig}
\usepackage{graphicx}
\graphicspath{{figs/}}
\usepackage{bm}
\usepackage{array}
\usepackage{listings}
\usepackage{color}
\usepackage[normalem]{ulem}
\usepackage[dvipsnames]{xcolor}
\usepackage[colorlinks=true,breaklinks=true]{hyperref}

\definecolor{darkred}{rgb}{0.5,0,0}
\definecolor{darkblue}{rgb}{0,0,0.5}
\definecolor{firebrick}{rgb}{0.75,0.125,0.125}
\definecolor{darkgreen}{rgb}{0,0.5,0}
\hypersetup{urlcolor=darkblue,
	    citecolor=darkgreen,
	    linkcolor=firebrick}

\newcommand{\ie}{{\it i.e.}}

\newcommand{\eg}{{\it e.g.}}

\newcommand{\eq}{Eq.}

\newcommand{\fig}{Fig.}

\providecommand{\Ref}{}
\AtBeginDocument{\renewcommand{\Ref}{Ref.}}
\newcommand{\Refs}{Refs.}
\newcommand{\Tab}{Tab.}

\newcommand{\equ}[1]{\eq~(\ref{equ:#1})}
\newcommand{\figu}[1]{\fig~\ref{fig:#1}}
\newcommand{\tabl}[1]{\Tab~\ref{tab:#1}}

\definecolor{mygreen}{rgb}{0,0.6,0}
\definecolor{mygray}{rgb}{0.5,0.5,0.5}
\definecolor{mylightgray}{rgb}{0.937,0.937,0.937}
\definecolor{mymauve}{rgb}{0.58,0,0.82}

\newcounter{bla}

\journal{Computer Physics Communications}

\begin{document}

\begin{frontmatter}

\title{{\tt NuOscProbExact}: a fast, general-purpose code to compute\\
exact two-, three-, and four-flavor neutrino oscillation probabilities}

\author{Mauricio Bustamante\corref{author}}
\ead{mbustamante@nbi.ku.dk}
\cortext[author]{ORCID: \href{http://orcid.org/0000-0001-6923-0865}{0000-0001-6923-0865}.}
\address{Niels Bohr International Academy, Niels Bohr Institute,\\University of Copenhagen, DK-2100 Copenhagen, Denmark}

\date{\today}

\begin{abstract}

A neutrino created with one flavor can \textit{oscillate} and be detected later with another.  Ordinarily, the probability of this occurring---the \textit{oscillation probability}---is obtained by diagonalizing the Hamiltonian that drives the neutrino evolution.  We bypass the diagonalization: expanding the evolution operator in the generators of the groups SU(2), SU(3), and SU(4) yields exact closed-form probabilities for two, three, and four neutrino flavors that need the eigenvalues of the Hamiltonian and never its eigenvectors.  As a result, every probability is then the same short run of arithmetic, whatever the Hamiltonian contains.  We implement these expressions in {\tt NuOscProbExact}, which we make publicly available.  Any scenario with a time- or position-independent Hermitian Hamiltonian is in scope: oscillations in vacuum, matter of constant density, with non-standard interactions, in a Lorentz-violating background, with sterile states, combinations of them, or completely novel theory proposals.  Matter of piecewise constant density, including the density profile of Earth, follows by composing the exact solutions for consecutive matter slabs.  {\tt NuOscProbExact} offers three properties that are seldom found in one program.  It is \emph{flexible}: computing probabilities for a new oscillation scenario needs no new solver, only a new Hamiltonian matrix.  It is \emph{fast}: among the surveyed existing public oscillation codes, it is the quickest to compute probabilities for neutrinos going through the Earth at every accuracy but the coarsest.  It is \emph{accurate}: for the Hamiltonian it is given, only machine round-off limits the probability; at constant density, no other surveyed code comes within four decades of it.  Deriving only a Hamiltonian, and evaluating a million of them in one call, puts whole parameter spaces within reach in scenarios that no approximation covers.

\end{abstract}


\end{frontmatter}

\noindent
{\bf Program summary}
\smallskip

\begin{small}

\noindent
{\em Program title:} NuOscProbExact v1.13.1    \\ \\
\smallskip
%%%%
{\em Licensing provisions:} MIT License  \\ \\
\smallskip
%%%%
{\em Program obtainable from:} \\
\href{https://github.com/mbustama/NuOscProbExact}{\tt https://github.com/mbustama/NuOscProbExact}  \\
\href{https://pypi.org/project/nuoscprobexact/}{\tt https://pypi.org/project/nuoscprobexact/} \\ 
{\tt pip install nuoscprobexact}  \\ \\
\smallskip
%%%%
{\em Distribution format:} {\tt git} repository; source distribution and built wheel on the Python Package Index (PyPI), installable with {\tt pip} \\ \\
\smallskip
%%%%
{\em Programming language:} Python 3.9 or later; tested on 3.9, 3.10, 3.11, 3.12, and 3.13   \\ \\
\smallskip
%%%%
{\em Classification:} 11.1 General, High Energy Physics and Computing   \\ \\
\smallskip
%%%%
{\em External routines:} {\tt NumPy} and {\tt Numba}, both installed automatically.  The library imports nothing else; {\tt SciPy} and {\tt matplotlib} are needed only to run the test suite and the notebooks   \\ \\
\smallskip
%%%%
{\em Nature of problem:} \\
Computation of exact neutrino flavor oscillation probabilities in arbitrary physical systems described by time- or position-independent Hamiltonians.  Applicable to closed systems of two, three, and four neutrino flavors, in vacuum, in matter of constant density, and in matter whose density varies along the trajectory, including the Earth. \\ \\
\smallskip
%%%%
{\em Solution method:}\\
Closed-form expansions of time-evolution quantum operators using SU(2), SU(3), and SU(4) identities.  For two and three neutrino flavors the program is a lightweight implementation of the method of Refs.\ [1, 2].  The four-flavor case is obtained instead from the spectral form of the evolution operator, Sylvester's formula, with the eigenvalues written in radicals; four flavors is the largest case for which that is possible.  A varying density profile is cut into slabs of constant density, each solved exactly; the operators are then composed in order. \\ \\
\smallskip
%%%%
{\em Additional comments, including restrictions, unusual features:}\\
The Hamiltonian must be Hermitian and time- or position-independent, so that the evolution is unitary and the system closed: interactions, absorption, decay, and other non-unitary terms fall outside the method and require a code that integrates the density matrix instead.  Where the density varies continuously, as in the Earth, the profile is discretized and the answer is exact only for the slabs it is given, so the slab count is a precision setting; it can be set directly or, preferably, chosen by the routine from a requested tolerance.  Unusually for a code of this kind, no approximation is made in the probability itself: the result is exact to round-off for whatever Hermitian Hamiltonian is supplied, at two, three, or four flavors, including Hamiltonians for non-standard neutrino physics. \\ \\
%%%%
{\em Running time:}\\
Microseconds to milliseconds per neutrino-oscillation probability, depending on requested precision.  For many evaluations, batched execution flattens the growth curve of execution time.  

\medskip
  
\end{small}

%%%%%%%%%%%%%%%%%%%%%%%%%%%%%%%%%%%%%%%%%%%%%%%%%%%%%
%%%%%%%%%%%%%%%%%%%%%%%%%%%%%%%%%%%%%%%%%%%%%%%%%%%%%

\section{Introduction}\label{sec:introduction}

Neutrinos are created and detected in weak interactions as flavor states---$\nu_e$, $\nu_\mu$, $\nu_\tau$---but they propagate as superpositions of propagation states---in vacuum, these are the mass eigenstates $\nu_1$, $\nu_2$, $\nu_3$.  Because the superposition evolves with time, a neutrino created with a certain flavor has a non-zero probability of being detected later with a different flavor\ \cite{Pontecorvo:1967fh, Barger:1980hs, Bilenky:2004sn, Fantini:2018itu}.  The observation of oscillations in solar, atmospheric, reactor, and accelerator neutrinos has led to the momentous discovery of neutrino mass and of flavor mixing in leptons\ \cite{Kajita:2016cak, McDonald:2016ixn}.

Computing the probabilities of flavor transition is integral to studying oscillations, and computing them exactly typically involves diagonalizing the Hamiltonian that drives the time-evolution.  Because the expressions involved are often complex, it is notoriously hard to produce exact analytical expressions that also provide physical insight.  Oscillations in vacuum and in matter of constant density are the exception\ \cite{Kayser:1981ye, Giunti:2007ry, Tanabashi:2018oca}.  Beyond that, a large body of work derives exact probabilities for particular scenarios---see, \eg, \Refs\ \cite{Barger:1980tf, Petcov:1987zj, Zaglauer:1988gz, TorrenteLujan:1995qi, Balantekin:1997jp, Coleman:1998ti, Gago:2001xg, Kimura:2002wd, Harrison:2003fi, Kostelecky:2003xn, Kostelecky:2003cr, GonzalezGarcia:2004wg, Blennow:2004qd, Meloni:2009ia, Ando:2009ts, Abe:2014wla, Flores:2015mah, Popov:2018seq}---but they are long, and beyond two-flavor treatments their complexity can defeat the insight they were derived for.

More often, carefully selected perturbative expansions are used instead, to cast the probabilities in interpretable form.  Many such approximations exist\ \cite{Gago:2002na, Blennow:2008eb, Diaz:2009qk, Meloni:2009ia, Chaves:2018sih}, especially for oscillations in matter\ \cite{Petcov:1986qg, Hirota:1998wv, Cervera:2000kp, Freund:2001pn, Akhmedov:2004ny, Friedland:2006pi, Ioannisian:2008ve, Supanitsky:2008eq, Minakata:2015gra, Flores:2015mah, Li:2016pzm, Denton:2016wmg, Parke:2019vbs}, reaching the percent level or better.  However, there is no systematic way to produce them (nonetheless, see the calculation program proposed by \Ref\ \cite{Denton:2018fex}): each is tailored to a specific Hamiltonian, its derivation is not trivial, and its validity is limited to a range of some perturbative parameter.  Where high precision is wanted, the better course is to compute the probabilities exactly, numerically.

%%%%%%%%%%%%%%%%%%%%%%%%%%%%%%%%%%%%%%%%%%%%%%%%%%%%%
\begin{figure}[t!]
 \centering
 \includegraphics[width=\columnwidth]{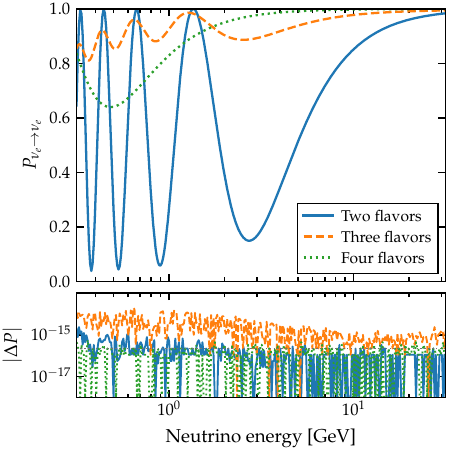}
 \caption{\label{fig:validation}{\it Top}: Survival probability $P_{\nu_e \to \nu_e}$ computed with {\tt NuOscProbExact}\ \cite{NuOscProbExact}.  For each case the same probability was also obtained by direct numerical exponentiation of the same Hamiltonian (using {\tt scipy.linalg.expm} from the SciPy Python library), as reference; the two curves are indistinguishable at this scale, so only one is drawn.  The two-flavor case is in matter, the three-flavor case with non-standard neutrino-matter interactions (NSI), and the four-flavor case is a 3+1 system (one sterile neutrino) in matter with $\Delta m_{41}^2 = 1$~eV$^2$ and $\sin^2\theta_{14} = \sin^2\theta_{24} = 0.1$.  Baselines are $L = 1300$~km for the first two and $L = 600$~m for the third, which is the scale an eV-mass sterile neutrino sets.  {\it Bottom}: The absolute difference between the {\tt NuOscProbExact} and reference calculations, with the dotted line marking the double-precision unit round-off.  The two agree to round-off, which is what ``exact'' means here in practice}
\end{figure}
%%%%%%%%%%%%%%%%%%%%%%%%%%%%%%%%%%%%%%%%%%%%%%%%%%%%%

Figure~\ref{fig:validation} shows what ``exactly'' means here: at two, three, and four flavors (when adding an additional, ``sterile'' neutrino) the probabilities agree with direct numerical exponentiation of the same Hamiltonian---which represents the time-evolution of the neutrino state---to round-off.  Yet exactness alone is not the point, since exponentiating the Hamiltonian numerically is also exact.  The point is what it costs and what it accepts.  We introduce {\tt NuOscProbExact}\ \cite{NuOscProbExact}, a lightweight code that evaluates exact, closed-form expressions for the probabilities; it takes an arbitrary Hermitian Hamiltonian without modification, encoding standard or new neutrino physics, and capable of handling static and varying environments (\ie, Hamiltonians that are time- and position-dependent and -independent).

Three properties recommend {\tt NuOscProbExact}, and they are hard to come by together.  It is \emph{flexible}: any Hermitian Hamiltonian at two, three, or four flavors goes through the same call, so what a probability costs is fixed by the size of the matrix and never by the physics inside it.  It is \emph{fast}: for calculation of oscillation probabilities of neutrinos going through the Earth it is the quickest of the codes we compare, at every accuracy finer than the coarsest setting they share.  And it is \emph{exact}: handed a Hamiltonian, it returns the probability to round-off, four decades finer for neutrinos traveling through constant matter density than the closest of the others reaches.  Only at the coarsest settings through the Earth does a specialized code come out ahead, and at constant density a closed form specialized to three flavors is faster still (Section \ref{sec:comparison}).  That combination suits the exploration of non-standard scenarios, where broad parameter spaces must be scanned without knowing \textit{a priori} which regions matter, and without waiting for a tailored approximation to be derived.

In lieu of diagonalizing the Hamiltonian, {\tt NuOscProbExact} computes exact probabilities by expanding the quantum operators that drive the neutrino time-evolution in terms of SU($n$) matrices, with $n$ the number of flavors. For two and three flavors this approach was developed by Ohlsson \& Snellman (hereafter, OS)\ \cite{Ohlsson:1999xb, Ohlsson:1999um, Ohlsson:2000zz, Ohlsson:2001et}.  More broadly, for any number of flavors, it rests on a nineteenth-century identity known as Sylvester's formula\ \cite{Sylvester:1883}, which bypasses the costly computation of eigenvectors. {\tt NuOscProbExact} implements this for two, three, and four flavors.  The method has two essential assumptions:
\begin{enumerate}
 \item
  The system must be closed, \ie, it must conserve the number of neutrinos summed over all flavors.  Equivalently, the Hamiltonian must be Hermitian for  the evolution to be unitary.  (Sylvester's formula holds without unitarity; the probabilities do not.)
 \item
  The Hamiltonian must be time- or position-independent along the neutrino path.
\end{enumerate}
Both conditions hold in many scenarios studied in the literature---oscillations in vacuum, in matter of constant density, and in diverse new-physics settings---but not in others, notably propagation through the Earth.  For those, we extend the method to Hamiltonians that vary along the path, by treating the density profile as piecewise constant: the approximation is in the profile, never in the propagation through it.

The one insurmountable limitation of the method is that it does not apply to scenarios where neutrinos ``leak out'' of the system.  This includes, \eg, neutrino decays into invisible products\ \cite{Joshipura:2002fb, Beacom:2002cb, Beacom:2002vi, GonzalezGarcia:2008ru, Baerwald:2012kc, Berryman:2014yoa} or open systems, like those with decoherence\ \cite{Benatti:2000ph, Lisi:2000zt, Ohlsson:2000mj, Carpio:2017nui, Coloma:2018idr}.  Whether a system leaks depends in part on how much of it one chooses to describe: written in three flavors a 3+1 sterile scenario appears to leak, and over all four it is closed and unitary. Adding the missing states works when they are neutrinos, and decoherence and invisible decay leave out an environment instead. Their evolution acts on the $n^2$ components of a density matrix, not on the $n$ amplitudes of a state, and no number of extra flavors bridges that. Section \ref{sec:what_to_use} says what to use instead of closed-form expansions.

Section \ref{sec:scope} sets the scope and context.  Section \ref{sec:general_remarks} recaps the basics of oscillations and fixes the goal of the computation.  Section \ref{sec:general_n} presents the construction for any number of flavors, of which the OS method is one reading. Sections \ref{sec:two_flavors}, \ref{sec:three_flavors} and \ref{sec:four_flavors} work it out for two, three, and four flavors.  Section \ref{sec:code} describes {\tt NuOscProbExact}.  Section \ref{sec:examples} shows examples of its use.  Sections \ref{sec:performance} and \ref{sec:comparison} measure its speed and accuracy, and compare it against six other public codes. Section \ref{sec:conclusions} concludes.  The appendices hold supporting detail.

%%%%%%%%%%%%%%%%%%%%%%%%%%%%%%%%%%%%%%%%%%%%%%%%%%%%%
%%%%%%%%%%%%%%%%%%%%%%%%%%%%%%%%%%%%%%%%%%%%%%%%%%%%%

\section{Overview and context}
\label{sec:scope}

%%%%%%%%%%%%%%%%%%%%%%%%%%%%%%%%%%%%%%%%%%%%%%%%%%%%%

\subsection{Overview of our methods}

We compute oscillation probabilities by expanding the evolution operator in the basis of SU($n$) generators: this is the OS method at two and three flavors, and the same construction at four.  References\ \cite{Ohlsson:1999xb, Ohlsson:1999um, Ohlsson:2000zz, Ohlsson:2001et} introduced expressions applicable to generic oscillation scenarios, and also found analytic expressions\ \cite{Ohlsson:1999xb} for the probabilities in the cases of two flavors in matter and three flavors in vacuum and matter.  Yet the problem has a longer history than the OS papers alone.  \Ref\ \cite{Barger:1980tf} gave the three-flavor probabilities in constant matter, \Ref\ \cite{Zaglauer:1988gz} the exact eigenvalues and mixing parameters in matter, and \Refs\ \cite{Kimura:2002wd, Harrison:2003fi} obtained exact three-flavor matter probabilities in a form that, like the one used here, does not require the eigenvectors to be constructed explicitly.  What the present paper adds to that literature is not a new closed form for a particular scenario: the expressions specialized to three flavors in vacuum and in matter are already available and are lengthy, and we neither reproduce nor extend them.  Instead, ours is a general-purpose implementation, at two, three, and four flavors, for a Hamiltonian that is arbitrary rather than chosen in advance.

Later, we work through the method.  Here, we give an overview.  We start by expanding the Hamiltonian in the basis of SU($n$) generators: of $2 \times 2$ Pauli matrices in the case of two flavors, of $3 \times 3$ Gell-Mann matrices in the case of three flavors, or of the $4 \times 4$ generators of SU(4) in the case of four flavors.  \ref{app:matrices} shows these matrices.  (When studying neutrino oscillations, these expansions are sometimes performed not on the Hamiltonian, but on the associated density matrix.  This approach is particularly useful to study oscillations in the early Universe\ \cite{McKellar:1992ja, Bell:1998ds, Hannestad:2001iy, Wong:2002fa, Hannestad:2012ky, Boriero:2017tkh} and in supernovae\ \cite{Pantaleone:1994ns, Sigl:1994hc, Pastor:2002we, Balantekin:2004ug, Fogli:2007bk, Dasgupta:2007ws, Duan:2010bg, Mirizzi:2015eza}.)

In the OS method, we instead first expand the Hamiltonian, $\mathbb{H}$, and then the associated evolution operator, $e^{-i\mathbb{H}L}$. The coefficients of that second expansion come, at two and three flavors, from the exponential identity for Pauli and Gell-Mann matrices \cite{MacFarlane:1968vc, Curtright:2015iba, VanKortryk:2015kua}, which follows from the Cayley-Hamilton theorem, that a matrix satisfies its own characteristic equation.  The consequence used here is that an analytic function of an $n \times n$ matrix is a polynomial of degree at most $n-1$ in that matrix. At four flavors they come from Sylvester's formula, \equ{sylvester}, instead.  Either way they are built from SU($n$) invariants, and that is what bypasses the explicit diagonalization the probabilities would otherwise need.

OS remark that beyond three flavors the characteristic equation must be solved numerically---though the theorem they cite for it allows radicals up to degree four~\cite{Kusnezov:1995, VanKortryk:2015kua}, so their own remark identifies $n=4$ as reachable without carrying it out.  It has been reached since: \Ref~\cite{Kamo:2002sj} obtained the four-flavor case for oscillations in matter.  Three items therefore need separating, so that our paper is not read as claiming more than it does.  The four-flavor expressions are not new, and are not an OS result; we do not present them as either.  The statement they follow from predates both, and the OS expansion is one reading of it among several; it holds at any $n$, and what ends at four is not the statement but the closed form for the eigenvalues it needs.  And what is new here is narrower and practical: an implementation that takes an arbitrary Hermitian Hamiltonian rather than one specialized to matter or to any other scenario fixed in advance.

%%%%%%%%%%%%%%%%%%%%%%%%%%%%%%%%%%%%%%%%%%%%%%%%%%%%%

\subsection{Relation to other oscillation codes}

Oscillation probabilities are already available from several mature codes: {\tt GLoBES}\ \cite{GLoBES}, an experiment-simulation framework; {\tt Prob3++}\ \cite{Prob3pp}, the analytic three-flavor treatment of \Ref\ \cite{Barger:1980tf}; {\tt nuCraft}\ \cite{nuCraft}, a numerical integration along a trajectory through Earth; {\tt NuFast-LBL}\ \cite{Denton:2024pzc} and {\tt NuFast-Earth}\ \cite{Denton:2025oxf}, fast specializations of the three-flavor case at constant density and through Earth; and {\tt nuSQuIDS}\ \cite{Delgado:2014kpa, NuSQuIDS}, the most general of them.  Section \ref{sec:comparison} takes each in turn and measures it, and Section \ref{sec:performance} shows how their runtime performance depends on how many probabilities are asked for at once.  What belongs here is only their relation to {\tt NuOscProbExact}.  {\tt nuSQuIDS} is the near relative: it expands in the same SU($n$) generator basis, but applies it to the density matrix rather than to the state and integrates the resulting equations of motion, a procedure that admits time dependence, interactions, and non-unitary terms, at the cost of solving a differential equation, which is, in general, computationally taxing.  The construction used by {\tt NuOscProbExact} is the analytic kernel of those equations, restricted to closed systems at constant density, as Section \ref{sec:general_n} makes precise.

The claim made for {\tt NuOscProbExact} is a strong one, and it is a claim about a class rather than about a case: inside the closed, constant-Hamiltonian class just described, there is no scenario it does not reach.  That generality costs nothing in accuracy---\textbf{none of the above codes is more accurate than ours anywhere, at any setting it exposes}---and, due to the batched evaluation and compiled kernels of Section \ref{sec:code}, it costs almost nothing in speed: \textbf{only {\tt NuFast-LBL}, on the standard three-flavor case at constant density, and {\tt NuFast-Earth}, at its coarsest setting, are ever quicker}.  One condition comes with all of that, as Section \ref{sec:introduction} states: the Hamiltonian must be Hermitian --- which is closedness in matrix form---and constant over the interval it is applied to; where a density varies continuously, as through Earth, {\tt NuOscProbExact} partitions the profile into piecewise-constant regions, and it is there---and only there---that an approximation enters.  Sections \ref{sec:performance} and \ref{sec:comparison} measure both the generality and its two exceptions.

\medskip

What follows is expository but condensed: enough to present and implement the method; earlier work is cited for the rest.

%%%%%%%%%%%%%%%%%%%%%%%%%%%%%%%%%%%%%%%%%%%%%%%%%%%%%
%%%%%%%%%%%%%%%%%%%%%%%%%%%%%%%%%%%%%%%%%%%%%%%%%%%%%

\section{Neutrino oscillation recap}
\label{sec:general_remarks}

Let $\nu$ represent the flavor state of a neutrino.  The state evolves according to the Schr\"odinger equation,
\begin{equation}
 i \frac{d\nu}{dt} = H \nu \;,
\end{equation}
where $t$ is the time elapsed since the creation of the neutrino and $\mathbb{H}$ is the Hamiltonian written in flavor space.  We use units where $c = \hbar = 1$.  We take $\mathbb{H}$ to be Hermitian, which is what makes the evolution unitary and conserves the number of neutrinos summed over flavors; Section \ref{sec:limitations} takes up what survives when it is not.  In a system of $n$ neutrinos, we represent $\mathbb{H}$ by an $n \times n$ matrix and $\nu$ by a column vector with $n$ entries.  Below, we consider the cases $n = 2$, 3, and 4; we write $\alpha, \beta$ for flavor labels, which run over $e, \mu, \tau$ at $n = 3$ and over $e, \mu, \tau, s$ at $n = 4$, the last being a sterile state.  At $n = 2$ the pair is a matter of naming: a two-flavor system is fixed by one mixing angle and one mass-squared splitting whichever two states are meant, and we call them $e$ and $\mu$ throughout, as the code does.

We restrict the discussion to time- or position-independent Hamiltonians, so that the corresponding evolution operator is $\mathbb{U}\left(t\right) = e^{-i \mathbb{H} t}$.  Hamiltonians of this type describe, for instance, neutrino propagation in vacuum and in matter of constant density; Section \ref{sec:composition} extends this to a path divided into regions, each with its own constant Hamiltonian.  Because neutrinos are relativistic, we approximate the propagated distance $L \simeq t$.  Thus, the evolved state of a neutrino born as $\nu_\alpha$ is
\begin{equation}
 \label{equ:evolved_state}
 \nu_\alpha\left(L\right) 
 = \mathbb{U}(L) \nu_\alpha
 = e^{-i \mathbb{H} L} \nu_\alpha \;.
\end{equation}
Since $\mathbb{H}$ is Hermitian, the evolution operator $\mathbb{U}$ is unitary, and so $\sum_\beta P_{\nu_\alpha \to \nu_\beta} = 1$ for every $\alpha$: the neutrino is found in one of the $n$ flavors, and does not abandon this system.

One simplification is available at the outset and is used throughout.  Split $\mathbb{H} = h_0 \mathbb{1} + \tilde{\mathbb{H}}$ into its trace and its traceless part, where $h_0 = \frac{1}{n}{\rm Tr}\,\mathbb{H}$.  Because $h_0 \mathbb{1}$ commutes with everything, the exponential factorizes exactly, $\mathbb{U}(L) = e^{-i h_0 L} e^{-i \tilde{\mathbb{H}} L}$, and the first factor is a phase common to every amplitude, which cancels in the operation $\lvert \cdot \rvert^2$ that renders the probability.  It may therefore be discarded, and the sections that follow work with $\tilde{\mathbb{H}}$ alone.

Because the Hamiltonian in flavor space is non-diagonal, \ie, because it mixes flavor states, after propagating for a distance $L$, the neutrino of initial flavor $\nu_\alpha$ becomes a superposition of neutrinos of all flavors, each with a different probability amplitude, $\nu_\beta^\dagger \nu_\alpha(L)$.  The probability of detecting the neutrino with flavor $\beta$ is $P_{\nu_\alpha \to \nu_\beta}(L) = \lvert \nu_\beta^\dagger \nu_\alpha(L) \rvert^2$.

Evaluating \equ{evolved_state} means exponentiating $\mathbb{H}$.  Were $\nu_\alpha$ an eigenstate the exponential would act as a phase and there would be nothing more to do, but in flavor space it is not, so the usual procedure is to diagonalize the Hamiltonian, evolve in the basis of its eigenvectors, and rotate back to flavor space to obtain $\nu_\alpha(L)$.  Those steps are often carried out numerically, especially for $n \geq 3$, because the closed forms become unwieldy; {\tt GLoBES}\ \cite{GLoBES, Kopp:2006wp, Huber:2007ji} is one code that does it efficiently, and Section \ref{sec:scope} overviews the others.  The method adopted by {\tt NuOscProbExact}, presented in Section \ref{sec:general_n} below, instead never builds the eigenvectors: it evaluates the exponential directly.

%%%%%%%%%%%%%%%%%%%%%%%%%%%%%%%%%%%%%%%%%%%%%%%%%%%%%
%%%%%%%%%%%%%%%%%%%%%%%%%%%%%%%%%%%%%%%%%%%%%%%%%%%%%

\section{The general-$n$ method}
\label{sec:general_n}

%%%%%%%%%%%%%%%%%%%%%%%%%%%%%%%%%%%%%%%%%%%%%%%%%%%%%

\subsection{The evolution operator}
\label{sec:evolution_operator}

Let $\mathbb{H}$ be an arbitrary Hermitian $n \times n$ Hamiltonian in the flavor basis, constant over a baseline $L$, and let $\tilde{\mathbb{H}}$ be its traceless part.  We write $\psi_m$ for the eigenvalues of $-\tilde{\mathbb{H}}$, the \emph{latent roots}, and $\chi(\psi) = \det(\psi \mathbb{1} + \tilde{\mathbb{H}})$ for their characteristic polynomial.  That sign is a convention and nothing more---it is the one used in the rest of this paper and in the code, and it turns the $e^{-i \tilde{\mathbb{H}} L}$ of Section \ref{sec:general_remarks} into phases of the form $e^{+i \psi_m L}$.  For a non-degenerate eigenvalue spectrum, the evolution operator is
\begin{equation}
 \label{equ:sylvester}
 \mathbb{U}(L)
 =
 \sum_{m=1}^n e^{i\psi_m L} \mathbb{P}_m \;,
 \qquad
 \mathbb{P}_m
 =
 \frac{{\rm adj}\left(\psi_m \mathbb{1} + \tilde{\mathbb{H}}\right)}{\chi^\prime\left(\psi_m\right)} \;,
\end{equation}
with $\mathbb{P}_m$ the orthogonal projector onto the $m$-th eigenspace, and ${\rm adj}(\mathbb{A})$ the adjugate of matrix $\mathbb{A}$ (\ie, the transpose of the cofactor matrix of $\mathbb{A}$).  This is Sylvester's interpolation formula\ \cite{Sylvester:1883} and $\mathbb{P}_m$ are the Frobenius covariants, both nineteenth-century developments in matrix theory.

%%%%%%%%%%%%%%%%%%%%%%%%%%%%%%%%%%%%%%%%%%%%%%%%%%%%%

\subsection{Coefficients from traces, not eigenvectors}
\label{sec:readings}

Equation (\ref{equ:sylvester}) is one object, and several methods in the oscillation literature are readings of it.  Contracting it with the trace metric, $u_0 = \frac{1}{n}{\rm Tr}\,\mathbb{U}$ and $u_a = -\frac{i}{2}{\rm Tr}(\lambda^a \mathbb{U})$---in the convention $\mathbb{U} = u_0 \mathbb{1} + i u_a \lambda^a$ used throughout, \equ{u3_def}---keeps the whole matrix in the generator basis and gives the OS expansion.  Keeping instead its columns gives the adjugate method\ \cite{Sylvester:1883}; keeping its diagonal gives the eigenvector-eigenvalue identity; writing it as a polynomial with Vandermonde-determined coefficients gives the Cayley-Hamilton form in \Ref\ \cite{Ohlsson:1999xb}.  What differs between them is which slice is kept, and in which basis.  Seen this way, the OS coefficients are the SU($n$)-basis components of the adjugate, and another value of $n$ requires no new idea.

Carrying this out, $u_0 = \frac{1}{n}\sum_m e^{i\psi_m L}$ and
\begin{equation}
 \label{equ:ua_general}
 u_a
 =
 \sum_{m=1}^n e^{i\psi_m L}
 \frac{N_a\left(\psi_m\right)}{\chi^\prime\left(\psi_m\right)} \;,~~
 N_a(\psi) \equiv -\tfrac{i}{2}{\rm Tr}\!\left[\lambda^a\,{\rm adj}\left(\psi\mathbb{1}+\tilde{\mathbb{H}}\right)\right] \;.
\end{equation}
Here, $h^a = \frac{1}{2}{\rm Tr}\left(\lambda^a \tilde{\mathbb{H}}\right)$ are the components of the Hamiltonian in the generator basis, so that $\tilde{\mathbb{H}} = h_a \lambda^a$, and $\ast$, the ``star product,'' is the symmetric product those components inherit from the generator algebra, \ie,
\begin{equation}
 \label{equ:star}
 \lambda^a \lambda^b
 =
 \frac{2}{n} \delta^{ab} \mathbb{1}
 + \left(d^{abc} + i f^{abc}\right) \lambda^c \;,
 \quad
 \left(x \ast y\right)^a
 \equiv
 d^{abc} x^b y^c \;,
\end{equation}
where repeated indices are summed over (also in later sections), $f^{abc}$ is the antisymmetric structure constants, and $d^{abc} = \frac{1}{4}{\rm Tr}\left(\left\{\lambda^a,\lambda^b\right\}\lambda^c\right)$ is the totally symmetric $d$ tensor.  For SU(2), the anticommutator of two generators is proportional to the identity, so $d^{abc}$ vanishes identically and the star product with it; this is why the two-flavor case of Section \ref{sec:two_flavors} carries none.

The numerator $N_a$ has degree $n-2$ in $\psi$---the leading $\psi^{n-1}$ of the adjugate multiplies the identity, which $\lambda^a$ traces away---and its coefficients are, up to normalization, $h$, $h \ast h$, $\left(h \ast h\right) \ast h$, and so on: the generator-basis components of successive powers of $\tilde{\mathbb{H}}$, a tower that Cayley-Hamilton closes at $n-1$ rungs, since $\tilde{\mathbb{H}}^n$ is a combination of the powers below it.  Everything needed is a trace: the latent roots follow from traces of powers of $\tilde{\mathbb{H}}$, through the invariants, and the coefficients from traces against the generators.  The $d$ tensor is fixed by the algebra, and mostly zeros (Table~\ref{tab:d_tensor} lists them at $n=3$): each star product takes a few products (three to five at $n = 3$) rather than the $\left(n^2-1\right)^2$ terms of the sum over $b$ and $c$.  And, above all, \textbf{no eigenvectors are ever required}.  This is what makes the method, implemented in {\tt NuOscProbExact}, fast: whatever the Hamiltonian, a probability is the same short run of arithmetic---traces, one root formula, a handful of products---so computationally cheap that Section \ref{sec:performance} finds the call pattern, not the algebra, sets what it costs.

One further reading fixes the relation to codes that propagate a density matrix rather than a state, which is the more general formulation referred to in Section \ref{sec:scope}.  With the evolution generated by the Hamiltonian alone, and that Hamiltonian constant over the interval, the Liouville-von Neumann equation, ${\rm d}\rho/{\rm d}L = -i\left[\mathbb{H}, \rho\right]$, has the closed-form solution $\rho(L) = \mathbb{U}(L) \, \rho(0) \, \mathbb{U}^\dagger(L)$, with $\mathbb{U}$ given by \equ{sylvester}.  What such a method obtains by integrating along the neutrino trajectory is, in that limit, what \equ{sylvester} returns in a single evaluation; what differs between the two is the terms that limit removes---interactions, attenuation, and the open-system terms mentioned in Section \ref{sec:limitations}---not the oscillation itself.

The eigenvalues are the only quantity in \equ{sylvester} that is not a polynomial in the Hamiltonian; with them in hand, the projectors are rational in $\tilde{\mathbb{H}}$ and the rest is the phases $e^{i \psi_m L}$.  That is why the construction extends to any $n$ for which the eigenvalues can be had, and why the accuracy of the whole procedure is set by the accuracy of the roots, $\psi_m$.  Section \ref{sec:beyond_four} takes up what happens where they cannot be had in closed form, at $n \geq 5$.

%%%%%%%%%%%%%%%%%%%%%%%%%%%%%%%%%%%%%%%%%%%%%%%%%%%%%
\begin{figure}[t!]
 \centering
 \includegraphics[width=\columnwidth]{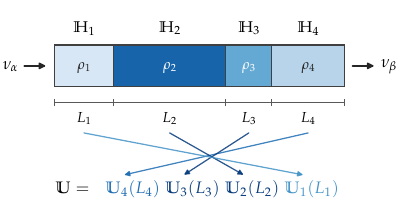}
 \caption{\label{fig:slabs}Composition of the evolution operators of \equ{composition}, for a neutrino crossing a path divided into regions of constant Hamiltonian.  A neutrino enters as $\nu_\alpha$ on the left and leaves as $\nu_\beta$ on the right.  Each region $k$ has its own width $L_k$ and Hamiltonian $\mathbb{H}_k$, and is solved exactly to give its own evolution operator $\mathbb{U}_k(L_k)$; the operators are then multiplied.  Drawn here for the commonest case, in which what changes from one region to the next is the matter density $\rho_k$, though nothing in the construction requires that.  Because the operators act to the left on the initial state, the region crossed \emph{first} is applied first and so appears \emph{rightmost} in the product, $\mathbb{U} = \mathbb{U}_4(L_4) \, \mathbb{U}_3(L_3) \, \mathbb{U}_2(L_2) \, \mathbb{U}_1(L_1)$.  Section \ref{sec:layered_earth} applies this to the internal matter profile of Earth.}
\end{figure}
%%%%%%%%%%%%%%%%%%%%%%%%%%%%%%%%%%%%%%%%%%%%%%%%%%%%%

%%%%%%%%%%%%%%%%%%%%%%%%%%%%%%%%%%%%%%%%%%%%%%%%%%%%%

\subsection{Conjugate Hamiltonians}
\label{sec:conjugation}

One symmetry of the construction saves repeating the whole of it for conjugated Hamiltonians.  Under $\mathbb{H} \to \mathbb{H}^\ast$ the generators divide: those that are real and symmetric are unchanged, those that are imaginary and antisymmetric change sign.  The coefficients of the SU($n$) expansion of the Hamiltonian follow: $h_a \to h_a$ on the first set and $h_a \to -h_a$ on the second.

The latent roots, $\psi_m$, do not change.  The characteristic polynomial of a Hermitian matrix has real coefficients, and conjugation leaves them alone, so $\chi$ and every $\psi_m$ are the same for $\mathbb{H}$ and $\mathbb{H}^\ast$; only the numerators $N_a$ of \equ{ua_general} feel the change, and they feel it as those signs.  As a result, the probabilities for $\mathbb{H}^\ast$ are those for $\mathbb{H}$ with each channel exchanged for its reverse,
\begin{equation}
 \label{equ:conjugation}
 P_{\nu_\alpha \to \nu_\beta}\!\left(\mathbb{H}^\ast\right)
 =
 P_{\nu_\beta \to \nu_\alpha}\!\left(\mathbb{H}\right) \;,
\end{equation}
so a conjugated Hamiltonian needs no new calculation: the tables of Sections \ref{sec:three_flavors} and \ref{sec:four_flavors} are read transposed.  Section \ref{sec:examples} puts this to use, since conjugation is one of the two operations that turn a neutrino Hamiltonian into an antineutrino one.

%%%%%%%%%%%%%%%%%%%%%%%%%%%%%%%%%%%%%%%%%%%%%%%%%%%%%

\subsection{Piece-wise constant Hamiltonians}
\label{sec:composition}

Equation (\ref{equ:sylvester}) does require a constant $\mathbb{H}$, as Section \ref{sec:introduction} states, but only over the interval to which it is applied, and nothing ties that interval to the whole path traveled by the neutrino.  If the path divides into $N$ regions, in the $j$-th of which a Hamiltonian $\mathbb{H}_j$ is constant over a length $L_j$, then \equ{sylvester} applies within each region, or ``slab,'' separately, and the evolution operator for the whole path is their ordered product,
\begin{equation}
 \label{equ:composition}
 \mathbb{U}\left(\left\{L_j\right\}\right)
 =
 \mathbb{U}_N\left(L_N\right) \cdots
 \mathbb{U}_2\left(L_2\right) \,
 \mathbb{U}_1\left(L_1\right) \;,
\end{equation}
so that a neutrino created as $\nu_\alpha$ is detected as $\nu_\beta$ with probability $P_{\nu_\alpha \to \nu_\beta}(\{L_j\}) = \lvert \nu_\beta^\dagger \, \mathbb{U}(\{L_j\}) \, \nu_\alpha \rvert^2$.  

Figure~\ref{fig:slabs} illustrates the composition for four slabs: a neutrino enters as $\nu_\alpha$ on the left, crosses widths $L_1$ to $L_4$ whose densities are constant but different, and leaves as $\nu_\beta$ on the right.  The crossing dashed lines carry the ordering: the region crossed \emph{first} stands \emph{rightmost} in \equ{composition}, since the operators act on the state to their right.  Because the $\mathbb{U}_j$ do not commute in general, writing the product in the opposite order describes a neutrino crossing the same path backwards, and gives a different answer.

Two features of \equ{composition} make it more than a convenience.  It is \emph{exact}: each region is solved exactly by \equ{sylvester} and the operators are composed exactly, so nothing is approximated in the stitching.  And it is indifferent to \emph{why} the Hamiltonians differ.  The familiar case is a change in matter density, but a non-standard interaction that switches on in part of the path, a long-range potential that varies with radius, or any combination of these is the same statement and the same product.

An approximation enters only when the true Hamiltonian varies \emph{continuously} and is replaced by a piecewise constant one---as for a neutrino crossing Earth, whose density profile is smoothly varying within each shell~\cite{Dziewonski:1981xy}.  The approximation is then in the profile, never in the propagation through it, which is why the number of regions is a precision setting rather than a modeling choice, and why refining it converges.  Section \ref{sec:examples} treats that case, which is by far the most common in practice.

%%%%%%%%%%%%%%%%%%%%%%%%%%%%%%%%%%%%%%%%%%%%%%%%%%%%%
%%%%%%%%%%%%%%%%%%%%%%%%%%%%%%%%%%%%%%%%%%%%%%%%%%%%%

\subsection{Why the construction stops at four flavors}
\label{sec:beyond_four}

At $n \geq 5$, the Abel-Ruffini theorem removes the closed-form-in-radicals eigenvalues: the characteristic polynomial, $\chi$, is then a quintic or higher-order polynomial, and the general quintic is not soluble in radicals because the symmetric group $S_5$ is not soluble, while $S_2$, $S_3$, and $S_4$ are.  The obstruction is a theorem, external to neutrino physics entirely.

Three remarks calibrate what is actually lost.  First, a closed-form solution survives, but uselessly: the general quintic is solvable beyond radicals, through the Bring radical and elliptic modular functions or through hypergeometric functions, but those special functions are themselves evaluated iteratively, so nothing is gained over applying Newton's method to the solution of $\chi$.  The radicals were never the point; the eigenvalue-only structure was.  What they give at $n \leq 4$ is a method that can be written down: the roots, and the expressions built on them in Sections \ref{sec:two_flavors}--\ref{sec:four_flavors}, are implementable in any programming language without a linear-algebra library, and checkable by hand.  

Second, and consequently, the practical scheme at $n \geq 5$ is to seed the eigenvalues---from an approximation, or from the closed-form $n \leq 4$ solution of a decoupled sub-block when a hierarchy permits---and polish them with Newton's method; the assembly in \equ{sylvester} then proceeds unchanged.  Third, everything structural continues: \equ{ua_general} needs only that $\psi_m$ be a root of $\chi$, not that it be available in radicals.  What ends at four is the closed-form formula, not the method.

%%%%%%%%%%%%%%%%%%%%%%%%%%%%%%%%%%%%%%%%%%%%%%%%%%%%%
%%%%%%%%%%%%%%%%%%%%%%%%%%%%%%%%%%%%%%%%%%%%%%%%%%%%%

\section{Two-flavor oscillations}
\label{sec:two_flavors}

Two flavors is the natural place to start: it is a good approximation wherever one mass-squared splitting dominates, as in short-baseline reactor experiments and in accelerator and atmospheric analyses restricted to the leading oscillation, and it is the case in which the construction of Section \ref{sec:general_n} is shortest.  Sections \ref{sec:three_flavors} and \ref{sec:four_flavors} carry the same construction to three flavors and to four.  We represent the two-neutrino Hamiltonian operator by a $2 \times 2$ matrix $\mathbb{H}_2$.  The three traceless, Hermitian Pauli matrices $\sigma^k$ ($k=1,2,3$)---with $\sigma^k/2$ the generators of the SU(2) algebra, and written out in \ref{app:matrices}---plus the identity matrix $\mathbb{1}$ form an orthogonal basis of the Hermitian $2 \times 2$ matrices.  Thus, we expand the Hamiltonian as
\begin{equation}
 \label{equ:h2_expansion}
 \mathbb{H}_2 = h_0 \mathbb{1} + h_k \sigma^k \;.
\end{equation}
The coefficients $h_0$ and $h_k$ are functions of the components of the Hamiltonian, and are real because $\mathbb{H}_2$ is Hermitian; we show their explicit expressions in Table \ref{tab:h2_coefficients}.  (For non-Hermitian matrices, the coefficients may be complex, and some of the expressions below would have to exchange $h_i$ for $|h_i|^2$.)  In the two-neutrino case, the neutrino state at any time is $\nu(L) = f_\alpha(L) \nu_\alpha + f_\beta(L) \nu_\beta$, where $f_\alpha$ and $f_\beta$ are, respectively, the probability amplitudes of measuring the state to be a $\nu_\alpha$ or a $\nu_\beta$ (with $\alpha \neq \beta$).  We represent the neutrino state as a two-component column vector; the pure states are $\nu_\alpha = \left(1~0\right)^{\rm T}$ and $\nu_\beta = \left(0~1\right)^{\rm T}$.

%%%%%%%%%%%%%%%%%%%%%%%%%%%%%%%%%%%%%%%%%%%%%%%%%%%%%
{\def\arraystretch{1.2}
\begin{table}[t!]
 \begin{tabular}{ll}
  Coefficient ($2\nu$) & Expression \\
  \hline
  $h_0$ & $\frac{1}{2}\left[(\mathbb{H}_2)_{11}+(\mathbb{H}_2)_{22}\right]$ \\
  $h_1$ & ${\rm Re}\left[ (\mathbb{H}_2)_{12} \right]$ \\
  $h_2$ & $-{\rm Im}\left[ (\mathbb{H}_2)_{12} \right]$ \\
  $h_3$ & $\frac{1}{2}\left[(\mathbb{H}_2)_{11}-(\mathbb{H}_2)_{22}\right]$ \\
 \end{tabular}
 \caption{\label{tab:h2_coefficients}Coefficients in the expansion of the two-neutrino Hamiltonian $\mathbb{H}_2$ in \equ{h2_expansion}.  The coefficient $h_0$ contributes only the global phase, and is included for completeness.}
\end{table}
}
%%%%%%%%%%%%%%%%%%%%%%%%%%%%%%%%%%%%%%%%%%%%%%%%%%%%%

Thus, the evolution operator is $\mathbb{U}_2\left(L\right) = e^{-i \mathbb{H}_2 L}$, and, discarding the global phase as in Section \ref{sec:general_remarks}, $\mathbb{U}_2\left(L\right) = e^{-ih_k\sigma^kL}$.  To compute the action of $\mathbb{U}_2$, we use a well-known identity of Pauli matrices that generalizes Euler's formula, 
\begin{equation}
 \label{equ:pauli_identity}
 e^{\pm i a_k \sigma^k}
 =
 \cos\left(\left\vert a \right\vert \right) \mathbb{1}
 \pm
 i \hat{a}_k\sigma^k\sin\left(\left\vert a\right\vert \right) \;,
\end{equation}
valid for real $a_k$, where $\hat{a}$ is a unit vector in the direction of the vector $a = (a_1, a_2, a_3)$ and $\lvert a \rvert$ is its modulus.  [This is also \equ{sylvester} at $n = 2$, the two projectors being $\mathbb{P}_\pm = \frac{1}{2}\left(\mathbb{1} \pm \hat{a}_k \sigma^k\right)$.]  So we can write the evolution operator as in \Ref\ \cite{Ohlsson:1999xb},
\begin{equation}
 \label{equ:evol_op_2nu}
 \mathbb{U}_2\left(L\right)
 =
 \cos\left(\left\vert h\right\vert L\right) \mathbb{1}
 -
 i \frac{\sin\left(\left\vert h\right\vert L\right)}{\lvert h \rvert} h_k\sigma^k  \;,
\end{equation}
where $\lvert h \rvert^2 \equiv h_1^2 + h_2^2 + h_3^2$ is the squared modulus of the vector $h$.  At $n = 2$ the latent roots of Section \ref{sec:evolution_operator} are $\psi_\pm = \pm \lvert h \rvert$, the eigenvalues of $-\tilde{\mathbb{H}}_2 = -h_k \sigma^k$, and \equ{evol_op_2nu} is the expansion $\mathbb{U}_2 = u_0 \mathbb{1} + i u_k \sigma^k$ [\equ{u3_def} at $n = 2$] with
\begin{equation}
 u_0 = \cos\left(\lvert h \rvert L\right) \;, \qquad
 u_k = -\frac{\sin\left(\lvert h \rvert L\right)}{\lvert h \rvert} h_k \;,
\end{equation}
which is \equ{ua_general} at $n=2$.

One condition attaches to these expressions, as it does to \equ{sylvester}: each divides by $\lvert h \rvert$, and the two latent roots coincide where it vanishes.  At two flavors, that happens only at $h = 0$, with $\tilde{\mathbb{H}}_2$ zero and no oscillation left to compute.  A matter resonance is not such a point, and Table \ref{tab:h2_coefficients} says why: the two-flavor matter potential $\mathbb{A}_2$---collected with the other two-flavor potentials in \ref{app:sample_2nu_hamiltonians}---is diagonal, so it enters $h_0$ and $h_3$ alone and leaves $h_1$ and $h_2$, which carry the off-diagonal entry that does the mixing, exactly where they were.  The resonance is the density at which the potential cancels $h_3$; the mixing is maximal there, but the separation between the roots bottoms out at $2\sqrt{h_1^2 + h_2^2}$ rather than closing.  Section \ref{sec:limitations} says what {\tt NuOscProbExact} returns at $h = 0$, and what the same condition costs at three flavors and four.

The evolved state $\nu_\alpha (L)$ of a neutrino that was created with flavor $\alpha$, \ie, with $f_\alpha(0) = 1$ and $f_\beta(0) = 0$, is $\nu_\alpha\left(L\right) = \mathbb{U}_2(L) \nu_\alpha$.  After some manipulation, the flavor-transition probability $P_{\nu_\alpha \to \nu_\beta}(L) = \lvert \nu_\beta^\dagger \mathbb{U}_2(L) \nu_\alpha \rvert^2$ becomes
\begin{equation}
 \label{equ:prob_2nu_general}
 P_{\nu_\alpha \to \nu_\beta}(L)
 = \frac{h_1^2 + h_2^2}
 {\left\vert h\right\vert^2} \sin^2\left(\left\vert h\right\vert L\right) \;\; (\alpha \neq \beta) \;,
\end{equation}
where $h_1^2 + h_2^2 = \lvert (\mathbb{H}_2)_{12} \rvert^2$ and $\lvert h \rvert^2 = \lvert (\mathbb{H}_2)_{12} \rvert^2 + \left[ (\mathbb{H}_2)_{11} - (\mathbb{H}_2)_{22} \right]^2 / 4$.  Because of the conservation of probability, $P_{\nu_\alpha \rightarrow \nu_\alpha}\left(L\right) = 1 - P_{\nu_\alpha \rightarrow \nu_\beta}\left(L\right)$.  Equation (\ref{equ:prob_2nu_general}) is our final result in the two-neutrino case.  \ref{app:derivation_prob_2nu_general} contains its derivation, \ref{app:2nu_vacuum} shows a simple application to two-flavor oscillations in vacuum, and \ref{app:sample_2nu_hamiltonians} gives the two-flavor Hamiltonians for vacuum, matter, non-standard interactions, and a Lorentz-violating background.

Equation (\ref{equ:prob_2nu_general}) does not distinguish $\alpha$ from $\beta$ on its right-hand side, so $P_{\nu_\alpha \to \nu_\beta} = P_{\nu_\beta \to \nu_\alpha}$ for any $\mathbb{H}_2$.  Nor is there a phase left to violate the CP symmetry with: a rephasing of the two states removes $h_2$, leaving $\mathbb{H}_2$ real and so unmoved by the conjugation of \equ{conjugation}, which returns the probabilities transposed.  That is why CP violation in oscillations needs three flavors.

What made \equ{prob_2nu_general} short was the Pauli-matrix identity, \equ{pauli_identity}.  Section \ref{sec:three_flavors} proceeds the same way, through the analogous identity for the Gell-Mann matrices.

%%%%%%%%%%%%%%%%%%%%%%%%%%%%%%%%%%%%%%%%%%%%%%%%%%%%%
%%%%%%%%%%%%%%%%%%%%%%%%%%%%%%%%%%%%%%%%%%%%%%%%%%%%%

\section{Three-flavor oscillations}
\label{sec:three_flavors}

%%%%%%%%%%%%%%%%%%%%%%%%%%%%%%%%%%%%%%%%%%%%%%%%%%%%%
{\def\arraystretch{1.2}
\begin{table}[t!]
 \begin{tabular}{ll}
  Coefficient ($3\nu$) & Expression \\
  \hline
  $h_0$ & $\frac{1}{3}\left[\left(\mathbb{H}_3\right)_{11}+\left(\mathbb{H}_3\right)_{22}+\left(\mathbb{H}_3\right)_{33}\right]$ \\
  $h_1$ & $\text{Re}\left[{\left(\mathbb{H}_3\right)_{12}}\right]$ \\
  $h_2$ & $-\text{Im}\left[{\left(\mathbb{H}_3\right)_{12}}\right]$ \\
  $h_3$ & $\frac{1}{2}\left[\left(\mathbb{H}_3\right)_{11}-\left(\mathbb{H}_3\right)_{22} \right]$ \\
  $h_4$ & $\text{Re}\left[{\left(\mathbb{H}_3\right)_{13}}\right]$ \\
  $h_5$ & $-\text{Im}\left[{\left(\mathbb{H}_3\right)_{13}}\right]$ \\
  $h_6$ & $\text{Re}\left[{\left(\mathbb{H}_3\right)_{23}}\right]$ \\
  $h_7$ & $-\text{Im}\left[{\left(\mathbb{H}_3\right)_{23}}\right]$ \\
  $h_8$ & $\frac{\sqrt{3}}{6}\left[\left(\mathbb{H}_3\right)_{11}+\left(\mathbb{H}_3\right)_{22}-2\left(\mathbb{H}_3\right)_{33}\right]$
 \end{tabular}
 \caption{\label{tab:h3_coefficients}Coefficients in the expansion of the three-neutrino Hamiltonian $\mathbb{H}_3$ in \equ{h3_expansion}.  The coefficient $h_0$ contributes only the global phase, and is included for completeness.}
\end{table}
}
%%%%%%%%%%%%%%%%%%%%%%%%%%%%%%%%%%%%%%%%%%%%%%%%%%%%%

The three-flavor case runs as the two-flavor one did, both being \equ{sylvester} at their own $n$.  We represent the three-neutrino Hamiltonian by a $3 \times 3$ matrix $\mathbb{H}_3$.  The eight traceless, Hermitian Gell-Mann matrices $\lambda^k$ ($k=1,\ldots,8$)---with $\lambda^k/2$ the generators of the SU(3) algebra, and written out in \ref{app:matrices}---plus the identity matrix $\mathbb{1}$ form an orthogonal basis of the Hermitian $3 \times 3$ matrices.  Thus, we expand the Hamiltonian as
\begin{equation}
 \label{equ:h3_expansion}
 \mathbb{H}_3 = h_0 \mathbb{1} + h_k \lambda^k \;,
\end{equation}
where $h_0$ and $h_k$ are now functions of the components of $\mathbb{H}_3$; we show their explicit expressions in Table \ref{tab:h3_coefficients}.  The neutrino state at any time is $\nu(L) = f_e(L) \nu_e + f_\mu(L) \nu_\mu + f_\tau(L) \nu_\tau$, where $f_e$, $f_\mu$, and $f_\tau$ are, respectively, the probability amplitudes of measuring the state to be a $\nu_e$, $\nu_\mu$, or $\nu_\tau$.  We represent the neutrino state as a three-component column vector; the pure states are $\nu_e = \left(1~0~0\right)^{\rm T}$, $\nu_\mu=\left(0~1~0\right)^{\rm T}$, and $\nu_\tau=\left(0~0~1\right)^{\rm T}$.

The evolution operator is $\mathbb{U}_3\left(L\right) = e^{-i \mathbb{H}_3 L}$, and, discarding the global phase as before, $\mathbb{U}_3\left(L\right) = e^{-ih_k\lambda^kL}$.  We compute the action of $\mathbb{U}_3$ on a neutrino state.  We wish to expand $\mathbb{U}_3$ using an identity for the Gell-Mann matrices that is similar to the identity for the Pauli matrices, \equ{pauli_identity}, and that allows us to write
\begin{equation}
 \label{equ:u3_def}
 \mathbb{U}_3\left(L\right)
 =
 u_0 \mathbb{1} + i u_k \lambda^k \;,
\end{equation}
where the complex coefficients $u_0$ and $u_k$ are functions of $L$ and the $h_k$.  Reference\ \cite{MacFarlane:1968vc} introduced and demonstrated such an identity; below, we make use of it, leaving most of the proofs to the reference.  See also \Refs\ \cite{Lehrer-Ilamed:1964, Rosen:1971nt, Torruella:1975, Kusnezov:1995, Curtright:2015iba, VanKortryk:2015kua} for further details.

The coefficients in \equ{u3_def} are the trace contractions of Section \ref{sec:readings}; next we unpack them.  An application of Sylvester's formula\ \cite{Sylvester:1883} to $3 \times 3$ matrices---that is, of \equ{ua_general} at $n = 3$---expresses them in terms of the two SU(3) invariants,
\begin{eqnarray}
 && I_2 \equiv \left\vert h \right\vert^2 = h_a h^a = \tfrac{1}{2}{\rm Tr}\,\tilde{\mathbb{H}}_3^2 \;, \\
 && I_3 \equiv \langle h \rangle = d^{abc}h_ah_bh_c = \tfrac{1}{2}{\rm Tr}\,\tilde{\mathbb{H}}_3^3 \;,
\end{eqnarray}
where the contractions are how {\tt NuOscProbExact} evaluates the two invariants, and the trace forms are the ones Section \ref{sec:four_flavors} carries over for four flavors.  The tensor is the $d^{abc}$ of the star product in \equ{star}; Table \ref{tab:d_tensor} gives its non-zero components.

Next, we solve the characteristic equation of $-\tilde{\mathbb{H}}_3$, which the invariants write as $\chi\left(\psi\right) = \psi^3 - \left\vert h \right\vert^2 \psi + \frac{2}{3} \langle h \rangle = 0$, with $\chi$ the characteristic polynomial of Section \ref{sec:evolution_operator}.  That form follows from the Cayley-Hamilton theorem, written conveniently in terms of invariants\ \cite{MacFarlane:1968vc, Curtright:2015iba}.  Solving it is a separate step: because $\tilde{\mathbb{H}}_3$ is Hermitian, its three roots are real, and Vi\`ete's trigonometric solution of the depressed cubic returns them as the latent roots $\psi_m$ ($m=1,2,3$),
\begin{equation}
 \label{equ:psi_m}
 \psi_m
 =
 \frac{2 \left\vert h \right\vert}{\sqrt{3}} \cos\left[\frac{1}{3}\left(\varphi+2\pi m \right)\right] \;,
\end{equation}
where $\cos\varphi = -\sqrt{3} \langle h \rangle / \lvert h \rvert^3$, an expression the reality of the roots keeps between $-1$ and $1$.

%%%%%%%%%%%%%%%%%%%%%%%%%%%%%%%%%%%%%%%%%%%%%%%%%%%%%
{\def\arraystretch{1.2}
\begin{table}[t!]
 \begin{tabular}{lc}
  Tensor component ($3\nu$) & Value \\
  \hline
  $d^{(118)} = d^{(228)} = d^{(338)}$                         & $\frac{1}{\sqrt{3}}$ \\
  $d^{(146)} = d^{(157)} = d^{(256)} = d^{(344)} = d^{(355)}$ & $\frac{1}{2}$ \\
  $d^{(247)} = d^{(366)} = d^{(377)}$                         & $-\frac{1}{2}$ \\
  $d^{(448)} = d^{(558)} = d^{(668)} = d^{(778)}$             & $-\frac{1}{2\sqrt{3}}$ \\
  $d^{888}$                                                   & $-\frac{1}{\sqrt{3}}$
 \end{tabular}
 \caption{\label{tab:d_tensor}All of the non-zero components of the totally symmetric tensor $d^{abc}$ of \equ{star}, at $n = 3$.  Here, $a$, $b$, and $c$ can each take integer values between 1 and 8, and the notation $\left( abc \right)$ represents all permutations of the indices in parentheses.  A component vanishes if the number of indices in the set $\left\{ 2,5,7 \right\}$ is odd.}
\end{table}
}
%%%%%%%%%%%%%%%%%%%%%%%%%%%%%%%%%%%%%%%%%%%%%%%%%%%%%

With this, the coefficients in \equ{u3_def} are
\begin{eqnarray}
 && \label{equ:u0}
 u_0=\frac{1}{3}\sum_{m=1}^3 e^{i L \psi_m} \;, \\
 \label{equ:uk_expanded}
 && u_k
 =
 i\sum_{m=1}^3
 e^{i L \psi_m} \frac{\psi_m h_k - \left(h \ast h\right)_k}{3\psi_m^2 - \left\vert h \right\vert^2} \;,
\end{eqnarray}
where $h \ast h$ is the star product of \equ{star}.  \ref{app:derivation_uk_expanded} contains the derivation of \equ{uk_expanded}.  Using Eqs.\ (\ref{equ:psi_m}), (\ref{equ:u0}), and (\ref{equ:uk_expanded}), we write the evolution operator concisely as in \Ref\ \cite{Ohlsson:1999xb},
\begin{equation}
 \label{equ:u3_expanded}
 \mathbb{U}_3(L)
 =
 \sum_{m=1}^3
 e^{i L \psi_m}
 \left[
 \frac{1}{3} \mathbb{1}
 -
 \frac{\psi_m h_k - \left( h \ast h \right)_k}{3\psi_m^2-\lvert h \rvert^2} \lambda^k
 \right] \;.
\end{equation}
Equation~(\ref{equ:u3_expanded}) was introduced in \Refs\ \cite{Ohlsson:1999xb, Ohlsson:1999um, Ohlsson:2000zz}, and applied to find analytic expressions of the probabilities in the cases of oscillations in vacuum and matter.  For a numerical implementation, it is convenient to calculate the coefficients $u_0$ and $u_k$ with Eqs.\ (\ref{equ:u0}) and (\ref{equ:uk_expanded}), and to build the entries of $\mathbb{U}_3$ from them as in \equ{u3_def}.  This is the strategy that we adopt in {\tt NuOscProbExact}\ \cite{NuOscProbExact}: where the evolution operator itself is wanted, as in the composition of \equ{composition}, it is assembled; where only the probabilities are, the entries are squared as they are produced, which is what Table \ref{tab:prob_3nu} calls for.

One condition attaches to Eqs.~(\ref{equ:uk_expanded}) and (\ref{equ:u3_expanded}), as it does to \equ{sylvester}.  Their denominator, $3\psi_m^2 - \lvert h \rvert^2$, is $\chi^\prime(\psi_m)$, and it vanishes when two latent roots coincide, so they hold for a non-degenerate eigenvalue spectrum.  That is not only the trivial case $\lvert h \rvert = 0$, where all three roots vanish together: any two eigenvalues of $\tilde{\mathbb{H}}_3$ may meet at any $\lvert h \rvert$, and a matter resonance is an approach to it.  A third flavor is what opens that possibility: with only two roots, the gap between them is the entire spectral spread, and can close only by the spectrum collapsing altogether, which is why the resonance of Section \ref{sec:two_flavors} bottoms the gap out rather than approaching a crossing.  Where roots do coincide, the interpolation is taken over the distinct ones instead, and Section \ref{sec:limitations} describes how {\tt NuOscProbExact} does that, and what a near-degenerate spectrum costs in accuracy.

%%%%%%%%%%%%%%%%%%%%%%%%%%%%%%%%%%%%%%%%%%%%%%%%%%%%%
{\def\arraystretch{1.3}
\begin{table}[t!]
 \begin{tabular}{lc}
  Three-flavor probability & Expression \\
  \hline
  $P_{\nu_e \to \nu_e}$        & $\left\vert u_0 + i u_3 + i \frac{u_8}{\sqrt{3}} \right\vert^2$ \\
  $P_{\nu_e \to \nu_\mu}$      & $\left\vert i u_1 - u_2 \right\vert^2$ \\
  $P_{\nu_e \to \nu_\tau}$     & $\left\vert i u_4 - u_5 \right\vert^2$ \\
  $P_{\nu_\mu \to \nu_e}$      & $\left\vert i u_1 + u_2 \right\vert^2$ \\
  $P_{\nu_\mu \to \nu_\mu}$    & $\left\vert u_0 - i u_3 + i \frac{u_8}{\sqrt{3}} \right\vert^2$ \\
  $P_{\nu_\mu \to \nu_\tau}$   & $\left\vert i u_6 - u_7 \right\vert^2$ \\
  $P_{\nu_\tau \to \nu_e}$     & $\left\vert i u_4 + u_5 \right\vert^2$ \\
  $P_{\nu_\tau \to \nu_\mu}$   & $\left\vert i u_6 + u_7 \right\vert^2$ \\
  $P_{\nu_\tau \to \nu_\tau}$  & $\left\vert u_0 - i \frac{2 u_8}{\sqrt{3}} \right\vert^2$ \\
 \end{tabular}
 \caption{\label{tab:prob_3nu}Exact three-neutrino oscillation probabilities, for an arbitrary time- or position-independent Hamiltonian.  The complex coefficients $u_0$ and $u_k$ are computed in Eqs.\ (\ref{equ:u0}) and (\ref{equ:uk_expanded}).}
\end{table}
}
%%%%%%%%%%%%%%%%%%%%%%%%%%%%%%%%%%%%%%%%%%%%%%%%%%%%%

The evolved state of a neutrino created as $\nu_\alpha$ is $\nu_\alpha(L) = \mathbb{U}_3(L) \nu_\alpha$.  Therefore, the flavor-transition probability is $P_{\nu_\alpha \to \nu_\beta}(L) = \lvert \nu_\beta^\dagger \mathbb{U}_3(L) \nu_\alpha \rvert^2$.   Table \ref{tab:prob_3nu} shows the expressions for the probabilities in terms of the coefficients $u_0$ and $u_k$.  These are our final results in the three-neutrino case.

The identity that expands the exponential of the Gell-Mann matrices, \equ{u3_expanded}, is longer than the Pauli one it generalizes, \equ{pauli_identity}, and \Ref\ \cite{MacFarlane:1968vc} gives the reason it must be: an exponential parametrization of SU(3) cannot avoid solving a cubic, and \equ{psi_m} is that cubic solved.  What the extra length buys is the whole of the three-flavor case at once, \equ{u3_expanded} holding for any Hermitian $\mathbb{H}_3$.  At four flavors the same steps meet a quartic instead of a cubic, and Section \ref{sec:four_flavors} solves it.

%%%%%%%%%%%%%%%%%%%%%%%%%%%%%%%%%%%%%%%%%%%%%%%%%%%%%
%%%%%%%%%%%%%%%%%%%%%%%%%%%%%%%%%%%%%%%%%%%%%%%%%%%%%

\section{Four-flavor oscillations}
\label{sec:four_flavors}

Sections \ref{sec:two_flavors} and \ref{sec:three_flavors} are \equ{sylvester} at $n=2$ and $n=3$; the expansions there, and the closed forms built on them, are due to OS.  This section is the same construction at $n=4$, which their own remark identifies as reachable without carrying it out (see Section \ref{sec:scope}), and which is the last value at which the eigenvalues can be written in radicals.

%%%%%%%%%%%%%%%%%%%%%%%%%%%%%%%%%%%%%%%%%%%%%%%%%%%%%

\subsection{The four-flavor case}
\label{sec:4nu_case}

%%%%%%%%%%%%%%%%%%%%%%%%%%%%%%%%%%%%%%%%%%%%%%%%%%%%%
{\def\arraystretch{1.2}
\begin{table}[t!]
 \begin{tabular}{ll}
  Coefficient ($4\nu$) & Expression \\
  \hline
  $h_0$    & $\frac{1}{4}\left[\left(\mathbb{H}_4\right)_{11}+\left(\mathbb{H}_4\right)_{22}+\left(\mathbb{H}_4\right)_{33}+\left(\mathbb{H}_4\right)_{44}\right]$ \\
  $h_1$    & $\text{Re}\left[\left(\mathbb{H}_4\right)_{12}\right]$ \\
  $h_2$    & $-\text{Im}\left[\left(\mathbb{H}_4\right)_{12}\right]$ \\
  $h_3$    & $\text{Re}\left[\left(\mathbb{H}_4\right)_{13}\right]$ \\
  $h_4$    & $-\text{Im}\left[\left(\mathbb{H}_4\right)_{13}\right]$ \\
  $h_5$    & $\text{Re}\left[\left(\mathbb{H}_4\right)_{14}\right]$ \\
  $h_6$    & $-\text{Im}\left[\left(\mathbb{H}_4\right)_{14}\right]$ \\
  $h_7$    & $\text{Re}\left[\left(\mathbb{H}_4\right)_{23}\right]$ \\
  $h_8$    & $-\text{Im}\left[\left(\mathbb{H}_4\right)_{23}\right]$ \\
  $h_9$    & $\text{Re}\left[\left(\mathbb{H}_4\right)_{24}\right]$ \\
  $h_{10}$ & $-\text{Im}\left[\left(\mathbb{H}_4\right)_{24}\right]$ \\
  $h_{11}$ & $\text{Re}\left[\left(\mathbb{H}_4\right)_{34}\right]$ \\
  $h_{12}$ & $-\text{Im}\left[\left(\mathbb{H}_4\right)_{34}\right]$ \\
  $h_{13}$ & $\frac{1}{2}\left[\left(\mathbb{H}_4\right)_{11}-\left(\mathbb{H}_4\right)_{22}\right]$ \\
  $h_{14}$ & $\frac{\sqrt{3}}{6}\left[\left(\mathbb{H}_4\right)_{11}+\left(\mathbb{H}_4\right)_{22}-2\left(\mathbb{H}_4\right)_{33}\right]$ \\
  $h_{15}$ & $\frac{\sqrt{6}}{12}\left[\left(\mathbb{H}_4\right)_{11}+\left(\mathbb{H}_4\right)_{22}+\left(\mathbb{H}_4\right)_{33}-3\left(\mathbb{H}_4\right)_{44}\right]$
 \end{tabular}
 \caption{\label{tab:h4_coefficients}Coefficients in the expansion $\mathbb{H}_4 = h_0 \mathbb{1} + h_k \lambda^k$ of the four-neutrino Hamiltonian, in the generator ordering of \ref{app:matrices_4nu}.  The coefficient $h_0$ contributes only the global phase, and is included for completeness.  Two coefficients belong to each flavor pair, the real and the negated imaginary part of one off-diagonal entry; of the three Cartan coefficients, $h_{13}$ and $h_{14}$ are the $h_3$ and $h_8$ of Table \ref{tab:h3_coefficients} unchanged, and $h_{15}$ is the one the fourth flavor adds.}
\end{table}
}
%%%%%%%%%%%%%%%%%%%%%%%%%%%%%%%%%%%%%%%%%%%%%%%%%%%%%

We represent the four-neutrino Hamiltonian by a $4 \times 4$ matrix $\mathbb{H}_4$ and expand it as $\mathbb{H}_4 = h_0 \mathbb{1} + h_k \lambda^k$ ($k = 1, \ldots, 15$) in the SU(4) generators of \ref{app:matrices_4nu}, with the coefficients of Table \ref{tab:h4_coefficients}.  SU(4) has rank three, so $\tilde{\mathbb{H}}_4$ carries three invariants, all obtainable from traces,
\begin{equation}
 \label{equ:invariants_4nu}
 I_2 = \tfrac{1}{2}{\rm Tr}\,\tilde{\mathbb{H}}_4^2 \;, \quad I_3 = \tfrac{1}{2}{\rm Tr}\,\tilde{\mathbb{H}}_4^3 \;, \quad I_4 = \tfrac{1}{2}\left({\rm Tr}\,\tilde{\mathbb{H}}_4^4 - I_2^2\right) \;,
\end{equation}
the first two being the $\lvert h \rvert^2$ and $\langle h \rangle$ of Section \ref{sec:three_flavors}, and $I_4$ the one that SU(4) adds.  The characteristic equation is a quartic, and a depressed one: $\tilde{\mathbb{H}}_4$ is traceless, so its latent roots sum to zero and the cubic term is absent, \ie,
\begin{equation}
 \chi\left(\psi\right) = \psi^4 - I_2\psi^2 + \tfrac{2}{3}I_3\psi + \tfrac{1}{4}\left(I_2^2 - 2I_4\right) = 0 \;.
\end{equation}

Ferrari-Euler reduction builds the roots of such a quartic out of those of an auxiliary cubic, the resolvent,
\begin{equation}
 \label{equ:resolvent_cubic}
 z^3 - 2I_2 z^2 + 2I_4 z - \tfrac{4}{9}I_3^2 = 0 \;.
\end{equation}
Its three roots are the $\left(\psi_j+\psi_k\right)^2$ formed by the three ways of splitting the four latent roots into two pairs.  Each splitting gives two sums, one from each pair, and they add to zero because the four roots do; the two are therefore negatives of each other, and only the square they share is well defined.  Six sums collapse to three squares, which is why the resolvent is a cubic.  Squaring is what makes them well defined: the roots sum to zero, so each splitting gives one pair sum that is minus the other, and only the square is common to the two.  They are real and non-negative  because $\tilde{\mathbb{H}}_4$ is Hermitian, which makes the $\psi_m$ real and the $z_i$ squares of real numbers, so the $\sqrt{z_i}$ below are real as well.  

Depressing \equ{resolvent_cubic} in turn, by $z \to z + \frac{2}{3}I_2$, removes its quadratic term and leaves a cubic of the form already solved at three flavors, so the same trigonometric formula, \equ{psi_m}, returns the three $z_i$, read with $\lvert h \rvert^2 \to \frac{4}{3}I_2^2 - 2I_4$ and $\langle h \rangle \to 2I_2I_4 - \frac{8}{9}I_2^3 - \frac{2}{3}I_3^2$, and with $\frac{2}{3}I_2$ added back at the end.  Each $\sqrt{z_i}$ recovers the pair sum that $z_i$ came from, up to the sign that squaring discarded, and $s_i$ is that sign.  Adding the three pair sums that contain a given latent root returns twice that root---each carries it once, and the other three sum to minus it---so half of their sum is the root itself,
\begin{equation}
 \label{equ:psi_m_4nu}
 \psi_m = \tfrac{1}{2}\left(s_1\sqrt{z_1}+s_2\sqrt{z_2}+s_3\sqrt{z_3}\right) \;,
 \quad
 s_i = \pm 1 \;.
\end{equation}
Not every sign pattern is admissible: $s_1s_2s_3\sqrt{z_1z_2z_3} = -\frac{2}{3}I_3$ pins the product $s_1s_2s_3$, leaving four of the eight, one for each latent root.  The three-flavor machinery sits inside \equ{psi_m_4nu}, which is why the extension needs no new idea, only more of the same arithmetic.

Two accidents that make $n=3$ compact both fail here.  The rank is three rather than two, so $I_4$ appears alongside $I_2$ and $I_3$.  And the SU(3) identity $\left(h \ast h\right) \ast h = \frac{1}{3}\lvert h \rvert^2 h$, which truncates the tower of \equ{ua_general} at one rung, is a Cayley-Hamilton accident of $n=3$: at $n=4$ the theorem constrains $\tilde{\mathbb{H}}_4^4$, and hence $(h \ast h) \ast (h \ast h)$, but leaves $(h \ast h) \ast h$ free, so it must be included.  With that, Eqs.\ (\ref{equ:u0}) and (\ref{equ:uk_expanded}) become
\begin{eqnarray}
 \label{equ:u0_4nu}
 \!\!\!\!\!\!\!\!\!\!
 u_0
 \!\!\!\!&=&\!\!\!\!
 \frac{1}{4}\sum_{m=1}^4 e^{iL\psi_m} \;, \\
 \label{equ:uk_4nu}
 \!\!\!\!\!\!\!\!\!\!
 u_k
 \!\!\!\!&=&\!\!\!\!
 i\sum_{m=1}^4
 e^{iL\psi_m}
 \frac{\left(\psi_m^2 - \frac{1}{2}I_2\right)h_k - \psi_m \left(h \ast h\right)_k + \left(\left(h \ast h\right)\ast h\right)_k}
      {4\psi_m^3 - 2I_2\psi_m + \frac{2}{3}I_3} \;,
\end{eqnarray}
in the convention of Section \ref{sec:evolution_operator}, kept throughout: the $\psi_m$ are the latent roots, the eigenvalues of $-\tilde{\mathbb{H}}_4$, and $\mathbb{U}_4 = u_0\mathbb{1} + iu_k\lambda^k$, which is \equ{u3_def} at $n = 4$.  The denominator is $\chi^\prime(\psi_m)$, exactly as the $3\psi_m^2 - \lvert h \rvert^2$ of \equ{uk_expanded} is.  Comparing the two makes the pattern of \equ{ua_general} concrete: one more power of $\psi_m$ in the numerator, and one more rung of the star tower.

The closed-form SU(4) exponential is not new outside neutrino physics: \Ref\ \cite{VanKortryk:2015kua} gives the SU(4) and SU(5) exponentials directly in the trace invariants of \equ{invariants_4nu}.  Within the four-flavor oscillation problem, the characteristic quartic has been obtained before\ \cite{Kamo:2002sj}.  What we add is the implementation, and the probability table of Table \ref{tab:prob_4nu}.

%%%%%%%%%%%%%%%%%%%%%%%%%%%%%%%%%%%%%%%%%%%%%%%%%%%%%
{\def\arraystretch{1.3}
\begin{table}[t!]
 \begin{tabular}{lc}
  Four-flavor probability & Expression \\
  \hline
  $P_{\nu_e \to \nu_e}$        & $\left\vert u_0 + i u_{13} + i \frac{u_{14}}{\sqrt{3}} + i \frac{u_{15}}{\sqrt{6}} \right\vert^2$ \\
  $P_{\nu_e \to \nu_\mu}$      & $\left\vert i u_1 - u_2 \right\vert^2$ \\
  $P_{\nu_e \to \nu_\tau}$     & $\left\vert i u_3 - u_4 \right\vert^2$ \\
  $P_{\nu_e \to \nu_s}$        & $\left\vert i u_5 - u_6 \right\vert^2$ \\
  $P_{\nu_\mu \to \nu_e}$      & $\left\vert i u_1 + u_2 \right\vert^2$ \\
  $P_{\nu_\mu \to \nu_\mu}$    & $\left\vert u_0 - i u_{13} + i \frac{u_{14}}{\sqrt{3}} + i \frac{u_{15}}{\sqrt{6}} \right\vert^2$ \\
  $P_{\nu_\mu \to \nu_\tau}$   & $\left\vert i u_7 - u_8 \right\vert^2$ \\
  $P_{\nu_\mu \to \nu_s}$      & $\left\vert i u_9 - u_{10} \right\vert^2$ \\
  $P_{\nu_\tau \to \nu_e}$     & $\left\vert i u_3 + u_4 \right\vert^2$ \\
  $P_{\nu_\tau \to \nu_\mu}$   & $\left\vert i u_7 + u_8 \right\vert^2$ \\
  $P_{\nu_\tau \to \nu_\tau}$  & $\left\vert u_0 - i \frac{2 u_{14}}{\sqrt{3}} + i \frac{u_{15}}{\sqrt{6}} \right\vert^2$ \\
  $P_{\nu_\tau \to \nu_s}$     & $\left\vert i u_{11} - u_{12} \right\vert^2$ \\
  $P_{\nu_s \to \nu_e}$        & $\left\vert i u_5 + u_6 \right\vert^2$ \\
  $P_{\nu_s \to \nu_\mu}$      & $\left\vert i u_9 + u_{10} \right\vert^2$ \\
  $P_{\nu_s \to \nu_\tau}$     & $\left\vert i u_{11} + u_{12} \right\vert^2$ \\
  $P_{\nu_s \to \nu_s}$        & $\left\vert u_0 - i \frac{3 u_{15}}{\sqrt{6}} \right\vert^2$ \\
 \end{tabular}
 \caption{\label{tab:prob_4nu}Exact four-neutrino oscillation probabilities, for an arbitrary time- or position-independent Hamiltonian.  The complex coefficients $u_0$ and $u_k$ are computed in Eqs.\ (\ref{equ:u0_4nu}) and (\ref{equ:uk_4nu}), from the $h_k$ of Table \ref{tab:h4_coefficients}.  This is the four-flavor counterpart of Table \ref{tab:prob_3nu}; dropping the entries that involve $\nu_s$ does \emph{not} recover it, because the $u_k$ are different objects at each $n$, even where the $h_k$ are not.  The generator ordering is that of \ref{app:matrices_4nu}.  Reverse channels, and antineutrinos, follow by the exchange of \equ{conjugation}.  Every entry has been verified against direct diagonalization, to $10^{-14}$ over three hundred random Hamiltonians.}
\end{table}
}
%%%%%%%%%%%%%%%%%%%%%%%%%%%%%%%%%%%%%%%%%%%%%%%%%%%%%

%%%%%%%%%%%%%%%%%%%%%%%%%%%%%%%%%%%%%%%%%%%%%%%%%%%%%

\subsection{The probabilities at any $n$}

Tables \ref{tab:prob_3nu} and \ref{tab:prob_4nu} hold just two shapes between them, and fixing the generator basis is what fixes them, at any $n$.  Off the diagonal ($\alpha \neq \beta$), every entry is $\lvert i u_a \mp u_b \rvert^2$: only two generators have an entry in a given off-diagonal slot, one symmetric and one antisymmetric, and $u_a$ and $u_b$ are their coefficients.  The two signs are the two directions: $\lvert i u_a - u_b \rvert^2$ for one channel of that pair ($\nu_\alpha \to \nu_\beta$), and $\lvert i u_a + u_b \rvert^2$ for its reverse ($\nu_\beta \to \nu_\alpha$).  On the diagonal ($\alpha = \beta$), every entry is $\lvert u_0 + i \, c_\alpha \rvert^2$, with $c_\alpha$ a combination of the $n-1$ Cartan coefficients fixed by the flavor $\alpha$.

Three features hold at any $n$.  First, a channel and its reverse differ by $P_{\nu_\alpha \to \nu_\beta} - P_{\nu_\beta \to \nu_\alpha} \equiv 4\,{\rm Im}\left(u_a \bar{u}_b\right)$, so T-violation is carried by the relative phase of the pair's two coefficients, and vanishes when they share one.  (In vacuum, the CP asymmetry equals the T asymmetry, by CPT conservation.  In matter it does not: CP would replace electrons with positrons, and an experiment sends its antineutrinos through the same matter instead.)  Second, unitarity is inherited from $\mathbb{U}$: the rows sum to one, $\sum_\beta P_{\nu_\alpha \to \nu_\beta} = 1$, as do the columns, $\sum_\alpha P_{\nu_\alpha \to \nu_\beta} = 1$.  And, third, the off-diagonal entries can carry $(n-1)(n-2)/2$ physical Dirac phases: none at $n=2$, which is why Section \ref{sec:two_flavors} had none to report, one at $n=3$, and three at $n=4$.

Nothing above singled out a scenario, so a 3+1 system with non-standard interactions or a Lorentz-violating term needs no further derivation, only its own $\mathbb{H}_4$.  \ref{app:sample_2nu_hamiltonians} collects those, Section \ref{sec:sterile_earth} runs one through Earth, and Section \ref{sec:code} turns to the code that evaluates them.

%%%%%%%%%%%%%%%%%%%%%%%%%%%%%%%%%%%%%%%%%%%%%%%%%%%%%
%%%%%%%%%%%%%%%%%%%%%%%%%%%%%%%%%%%%%%%%%%%%%%%%%%%%%

\section{Code description}
\label{sec:code}

%%%%%%%%%%%%%%%%%%%%%%%%%%%%%%%%%%%%%%%%%%%%%%%%%%%%%

\subsection{Scope and interface}

The code {\tt NuOscProbExact} that we provide is a lightweight numerical implementation of the construction of Section \ref{sec:general_n}, \equ{sylvester}: its coefficients come from the OS reading at two (Section~\ref{sec:two_flavors}) and three (Section~\ref{sec:three_flavors}) neutrino flavors, and from \equ{sylvester} itself at four (Section \ref{sec:four_flavors}), where OS does not reach.   It computes exact oscillation probabilities in every one of those cases, for an arbitrary Hermitian Hamiltonian, time- or position-independent.   A Hamiltonian that varies along the path of the neutrino is handled by composing regions of constant Hamiltonian, \equ{composition}, each solved exactly; Sections \ref{sec:layered_earth} and \ref{sec:sterile_earth} do this for the profile of Earth.  The open-system content falls outside---interactions, absorption, and other non-unitary terms---and for it {\tt nuSQuIDS}\ \cite{NuSQuIDS} is the tool; Section \ref{sec:limitations} sets out the rest.  {\tt NuOscProbExact} is fully written in {\tt Python}, and requires version 3.9 or later; it is open source, and publicly available in a GitHub repository\ \cite{NuOscProbExact}.  \ref{app:structure} shows how its modules fit together.

The main input to {\tt NuOscProbExact} is the Hamiltonian matrix $\mathbb{H}_2$, $\mathbb{H}_3$, or $\mathbb{H}_4$, provided as a $2 \times 2$, $3 \times 3$, or $4 \times 4$ array or nested list.  The code internally computes the coefficients $h_a = \frac{1}{2}{\rm Tr}(\lambda^a \tilde{\mathbb{H}})$ of Section \ref{sec:readings}: written out, these are Tables \ref{tab:h2_coefficients}, \ref{tab:h3_coefficients}, and \ref{tab:h4_coefficients} at two, three, and four flavors, respectively.  It then evaluates \equ{prob_2nu_general} at two flavors, and the expressions of Tables \ref{tab:prob_3nu} and \ref{tab:prob_4nu} at three and four.

The same routines also accept a stack of Hamiltonians, an array of baselines, or both, broadcast against each other, and evaluate them without a {\tt Python}-level loop, using instead the fast broadcasting provided by the {\tt NumPy} library.  What that saves depends on how many probability calculations are asked for at once: the energy scans behind Figs.~\ref{fig:prob_3nu} and \ref{fig:prob_2nu} ask for a few thousand, and the oscillograms in energy and arrival direction for tens of thousands (\figu{earth_oscillogram}), where the saving is a factor of thirty to forty over calling the scalar routine in a loop.  A compiled backend based on the {\tt Numba} just-in-time compiler, and installed with the code, gives a further factor of four to seven on top of that, and agrees with the pure-{\tt NumPy} path to round-off; it is used only above the stack sizes where it has been measured to win, which is every stack at three and four flavors and only large ones at two.  Section \ref{sec:performance} quantifies all of this.  By default, the code also checks that the Hamiltonian it is given is Hermitian, and raises an error if it is not, rather than silently returning quantities that still sum to one.

%%%%%%%%%%%%%%%%%%%%%%%%%%%%%%%%%%%%%%%%%%%%%%%%%%%%%

\subsection{Documentation and executable examples}
\label{sec:documentation}

The main source of documentation is the accompanying site\ \cite{NuOscProbExactDocs}, whose source is in the repository beside the code.  Besides an API (application programming interface) reference generated from the source, it carries three things that do not fit in a paper: a methodology section that works through the construction of Sections \ref{sec:general_n}--\ref{sec:four_flavors} in the form in which it is implemented, including the conditioning behavior discussed below; a set of numerical recipes for the cases users actually meet, including layered profiles, oscillograms, antineutrinos, parameter scans; and twenty executable notebooks, one of which generates every figure in this paper, listed in \ref{app:notebooks}.  

%%%%%%%%%%%%%%%%%%%%%%%%%%%%%%%%%%%%%%%%%%%%%%%%%%%%%

\subsection{Installation}

The latest released version of the code (v1.13.1 at the time of writing) is on the Python Package Index\ \cite{NuOscProbExactPyPI}, from which it is installed with
\begin{verbatim}
pip install nuoscprobexact
\end{verbatim}
Its dependencies are {\tt NumPy} and {\tt Numba}, so a plain install carries the compiled backend.

Installing from a clone of the repository\ \cite{NuOscProbExact} instead brings what the package omits --- the notebooks of \ref{app:notebooks} and the test suite of \ref{app:tests}:
\begin{verbatim}
git clone \
  https://github.com/mbustama/NuOscProbExact.git
cd NuOscProbExact
pip install -e .
\end{verbatim}
and is also the way to a version that has not yet been released.

%%%%%%%%%%%%%%%%%%%%%%%%%%%%%%%%%%%%%%%%%%%%%%%%%%%%%

\subsection{Code availability}

Development is version-controlled and happens in the public GitHub repository\ \cite{NuOscProbExact}, which carries the source, test suite, notebooks, and change log, and from which the code may also be installed (see above).  It is released under the MIT license.

%%%%%%%%%%%%%%%%%%%%%%%%%%%%%%%%%%%%%%%%%%%%%%%%%%%%%

\subsection{Code validation}
\label{sec:validation}

The repository carries a test suite of 882 tests, which cover every line and branch of the library in v1.13.1 of {\tt NuOscProbExact}; \ref{app:tests} groups them by what they cover.  Beyond internal consistency, the results are checked against independent sources: against {\tt nuSQuIDS}\ \cite{Delgado:2014kpa, NuSQuIDS}, an independently written code, with which the three-flavor probabilities agree to $10^{-11}$ and the four-flavor ones to $10^{-9}$ once conventions are matched; against the closed-form eigenvalues of Zaglauer \& Schwarzer\ \cite{Zaglauer:1988gz} for the three-flavor matter spectrum; against direct numerical exponentiation of the Hamiltonian, as in \figu{validation}; and, in the two-flavor vacuum case, against the standard analytic formula.

Where a residual remains---as at four flavors, where a large $\Delta m_{41}^2$ crowds three of the four latent roots together and leaves the fourth far off, the near-degeneracy that Section \ref{sec:limitations} quantifies---the wider tolerance it needs is not left to stand on its own.  A third reference settles where the residual belongs.  It could have either of two sources: the accuracy limit of the closed form near a near-degenerate spectrum, or a convention the two codes match imperfectly.  Against direct exponentiation of the same Hamiltonian, the error of the expansion and its disagreement with {\tt nuSQuIDS} come out the same size, which leaves only the first: {\tt nuSQuIDS} sits where the exponential does, and the whole residual is from the closed form.

%%%%%%%%%%%%%%%%%%%%%%%%%%%%%%%%%%%%%%%%%%%%%%%%%%%%%

\subsection{Limitations}
\label{sec:limitations}

Limitations are of two kinds.  Those of the method are assumptions that no better programming would remove; those of the implementation are contingent on choices made in the code.

%%%%%%%%%%%%%%%%%%%%%%%%%%%%%%%%%%%%%%%%%%%%%%%%%%%%%

\subsubsection{Limitations of the method}

The method assumes what Section \ref{sec:introduction} states: a closed system, and a Hamiltonian constant over each interval that \equ{sylvester} is applied to (which, by \equ{composition}, need not be the whole path).  What follows are the cases they exclude.

\smallskip

\textit{Open systems.---}Decay to invisible products and genuinely open systems--- decoherence, and any Lindblad term--- fall outside the method, and require the density-matrix formalism that {\tt nuSQuIDS}\ \cite{Delgado:2014kpa, NuSQuIDS} integrates, at a structural cost: a Lindbladian lives on the $n^2$-dimensional space of operators, so it is diagonalized in $O(n^6)$ against $O(n^3)$ for a Hamiltonian, and any integrator must resolve the oscillation phase, its step count growing with $\Delta m^2 L / 2E$ where \equ{sylvester} is evaluated once.

\smallskip

\textit{Non-Hermitian Hamiltonians.---}Used to describe absorption, these are a subtler case: \equ{sylvester} survives formally, but everything downstream that used the reality of the eigenvalue spectrum does not, including the trigonometric root formulas and the identification of the moduli with probabilities.

\smallskip

\textit{Continuously varying densities.---}A density that varies continuously is handled here only by approximating it as piecewise constant.  For a smooth profile, a solver based on the Magnus expansion of the evolution operator is the better tool, and the companion code {\tt Magnus}\ \cite{Magnus} is one.

\smallskip

\textit{Beyond four flavors.---}As Section \ref{sec:beyond_four} states, the eigenvalues can be written in radicals only up to four flavors.  What stops there is the closed form for the roots, not the construction: \equ{ua_general} asks only for a root of $\chi$, however obtained, and at $n \geq 5$ it is obtained numerically, though {\tt NuOscProbExact} implements two, three, and four ({\tt Magnus}\ \cite{Magnus} goes to five).

%%%%%%%%%%%%%%%%%%%%%%%%%%%%%%%%%%%%%%%%%%%%%%%%%%%%%

\subsubsection{Limitations of the implementation}

\textit{Accuracy near a level crossing.---}The closed form is not the most accurate route near an eigenvalue-level crossing, which is where resonances are.  Forming the characteristic polynomial from traces and then extracting its roots costs a power of the relative spectral gap: if $r$ eigenvalues cluster, the error in the roots grows as $g^{-(r-1)}$, with $g$ the gap in units of the spectral spread.  That is the classical instability of taking eigenvalues from a characteristic polynomial, and it is repairable neither by a more careful evaluation nor by the usual divided-difference remedy, the error being already in the roots.  What does remove it is carrying more digits into it.  

At four flavors, where a 3+1 spectrum crowds three roots together and $g^{-(r-1)}$ is correspondingly large, the code forms the invariants in $32$-digit arithmetic by default: measured at $2.3 \cdot 10^9$, that factor then multiplies $10^{-32}$ rather than $10^{-16}$ and lands at $10^{-23}$, so what limits the roots is the final rounding and not the algebra.  They are checked (on every push to the GitHub code repository) against fifty-digit reference values over spectra reaching down to a pair of roots separated by $10^{-16}$.  A backward-stable eigensolver, whose error does not depend on the gap at all, is carried beside it as a second route to the roots.  (Fewer flavors need less of this.  At three, only two roots can approach, which makes the exponent one rather than two, and the gap stays far wider than a large $\Delta m_{41}^2$ leaves it at four.  At two, the gap between the roots is the whole spectral spread however small the spread becomes, so $g = 1$ always and there is nothing to amplify: \equ{prob_2nu_general} holds to $8 \cdot 10^{-16}$ in relative terms over twelve orders of magnitude in $\lvert h \rvert L$.)

\smallskip

\textit{Exact degeneracy.---}This is handled by clustering roots that lie within a tolerance of one another and interpolating over distinct clusters; without that step the formulas divide by zero.  At two flavors, that is the whole of it: the latent roots $\pm \lvert h \rvert$ meet only at $h = 0$, where $\mathbb{H}_2$ is a multiple of the identity, nothing oscillates, and the code returns $P_{\nu_\alpha \to \nu_\beta} = 0$ without evaluating a formula that would divide by zero.

%%%%%%%%%%%%%%%%%%%%%%%%%%%%%%%%%%%%%%%%%%%%%%%%%%%%%
%%%%%%%%%%%%%%%%%%%%%%%%%%%%%%%%%%%%%%%%%%%%%%%%%%%%%

\section{{\tt NuOscProbExact} examples}
\label{sec:examples}

Listing\ \ref{lst:3nu_examples} shows a basic code example of how to use {\tt NuOscProbExact} to compute three-neutrino probabilities in four representative oscillation scenarios: in vacuum, in matter of constant density, with non-standard interactions in matter, and with Lorentz invariance violation.  Bundled with the code we provide further examples, also for two-neutrino oscillations.

Below, we introduce each of them briefly, with references for the phenomenology we do not explore.  As Section \ref{sec:scope} states, we derive no analytic expression for any of them; each is evaluated numerically, from its Hamiltonian.

We also go beyond those four scenarios: antineutrinos, which need no new construction, only the two operations of \equ{antineutrino}; matter of piecewise constant density and the profile of Earth, and a 3+1 sterile system, which extend the method's reach to matter that varies along the trajectory and to a fourth flavor; and a Hamiltonian of the reader's own, which is what the code was built for and for which no specialized expression exists in any of the codes of Section \ref{sec:comparison}.  Every figure that shows computed probabilities comes from notebook \#10 in the repository.

%%%%%%%%%%%%%%%%%%%%%%%%%%%%%%%%%%%%%%%%%%%%%%%%%%%%%
\begin{lstlisting}[language=Python, caption={Code snippet to use {\tt NuOscProbExact} to compute three-neutrino oscillation probabilities in vacuum, matter of constant density, with non-standard interactions, and with CPT-odd Lorentz-violating background, for fixed neutrino energy $E = 1$~GeV and baseline $L = 1300$~km.  Constants --- with variable names in capitals --- are pulled from the {\tt globaldefs} module; see the main text for their values.  The printed output is what v1.13.1 produces.}, label={lst:3nu_examples}, float=tbp]
import numpy as np

# NuOscProbExact modules
import oscprob3nu         # Core functionality
import hamiltonians3nu    # Sample Hamiltonians
from globaldefs import *  # Constants (in capitals)

energy = 1.e9     # Neutrino energy [eV]
baseline = 1.3e3  # Baseline [km]

# Vacuum Hamiltonian before multiplying by 1/energy
# NO: "normal ordering"; IO: inverted ordering
h_vacuum_energy_indep = hamiltonians3nu.\
    hamiltonian_3nu_vacuum_energy_independent( \
        S12_NO_BF, S23_NO_BF,
        S13_NO_BF, DCP_NO_BF,
        D21_NO_BF, D31_NO_BF)

# Hamiltonian for oscillations in vacuum
h_vacuum = np.multiply( 1./energy, 
                        h_vacuum_energy_indep)

# Hamiltonian for oscillations in matter                        
# VCC_EARTH_CRUST: Potential [eV], density 3 g cm^{-3}
h_matter = hamiltonians3nu.hamiltonian_3nu_matter( \
    h_vacuum_energy_indep, energy, VCC_EARTH_CRUST)

# Hamiltonian for non-standard interactions
# EPS_3 is the list of NSI strength parameters
# [EPS_EE, EPS_EM, EPS_ET, EPS_MM, EPS_MT, EPS_TT]
h_nsi = hamiltonians3nu.hamiltonian_3nu_nsi( \
    h_vacuum_energy_indep, energy, VCC_EARTH_CRUST, 
    EPS_3)

# Hamiltonian for Lorentz-invariance violation
# LIV parameters: SXI12, SXI23, SXI13, DXICP, B1,
# B2, B3, LAMBDA
h_liv = hamiltonians3nu.hamiltonian_3nu_liv( \
    h_vacuum_energy_indep, energy,
    SXI12, SXI23, SXI13, DXICP, B1, B2, B3, LAMBDA)

# The routine probabilities_3nu computes probabilities
for h_matrix in [h_vacuum, h_matter, h_nsi, h_liv]:
    # CONV_KM_TO_INV_EV converts km to eV^{-1}
    Pee, Pem, Pet, Pme, Pmm, Pmt, Pte, Ptm, Ptt = \
        oscprob3nu.probabilities_3nu( \
                            h_matrix,
                            baseline*CONV_KM_TO_INV_EV)

    print("Pee = %6.5f, Pem = %6.5f, Pet = %6.5f" \
        % (Pee, Pem, Pet))
    print("Pme = %6.5f, Pmm = %6.5f, Pmt = %6.5f" \
        % (Pme, Pmm, Pmt))
    print("Pte = %6.5f, Ptm = %6.5f, Ptt = %6.5f" \
        % (Pte, Ptm, Ptt))
    print()
    
# This returns:
# Pee = 0.92768, Pem = 0.01432, Pet = 0.05800
# Pme = 0.04023, Pmm = 0.37887, Pmt = 0.58090
# Pte = 0.03210, Ptm = 0.60680, Ptt = 0.36110

# Pee = 0.95262, Pem = 0.00623, Pet = 0.04114
# Pme = 0.02590, Pmm = 0.37644, Pmt = 0.59766
# Pte = 0.02148, Ptm = 0.61733, Ptt = 0.36119

# Pee = 0.92494, Pem = 0.01758, Pet = 0.05749
# Pme = 0.03652, Pmm = 0.32523, Pmt = 0.63826
# Pte = 0.03855, Ptm = 0.65719, Ptt = 0.30426

# Pee = 0.92721, Pem = 0.05299, Pet = 0.01980
# Pme = 0.05609, Pmm = 0.25288, Pmt = 0.69103
# Pte = 0.01670, Ptm = 0.69412, Ptt = 0.28917
\end{lstlisting}
%%%%%%%%%%%%%%%%%%%%%%%%%%%%%%%%%%%%%%%%%%%%%%%%%%%%%

%%%%%%%%%%%%%%%%%%%%%%%%%%%%%%%%%%%%%%%%%%%%%%%%%%%%%
\begin{figure}[t!]
 \centering
 \includegraphics[width=\columnwidth]{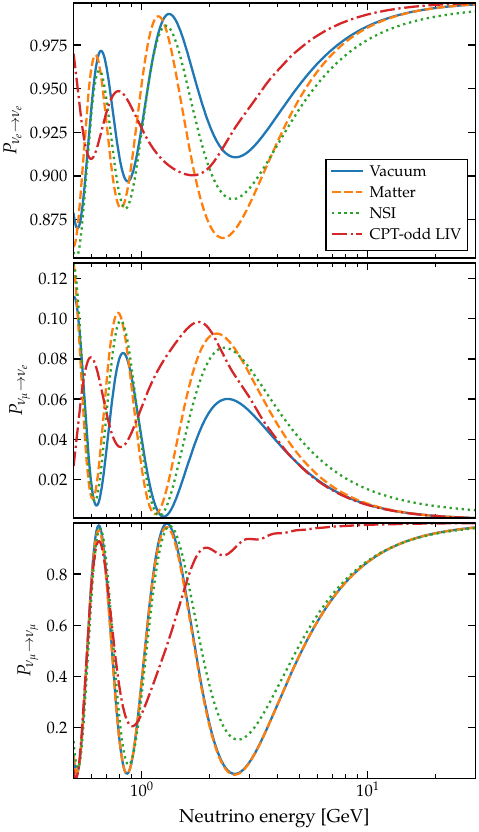}
 \caption{\label{fig:prob_3nu}Three-neutrino oscillation probabilities $P_{\nu_e \to \nu_e}$ ({\it top}), $P_{\nu_\mu \to \nu_e}$ ({\it center}), and $P_{\nu_\mu \to \nu_\mu}$ ({\it bottom}), computed using {\tt NuOscProbExact}\ \cite{NuOscProbExact}.  The scenarios shown are for oscillations in vacuum, in matter of constant density, with non-standard interactions (NSI), and in a CPT-odd Lorentz-violating background (LIV).  In all cases, the baseline is $L = 1300$~km.  See the main text for details. \vspace*{-0.5cm}}
 \vspace*{0.03cm}
\end{figure}
%%%%%%%%%%%%%%%%%%%%%%%%%%%%%%%%%%%%%%%%%%%%%%%%%%%%%

Figure \ref{fig:prob_3nu} shows the probabilities $P_{\nu_e \to \nu_e}$, $P_{\nu_\mu \to \nu_e}$, and $P_{\nu_\mu \to \nu_\mu}$ for the four scenarios, as a function of neutrino energy, computed using {\tt NuOscProbExact}\ \cite{NuOscProbExact}.  We set the baseline to $L = 1300$~km to match that of the far detector of the planned DUNE experiment\ \cite{Abi:2018dnh}.  The parameters and their values used in each example case are introduced below.  All of the Hamiltonians below are written in the flavor basis.

\ref{app:sample_2nu_hamiltonians} shows the two- and four-flavor counterparts of the example three-neutrino scenarios presented below.  We provide implementations of all these scenarios as part of {\tt NuOscProbExact}\ \cite{NuOscProbExact}.  

%%%%%%%%%%%%%%%%%%%%%%%%%%%%%%%%%%%%%%%%%%%%%%%%%%%%%

\subsection{Oscillations in vacuum}

The Hamiltonian that drives oscillations in vacuum is
\begin{equation}
 \label{equ:h3_vacuum}
 \mathbb{H}_3^{\rm vac}(E)
 = 
 \frac{1}{2E} \left( \mathbb{R}_{3,\theta} \mathbb{M}_3^2 \mathbb{R}_{3,\theta}^\dagger \right) \;,
\end{equation}
where $\mathbb{M}_3^2 \equiv {\rm diag}(0, \Delta m_{21}^2, \Delta m_{31}^2)$ is the mass-squared matrix, with $\Delta m_{21}^2 \equiv m_2^2-m_1^2$ and $\Delta m_{31}^2 \equiv m_3^2-m_1^2$, and the $3 \times 3$ complex rotation matrix $\mathbb{R}_{3,\theta}$ is the Pontecorvo-Maki-Nakagawa-Sakata (PMNS) mixing matrix.  We express it in terms of three mixing angles, $\theta_{13}$, $\theta_{12}$, $\theta_{23}$, and one CP-violation phase, $\delta_{\rm CP}$.  Conventions may differ between codes, so we write ours out.  With $c_{ij} \equiv \cos\theta_{ij}$ and $s_{ij} \equiv \sin\theta_{ij}$, and $\delta \equiv \delta_{\rm CP}$,
\begin{equation}
 \label{equ:pmns}
 \mathbb{R}_{3,\theta}
 =
 \setlength{\arraycolsep}{2pt}
 \left(
 \begin{array}{ccc}
  c_{12}c_{13} & s_{12}c_{13} & s_{13}e^{-i\delta} \\
  -s_{12}c_{23}-c_{12}s_{23}s_{13}e^{i\delta} & c_{12}c_{23}-s_{12}s_{23}s_{13}e^{i\delta} & s_{23}c_{13} \\
  s_{12}s_{23}-c_{12}c_{23}s_{13}e^{i\delta} & -c_{12}s_{23}-s_{12}c_{23}s_{13}e^{i\delta} & c_{23}c_{13}
 \end{array}
 \right) \;,
\end{equation}
which is the standard parametrization of \Ref\ \cite{Tanabashi:2018oca}.  In {\tt NuOscProbExact}, every routine that takes an angle takes an {\tt angles} argument, so that a published parameter set can be given in the form it was published in: {\tt sin} (sine of the angle) by default, or {\tt sin2} (sine squared), {\tt rad}, or {\tt deg}.  The last two affect $\delta_{\rm CP}$; the first two leave it in radians, a phase having no sine to pass, since $\sin\delta_{\rm CP}$ would lose its quadrant.  In Listing\ \ref{lst:3nu_examples}, what {\tt hamiltonian\_3nu\_vacuum\_energy\_independent} returns is $\tfrac{1}{2} \mathbb{R}_{3,\theta} \mathbb{M}_3^2 \mathbb{R}_{3,\theta}^\dagger$, with the factor of $\tfrac{1}{2}$ in \equ{h3_vacuum} folded in, which is why the listing divides it by $E$ and not by $2E$.

To compute the probabilities in \figu{prob_3nu}, and in the rest of the figures in this paper, we fix the mixing parameters to their best-fit values from the NuFit 4.0 global fit to oscillation data\ \cite{Esteban:2018azc, NuFit4.0}, assuming normal mass hierarchy, and including Super-Kamiokande atmospheric neutrino data: $\sin^2 \theta_{12} = 0.310$, $\sin^2 \theta_{23} = 0.582$, $\sin^2 \theta_{13} = 0.02240$, $\delta_{\rm CP} = 217^\circ$, $\Delta m_{21}^2 = 7.39 \cdot 10^{-5}$~eV$^2$, $\Delta m_{31}^2 = 2.525 \cdot 10^{-3}$~eV$^2$.  Global fits have since moved on, but the default values carried by the code remain pinned to NuFit 4.0, so that the figures and examples in this paper stay reproducible.  Users are expected to supply current values for their own analyses.

%%%%%%%%%%%%%%%%%%%%%%%%%%%%%%%%%%%%%%%%%%%%%%%%%%%%%

\subsection{Oscillations in matter of constant density}
\label{sec:prob_3nu_matter}

When neutrinos propagate in matter, $\nu_e$ and $\bar{\nu}_e$ scatter on electrons via charged-current interactions.  The interactions introduce potentials that shift the energies of the neutrinos.  As a result, the values of the mass-squared differences and mixing angles in matter differ from their values in vacuum, and depend on the number density of electrons\ \cite{Wolfenstein:1977ue, Mikheev:1986gs, Bethe:1986ej, Parke:1986jy, Rosen:1986jy}.  Computing oscillation probabilities in constant matter is integral to long-baseline experiments, where neutrinos traverse hundreds of kilometers in the crust of the Earth to reach the detectors\ \cite{Feldman:2013vca, Diwan:2016gmz}.

The Hamiltonian that drives oscillations in matter is
\begin{equation}
 \mathbb{H}_3^{\rm matt}(E)
 = 
 \mathbb{H}_3^{\rm vac}(E) + \mathbb{A}_3 \;.
\end{equation}
The term $\mathbb{A}_3 \equiv {\rm diag}(V_{\rm CC}, 0 ,0)$ is due to interactions with matter, where $V_{\rm CC} = \sqrt{2} G_{\rm F} n_e$ is the charged-current potential, $G_F$ is the Fermi constant, and $n_e$ is the number density of electrons.

To compute the probabilities for oscillations in matter (and also with non-standard interactions) in \figu{prob_3nu}, we consider a constant matter density of $\rho = 3$~g~cm$^{-3}$, which is the mean density along a $1300$-km chord through the outer Earth: such a path reaches $33$~km down, below the crust, so about half of it lies in the upper mantle, and \Ref\ \cite{Dziewonski:1981xy} gives $2.6$ to $2.9$ for the crust against $3.4$ just beneath it.  The number density of electrons is $n_e = Y_e \rho / \bar{m}$, with $\bar{m} = Y_e m_p + (1 - Y_e) m_n$ the mean nucleon mass, $m_p$ and $m_n$ the masses of the proton and neutron, and $Y_e$ the electron fraction.  Here $Y_e = 0.5$: the crust is electrically neutral---so its electron fraction equals its proton fraction---and isoscalar, so that fraction is one half---for which $\bar{m}$ is $(m_p+m_n)/2$.\footnote{Dividing by a free-nucleon mass, rather than by the atomic mass unit, neglects the binding energy of nucleons inside nuclei.  This undercounts $n_e$, and hence $V_{\rm CC}$, by about $0.8\%$.  The code does the same, so that it matches the expression above; the effect is far below the precision at issue in the examples, but it should be known before the code is used where that level of accuracy matters.}  See \Refs\ \cite{Ohlsson:1999xb, Ohlsson:1999um} for the analytic form of the probabilities in matter, deduced with the OS method, and \Ref\ \cite{Freund:2001pn} for a related approximation.  In long-baseline experiments, even if there are density changes along the trajectory of the neutrino beam, using the average density is a good approximation\ \cite{Petcov:1998su, Freund:1999gy}---for a shallow chord through layers of similar density, which is what a long baseline is; Section \ref{sec:layered_earth} shows where averaging fails instead.

%%%%%%%%%%%%%%%%%%%%%%%%%%%%%%%%%%%%%%%%%%%%%%%%%%%%%

\subsection{Oscillations with non-standard interactions}

Oscillations in matter might receive sub-leading contributions due to new neutrino interactions with the fermions of the medium that they propagate in.  These are known as non-standard interactions (NSI); see \Refs\ \cite{Ohlsson:2012kf, Miranda:2015dra, Farzan:2017xzy, Esteban:2018ppq} for reviews.  

In this case, the Hamiltonian is
\begin{equation}
 \mathbb{H}_3^{\rm NSI}(E)
 =
 \mathbb{H}_3^{\rm vac}(E) + \mathbb{A}_3 + \mathbb{V}_3  \;,
\end{equation}
where $\mathbb{V}_3 \equiv V_{\rm CC} \epsilon_3$ is the matter potential due to NSI and $\epsilon_3$ is the matrix of NSI strength parameters, \ie,
\begin{equation}
 \label{equ:epsilon_3}
 \epsilon_3
 =
 \left(
  \begin{array}{ccc}
   \epsilon_{ee}         & \epsilon_{e\mu}         & \epsilon_{e\tau} \\
   \epsilon_{e\mu}^\ast  & \epsilon_{\mu\mu}       & \epsilon_{\mu\tau} \\
   \epsilon_{e\tau}^\ast & \epsilon_{\mu\tau}^\ast & \epsilon_{\tau\tau} \\
  \end{array}
 \right) \;.
\end{equation}
The parameters $\epsilon_{\alpha\beta}$ represent the total strength of the NSI between leptons of flavors $\alpha$ and $\beta$ interacting with the electrons, $u$ quarks, and $d$ quarks that make up standard matter.  Following \Ref\ \cite{Esteban:2018ppq}, we write $\epsilon_{\alpha\beta} = \epsilon_{\alpha\beta}^e + (2+ Y_n)\epsilon_{\alpha\beta}^u + (1+2Y_n)\epsilon_{\alpha\beta}^d$, with the ratio of the number densities of neutrons to electrons $Y_n \equiv n_n/n_e \approx 1$ in the Earth.  In our simplified treatment, we do not consider separately interactions with each fermion type or each chiral projection of the fermion\ \cite{Ohlsson:2012kf, Miranda:2015dra, Farzan:2017xzy, Esteban:2018ppq}.  The simplification is in the values rather than in the code, which takes the combined $\epsilon_{\alpha\beta}$: a reader who wants each fermion type, or each chiral projection, weighted separately forms the sum above themselves and passes the result.

To compute the probabilities for NSI in \figu{prob_3nu}, we again consider propagation at the same constant density as in \ref{sec:prob_3nu_matter}, with $V_{\rm CC}$ evaluated as explained there.  Because NSI have not been observed, we choose illustrative values for the strength parameters, of the order of those explored in global fits to oscillation plus COHERENT data\ \cite{Esteban:2018ppq} (see also \Refs\ \cite{Mitsuka:2011ty, Esmaili:2013fva, Salvado:2016uqu, Aartsen:2017xtt}): $\epsilon_{ee}^u = -\epsilon_{e\mu}^u = 0.01$, $\epsilon_{\mu\mu}^u = 0.2$, $\epsilon_{e\tau}^u = \epsilon_{\mu\tau}^u = \epsilon_{\tau\tau}^u = 0$, and the same for $d$ quarks.  Like in \Ref\ \cite{Esteban:2018ppq}, we set all $\epsilon_{\alpha\beta}^e = 0$.  Thus, for \figu{prob_3nu}, the NSI parameters in \equ{epsilon_3} are $\epsilon_{ee} = -\epsilon_{e\mu} = 0.06$, $\epsilon_{\mu\mu} = 1.2$, and $\epsilon_{e\tau} = \epsilon_{\mu\tau} = \epsilon_{\tau\tau} = 0$.

These values are illustrative, and should not be read as a fit.  The combination that matter oscillations are chiefly sensitive to is $\epsilon_{ee} - \epsilon_{\mu\mu} = -1.14$, which lies in the LMA-D, or ``dark-side,'' region rather than in the ordinary LMA one\ \cite{Esteban:2018ppq}.  We keep the values because they are deliberately large: their purpose is to make the NSI curves in \figu{prob_3nu} visibly different from the standard ones, not to represent a fit.  They should not be reused as though they were one.

%%%%%%%%%%%%%%%%%%%%%%%%%%%%%%%%%%%%%%%%%%%%%%%%%%%%%

\subsection{Antineutrinos}

Antineutrinos need no separate construction: conjugating the vacuum term and reversing the sign of every matter potential,
\begin{equation}
 \label{equ:antineutrino}
 \mathbb{H}\left(\bar{\nu}\right)
 =
 \left[\mathbb{H}^{\rm vac}\right]^\ast - \mathbb{A} - \mathbb{V} \;,
\end{equation}
gives a Hamiltonian that \equ{sylvester} then treats like any other; see Section~\ref{sec:conjugation}.  The two operations do different work and both are needed; applying only one is the commonest error in this kind of calculation.  The conjugation reverses the sign of $\delta_{\rm CP}$, and with it every T-odd term.  Its effect on the probabilities is the transposition of \equ{conjugation}, so in vacuum, where it is the whole of the story, no new calculation is required.  The reversal of $\mathbb{A}$ and $\mathbb{V}$ is what makes the matter resonance appear for one sign of $\Delta m_{31}^2$ in neutrinos and the other in antineutrinos; it changes the eigenvalue spectrum, so in matter the Hamiltonian of \equ{antineutrino} has to be built and expanded in its own right.  

For a Hamiltonian the user supplies, that is the user's job, and \equ{antineutrino} is the whole of the recipe.  For scenarios bundled with {\tt NuOscProbExact}, the routines build themselves---including a chord through Earth, where the potential varies from slab to slab, and where a fourth flavor adds a second potential to reverse---the {\tt earth} routines take an {\tt antineutrino} argument that does it.

Listing\ \ref{lst:antineutrinos} shows both routes: the flag, and the two operations written out for a Hamiltonian of one's own.  The cost of applying only one of them is not subtle.  At $L = 1300$~km in the crust and $E = 2.5$~GeV, $P_{\nu_\mu \to \nu_e}$ is $0.087$ for neutrinos and $0.024$ for antineutrinos; conjugating the vacuum term but leaving the potential alone gives $0.069$, which is neither, and is wrong by a factor of three.  Notebook \#13 in the GitHub repository works through both ways of getting it half right.

%%%%%%%%%%%%%%%%%%%%%%%%%%%%%%%%%%%%%%%%%%%%%%%%%%%%%
\begin{lstlisting}[language=Python, caption={Antineutrinos, by either route.  For the scenarios the code builds itself, {\tt antineutrino=True} conjugates the vacuum term and reverses every potential, slab by slab along a chord and for both potentials when a fourth flavor is present.  For a Hamiltonian the user supplies, \equ{antineutrino} is written out instead; both of its operations are needed, and the text gives the cost of doing one.}, label={lst:antineutrinos}, float=t]
import numpy as np

import earth
import hamiltonians3nu
import oscprob3nu
from globaldefs import *

energy = 2.5e9   # Neutrino energy [eV]
costhz = -0.8    # Arrival direction, through the mantle

h_vac = hamiltonians3nu.\
    hamiltonian_3nu_vacuum_energy_independent(
        S12_NO_BF, S23_NO_BF,
        S13_NO_BF, DCP_NO_BF,
        D21_NO_BF, D31_NO_BF)

# The scenarios the code builds itself: one argument
p_nu = earth.probabilities_3nu_earth(
    h_vac, energy, costhz)
p_nubar = earth.probabilities_3nu_earth(
    h_vac, energy, costhz, antineutrino=True)

# A Hamiltonian of your own: conjugate the vacuum term
# *and* reverse every potential.  Doing only the first
# is the error the main text prices
h_nubar = np.conj(h_vac)/energy \
    - np.diag([VCC_EARTH_CRUST, 0., 0.])
q_nubar = oscprob3nu.probabilities_3nu(
    h_nubar, 1300.*CONV_KM_TO_INV_EV)
\end{lstlisting}
%%%%%%%%%%%%%%%%%%%%%%%%%%%%%%%%%%%%%%%%%%%%%%%%%%%%%

\subsection{Oscillations in a Lorentz-violating background}

Lorentz invariance is one of the linchpins of the Standard Model (SM), but is violated in proposed extensions, some related to quantum gravity; see \Refs\ \cite{Mavromatos:2004sz, Liberati:2009pf, Liberati:2013xla, Tasson:2014dfa} for reviews.  There is no experimental evidence for Lorentz-invariance violation (LIV), but there are stringent constraints on it\ \cite{Kostelecky:2008ts, Abbasi:2010kx, Abe:2014wla, Aartsen:2017ibm}.  The effects of LIV are numerous, \eg, changes in the properties and rates of processes of particles versus their anti-particles, introduction of anisotropies in particle angular distributions, and, for neutrinos, changes to the effective mixing parameters and so to the probabilities.

To study LIV, we adopt the framework of the Standard Model Extension (SME)\ \cite{Colladay:1998fq}, an effective field theory that augments the SM by adding LIV parameters to all sectors, including neutrinos\ \cite{Coleman:1998ti, Kostelecky:2003xn, Kostelecky:2003cr, GonzalezGarcia:2004wg, Hooper:2005jp, Kostelecky:2011gq, Diaz:2014yva}.  In the SME, LIV is suppressed by a high energy scale $\Lambda$, still undetermined, and the operators that carry it are graded by their mass dimension $d$, an operator of dimension $d$ contributing to the Hamiltonian as $E^{d-3}$.  

Two restrictions fix the case we treat.  The first is the dimension: we keep $d = 4$, where the contribution grows linearly with neutrino energy, unlike every other case presented above, whose terms either fall with energy or are flat in it.  That operator is CPT-even, a tensor coupling of neutrinos to the LIV background (the CPT-odd operators sit at odd $d$, and the lowest of them, $d = 3$, carries no energy dependence at all).  The second restriction is isotropy: we keep only the part of the coupling that singles out no direction, so that its effect depends on the neutrino energy alone, where a general coefficient would make the probabilities depend on the direction of travel as well and, at a detector fixed to the Earth, on sidereal time~\cite{LSND:2005oop, MINOS:2008fnv, MINOS:2010kat, IceCube:2010fyu, MiniBooNE:2011pix, DoubleChooz:2012eiq, Telalovic:2023tcb, Telalovic:2025xor}.  Growth with energy is what makes high-energy atmospheric and astrophysical neutrinos ideal for testing LIV\ \cite{Coleman:1998ti, GonzalezGarcia:2004wg, GonzalezGarcia:2005xw, Anchordoqui:2005gj, Arguelles:2015dca, Bustamante:2015waa, Rasmussen:2017ert, Aartsen:2017ibm, Ahlers:2018mkf, Ackermann:2019cxh, IceCube:2021tdn}.

The Hamiltonian for isotropic, CPT-even LIV is\ \cite{Coleman:1998ti, Kostelecky:2003xn, Dighe:2008bu}
\begin{equation}
 \mathbb{H}_3^{\rm LIV}(E)
 =
 \mathbb{H}_3^{\rm vac}(E)
 +
 \frac{E}{\Lambda} \mathbb{R}_{3,\xi} \mathbb{B}_3 \mathbb{R}_{3,\xi}^\dagger \;,
\end{equation}
in vacuum.  The second term on the right-hand side is the effective Hamiltonian that introduces LIV.  Here, $\mathbb{B}_3 \equiv {\rm diag}(b_1, b_2, b_3)$, where $b_i$ ($i=1,2,3$) are the eigenvalues of the LIV operator $\mathbb{B}_3$, and $\mathbb{R}_{3,\xi}$ is the $3 \times 3$ mixing matrix that rotates it into the flavor basis.  It has the same structure as the PMNS matrix, but different values of the mixing angles and phase, and they are supplied the same way: {\tt hamiltonian\_3nu\_liv} takes the $\xi$ angles and the phase $\delta_{\xi{\rm CP}}$ through an {\tt angles} argument of its own, so they may be given in whichever of the four conventions introduced above they were published in, independently of how the PMNS angles were given.  In general, there is a relative phase between $\mathbb{R}_{3,\xi}$ and $\mathbb{R}_{3,\theta}$ that cannot be rotated away\ \cite{GonzalezGarcia:2004wg, GonzalezGarcia:2005xw}; in our simplified treatment, we set it to zero.

Because LIV has not been observed, the values of the eigenvalues $b_i$ and of the LIV mixing parameters $\xi$ are undetermined.  Current upper limits\ \cite{Aartsen:2017ibm} set using high-energy atmospheric neutrinos imply that $b_i/\Lambda \lesssim 10^{-28}$; this ratio is dimensionless because $d = 4$ is, and it is the amount by which the maximum attainable velocity of the $i$-th eigenstate departs from $c$.  It is $\mathring{c}^{(4)}$ in the notation of \Ref\ \cite{Aartsen:2017ibm}, whose superscript is that dimension and whose ring marks the isotropic part, so the limit is quoted for the same restrictions we have made.  The LIV energy scale is believed to be at least $\Lambda = 1$~TeV.  The values used in \figu{prob_3nu} are therefore illustrative rather than allowed, chosen so that the curves show features at the energies plotted, and are seven orders of magnitude above that limit: $b_1/\Lambda = b_2/\Lambda = 10^{-21}$ and $b_3/\Lambda = 2 \cdot 10^{-21}$.  For simplicity, we set all mixing angles to zero, so that $\mathbb{R}_{3,\xi} = \mathbb{1}$.  These three eigenvalues and these angles are the values carried by the {\tt globaldefs} module, and the ones with which both \figu{prob_3nu} and the output printed in Listing\ \ref{lst:3nu_examples} were produced.

%%%%%%%%%%%%%%%%%%%%%%%%%%%%%%%%%%%%%%%%%%%%%%%%%%%%%

\subsection{Layered matter and the profile of Earth}
\label{sec:layered_earth}

%%%%%%%%%%%%%%%%%%%%%%%%%%%%%%%%%%%%%%%%%%%%%%%%%%%%%
\begin{figure}[t!]
 \centering
 \includegraphics[width=\columnwidth]{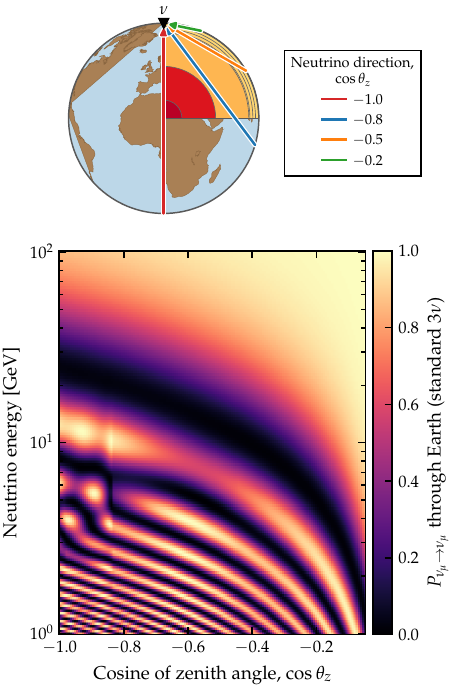}
 \caption{\label{fig:earth_oscillogram}{\it Top}: Earth, with a quarter cut away to show the interior.  The exposed layers are shaded by the PREM density\ \cite{Dziewonski:1981xy}, from $1$~g~cm$^{-3}$ in the ocean layer, pale, to $13$~g~cm$^{-3}$ in the inner core, dark red; the intact surface shows land and ocean.  The colored lines are the chords traveled by neutrinos reaching the detector (triangle) from four directions.  A neutrino with $\cos\theta_z = -1$ crosses the core; one with $\cos\theta_z = -0.2$ grazes the crust.  {\it Bottom}: Survival probability for $\nu_\mu$ crossing Earth, as a function of energy and arrival direction, computed with {\tt NuOscProbExact}\ \cite{NuOscProbExact}.  Each chord is decomposed into slabs following the PREM density profile\ \cite{Dziewonski:1981xy}, and each slab is solved exactly and composed as in \equ{composition}.  The sharp change in structure near $\cos\theta_z = -0.84$ is the core-mantle boundary, visible in the upper panel as the edge of the dark disc.}
\end{figure}
%%%%%%%%%%%%%%%%%%%%%%%%%%%%%%%%%%%%%%%%%%%%%%%%%%%%%

%%%%%%%%%%%%%%%%%%%%%%%%%%%%%%%%%%%%%%%%%%%%%%%%%%%%%
\begin{figure}[t!]
 \centering
 \includegraphics[width=\columnwidth]{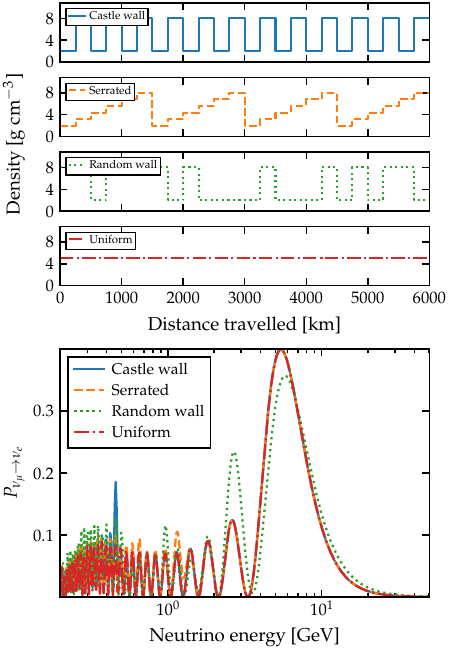}
 \caption{\label{fig:density_arrangement}Probability $P_{\nu_\mu \to \nu_e}$ through four matter profiles of $6000$~km, all of the same mean density, $5$~g~cm$^{-3}$.  Two of them go further than that: the ``random wall'' is a permutation of the ``castle wall'', so the two contain the identical multiset of slabs and differ only in the order in which those slabs are crossed --- and still give different probabilities.  Replacing a structured profile by its average, the ``uniform'' curve, is not in general a good approximation.}
\end{figure}
%%%%%%%%%%%%%%%%%%%%%%%%%%%%%%%%%%%%%%%%%%%%%%%%%%%%%

The case that the code meets most often is the one in which the regions of Section \ref{sec:composition} are slabs of matter, each of constant but different density.  See, \eg, \Refs\ \cite{Chizhov:1999az, Chizhov:1999he} for an overview of this scenario, and \Refs\ \cite{Ohlsson:1999um, Ohlsson:2001et, Merfeld:2014cha} for studies with the OS method.  It applies to long-baseline experiments that consider a non-uniform matter density profile\ \cite{Jacobsson:2001zk, Ohlsson:2003ip, Arguelles:2012nw, Roe:2017zdw, Kelly:2018kmb}, and to neutrinos crossing the several density layers inside Earth\ \cite{Freund:1999vc}; the amplitudes from the successive slabs are stitched together\ \cite{Giunti:2007ry} exactly as in \equ{composition}.  Each evolution operator, $\mathbb{U}^{(j)}$, is evaluated with the matter Hamiltonian at the density of the $j$-th slab.

The one practical warning is the order of the factors, which \equ{composition} fixes and \figu{slabs} draws: reversing the product sends the neutrino through the profile backwards.  The routines below compose in that order, so it is a warning for anyone implementing the composition by hand rather than for a user of the code.

The composition is implemented in the {\tt slabs} module, which takes arbitrary lists of widths and densities, solves each slab exactly, and composes the resulting operators in the order above.  The {\tt earth} module builds on it, using the Preliminary Reference Earth Model (PREM)\ \cite{Dziewonski:1981xy}, the standard spherically symmetric model of the Earth's interior: a fit to seismic data that gives the density, among other quantities, as a piecewise cubic in radius over ten shells, from the inner core to a global-average ocean layer.  Only the density is used here, to generate the slabs crossed along any chord through Earth.  Those two modules are what this subsection illustrates, for three flavors; Section \ref{sec:sterile_earth} then adds the fourth flavor.

One convention goes with them.  PREM is a density model and fixes no composition, so the electron fraction has to be supplied separately; every figure in this paper takes $Y_e = 0.5$ throughout, isoscalar matter, as Section \ref{sec:prob_3nu_matter} does.  The code carries a partition of $Y_e$ by PREM layer instead---core, mantle, crust and ocean, each with the value its material gives---which {\tt earth.electron\_fraction\_prem} supplies and any of whose four values may be replaced.  It is not what the routines assume unless they are asked, and the difference is not small: through the diameter it changes $P_{\nu_\mu \to \nu_e}$ by of order itself, through the mantle alone by a few percent.  PREM's ocean is a global average rather than a feature of any one baseline, so a chord ending at a detector under rock crosses none of it; passing the crust's value for {\tt ocean} is how that is said.

Figure~\ref{fig:earth_oscillogram} shows a standard three-flavor oscillogram for neutrinos crossing Earth: the survival probability $P_{\nu_\mu \to \nu_\mu}$ over energy and arrival direction, with each trajectory decomposed into slabs of the PREM profile and each slab solved exactly.  Computing it requires the slab composition of \equ{composition}, applied once per trajectory, and it is one call: the Earth routines take an array of energies together with an array of directions, indexed on separate axes, and return the whole oscillogram.

Figure \ref{fig:density_arrangement} makes a point that the slab machinery exists to capture.  It compares four density profiles that have the same mean density and shows that the resulting probabilities differ.  Two of the four make the point sharply: the ``random wall'' is a permutation of the ``castle wall'', so the two are built from exactly the same collection of slabs and differ only in the order in which the neutrino crosses them.  Because the evolution operators of the individual slabs do not commute, the arrangement of matter along the trajectory matters, not only its average.  

What decides how much it matters is the scale of the structure set against the oscillation length.  The castle wall alternates between $2$ and $8$~g~cm$^{-3}$ every $250$~km, which is short compared with the oscillation length: above $1$~GeV the latter is never below $10^3$~km and reaches tens of thousands.  The consequence is visible in \figu{density_arrangement}: above about $1$~GeV the castle-wall curve lies on the uniform one to three decimal places, despite a density that swings by a factor of four along the way.  Fine structure averages out, and the mean density is all that survives of it.  The permutation is what breaks that, for a reason worth stating: shuffling the same slabs gathers some of them into long runs of a single density, structure coarse enough to compare with the oscillation length, and the random wall parts from the uniform curve by as much as $0.12$.  Replacing a structured profile by a uniform one of the same mean density is therefore a good approximation exactly when the structure is finer than the oscillation length, and a poor one otherwise; \figu{density_arrangement} shows both cases at once, and the code makes no distinction between them.

Below $1$~GeV, where the layers and the oscillation length become comparable, the castle wall separates from the uniform curve as well, and the narrow feature near $0.5$~GeV---which the random-wall profile does not have---belongs to the arrangement rather than to the mean density.  A periodic profile can do more than this.  With the period tuned against the oscillation length, successive periods add coherently and the transition probability is driven above what any constant density can produce, a parametric resonance\ \cite{Chizhov:1999az, Chizhov:1999he}.  Reaching it takes few and wide periods rather than the twelve narrow ones drawn here, and Earth supplies the physical instance: along a chord through the core, $P_{\nu_\mu \to \nu_e}$ reaches $0.61$ near $5$~GeV, where the best that any constant density can do over the same $12742$~km is $0.20$.  That is beyond what \figu{density_arrangement} sets out to show, and it needs nothing new from the code: \equ{composition} is the same product of the same exact factors whether the profile averages away or resonates.

Listing \ref{lst:figures} shows how these two figures are produced, and how the one in the next subsection is.

%%%%%%%%%%%%%%%%%%%%%%%%%%%%%%%%%%%%%%%%%%%%%%%%%%%%%

\subsection{Four flavors: a 3+1 system through Earth}
\label{sec:sterile_earth}

%%%%%%%%%%%%%%%%%%%%%%%%%%%%%%%%%%%%%%%%%%%%%%%%%%%%%
\begin{figure}[t!]
 \centering
 \includegraphics[width=\columnwidth]{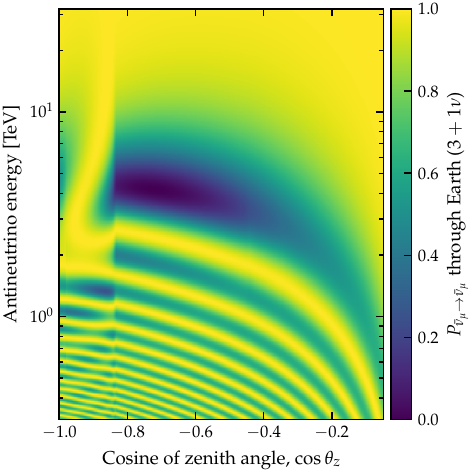}
 \caption{\label{fig:sterile_earth}Survival probability in a 3+1 sterile-neutrino scenario, for $\bar{\nu}_\mu$ crossing Earth along the PREM profile\ \cite{Dziewonski:1981xy}, computed with the four-flavor routines of {\tt NuOscProbExact}\ \cite{NuOscProbExact}.  Here $\Delta m_{41}^2 = 1$~eV$^2$ and $\sin^2\theta_{14} = \sin^2\theta_{24} = 0.1$, with $\theta_{34} = 0$.  Each chord is cut into slabs exactly as in \figu{earth_oscillogram}, and both $V_{\rm CC}$ and $V_{\rm NC}$ are evaluated per slab, since a sterile state feels neither and the neutral-current term therefore no longer cancels.  The broad absorption feature is the matter resonance in the sterile sector.}
\end{figure}
%%%%%%%%%%%%%%%%%%%%%%%%%%%%%%%%%%%%%%%%%%%%%%%%%%%%%

The fourth flavor divides the codes of Table~\ref{tab:codes} more sharply than anything else measured here.  A 3+1 sterile system is not a setting in {\tt NuFast-Earth}, {\tt GLoBES} or {\tt Prob3++}: none of the three admits a fourth state, {\tt GLoBES} only through the documented hook that Table~\ref{tab:codes} describes.  {\tt nuSQuIDS} and {\tt nuCraft} do accept one, and reach it as they reach everything else, by integrating the evolution numerically in a basis of whatever size is asked for.  What is distinctive here is neither: a sterile state is the same closed form as at three flavors, evaluated once, with a $4 \times 4$ Hamiltonian in place of a $3 \times 3$ one.

Figure~\ref{fig:sterile_earth} shows one, for antineutrinos crossing Earth along the full PREM profile.  Treated as the closed four-flavor system it is, it is solved by the construction of Section \ref{sec:four_flavors}, slab by slab and composed as in \equ{composition}, and the matter resonance driven by $V_s$ in \equ{a4} appears as a broad absorption feature.

Two features about it stand out.  The first is that the resonance is there at all: it is driven by the neutral-current potential, which cancels between the three active flavors and so is invisible to every three-flavor calculation, but which the sterile state does not feel and therefore cannot cancel.  A 3+1 system therefore probes the neutron density too, and \ref{app:sample_2nu_hamiltonians} sets out the $\mathbb{A}_4$ that follows.  The second is that the resonance follows the density along the chord rather than sitting at one energy.  For trajectories that stay in the mantle, the resonance is the broad feature near $4$~TeV.  Through the denser core, the resonance moves down, to between $1$ and $1.4$~TeV, since a larger potential sets a lower resonant energy.  It also breaks into finer structure, since a core-crossing chord accumulates more oscillation phase than any mantle one: it is longer, $12742$~km against $10673$ for the longest chord that stays in the mantle, and the potential along it is larger.  The core-mantle boundary near $\cos\theta_z = -0.84$ separates the two.  A calculation at a single averaged density cannot show that, which is the practical reason for carrying the profile.

%%%%%%%%%%%%%%%%%%%%%%%%%%%%%%%%%%%%%%%%%%%%%%%%%%%%%
\begin{lstlisting}[language=Python, caption={Producing Figures\ \ref{fig:earth_oscillogram}, \ref{fig:density_arrangement}, and \ref{fig:sterile_earth}.  Each snippet is the core of the calculation, with the geometry and the plotting left to the notebook \#10 of {\tt NuOscProbExact}.  The Earth routines take the energies and the directions on separate axes and return a whole oscillogram from one call, at three flavors and at four; the arbitrary profile of \figu{density_arrangement} is not one {\tt earth} knows about, so its slabs are composed directly, with the energies stacked the same way.}, label={lst:figures}, float=t]
import earth, slabs
import hamiltonians3nu, hamiltonians4nu
from globaldefs import *

# --- Fig. 4: an oscillogram through the Earth.
# Energies and directions on separate axes, so the
# whole grid comes back from one call; index 4 of
# the nine probabilities is P(mu->mu)
grid3 = earth.probabilities_3nu_earth(
    h_vacuum_energy_indep,
    energies[:, None], cos_zenith[None, :],
    n_slabs_per_segment=4)[..., 4]

# --- Fig. 5: an arbitrary layered profile.
# widths [km] and densities [g cm^-3] are free; each
# slab is solved exactly and composed in order.  The
# energies stack on a leading axis, so the whole scan
# crosses the profile in one call; index 3 of the
# nine probabilities is P(mu->e)
vcc = earth.matter_potential(densities)
h = hamiltonians3nu.hamiltonian_3nu_matter(
    h_vacuum_energy_indep, energies[:, None], vcc)
scan = slabs.probabilities_3nu_slabs(
    h, widths*CONV_KM_TO_INV_EV)[..., 3]

# --- Fig. 6: a 3+1 system, antineutrinos, on PREM.
# antineutrino=True conjugates the vacuum term and
# reverses both potentials.  A sterile state feels
# neither, so V_NC no longer cancels and is carried
# per slab beside V_CC.  Index 5 of the sixteen is
# P(mu->mu).  The sterile parameters are the reader's
# to choose; globaldefs carries only the active ones. 
DM41 = 1.0                 # Delta m^2_41 [eV^2]
S14 = S24 = 0.10**0.5
S34 = 0.0
h4_vacuum = hamiltonians4nu.\
    hamiltonian_4nu_vacuum_energy_independent( \
        S12_NO_BF, S23_NO_BF, S13_NO_BF,
        S14, S24, S34, DCP_NO_BF,
        D21_NO_BF, D31_NO_BF, DM41)

grid4 = earth.probabilities_4nu_earth(
    h4_vacuum, energies[:, None], cos_zenith[None, :],
    n_slabs_per_segment=4, antineutrino=True)[..., 5]
\end{lstlisting}
%%%%%%%%%%%%%%%%%%%%%%%%%%%%%%%%%%%%%%%%%%%%%%%%%%%%%

%%%%%%%%%%%%%%%%%%%%%%%%%%%%%%%%%%%%%%%%%%%%%%%%%%%%%

\subsection{Oscillations with arbitrary Hamiltonians}
\label{sec:arbitrary_hamiltonian}

%%%%%%%%%%%%%%%%%%%%%%%%%%%%%%%%%%%%%%%%%%%%%%%%%%%%%
\begin{lstlisting}[language=Python, caption={Using {\tt NuOscProbExact} with an arbitrary, user-supplied Hamiltonian, here a long-range $L_e - L_\mu$ interaction sourced by the electrons of the Earth.  The chord and $V_{\rm CC}$ follow PREM; only the source of $V_{e\mu}$ is idealized to a uniform ball, which is what puts \equ{yukawa} in closed form at any mediator mass and spares the quadrature.  Notebook 20 evaluates \equ{yukawa_radial} for the PREM distribution instead, checks the two against each other, and is what \figu{lri_earth} is drawn from.  What the new interaction costs the user is the last two lines.}, label={lst:3nu_arbitrary}, float=t]
import numpy as np

import earth
import slabs
import hamiltonians3nu
from globaldefs import *

energy = 5.e9    # Neutrino energy [eV]
costhz = -1.     # Straight through the core

h_vac = hamiltonians3nu.\
    hamiltonian_3nu_vacuum_energy_independent(
        S12_NO_BF, S23_NO_BF,
        S13_NO_BF, DCP_NO_BF,
        D21_NO_BF, D31_NO_BF)

# One Hamiltonian per PREM slab along the chord
widths, densities = earth.earth_slabs(costhz)
h = hamiltonians3nu.hamiltonian_3nu_matter(
        h_vac, energy,
        earth.matter_potential(densities))

# V_emu [eV], one value per slab.  For a uniform ball
# of electrons the Yukawa integral is elementary, so
# no quadrature is needed here.  n_e is routed through
# the library's own V_CC, so the two potentials cannot
# disagree about the electron fraction
alpha = 1.e-52                        # g'^2/(4 pi)
R = EARTH_RADIUS*CONV_KM_TO_INV_EV    # [eV^-1]
m_zp = 1./R                           # mediator range R
n_e = earth.matter_potential(5.51)/(np.sqrt(2.)*GF)
r = earth.earth_slab_radii(costhz)*CONV_KM_TO_INV_EV
v_emu = (4.*np.pi*alpha*n_e/m_zp**2.
         * (1. - (R + 1./m_zp)*np.exp(-m_zp*R)
            * np.sinh(m_zp*r)/r))

# nu_e and nu_mu carry opposite charge, nu_tau none
q = np.diag([1., -1., 0.])
h = h + v_emu[:, None, None]*q

Pee, Pem, Pet, Pme, Pmm, Pmt, Pte, Ptm, Ptt = \
    slabs.probabilities_3nu_slabs(
        h, widths*CONV_KM_TO_INV_EV)
\end{lstlisting}
%%%%%%%%%%%%%%%%%%%%%%%%%%%%%%%%%%%%%%%%%%%%%%%%%%%%%

%%%%%%%%%%%%%%%%%%%%%%%%%%%%%%%%%%%%%%%%%%%%%%%%%%%%%
\begin{figure}[t!]
 \centering
 \includegraphics[width=\columnwidth]{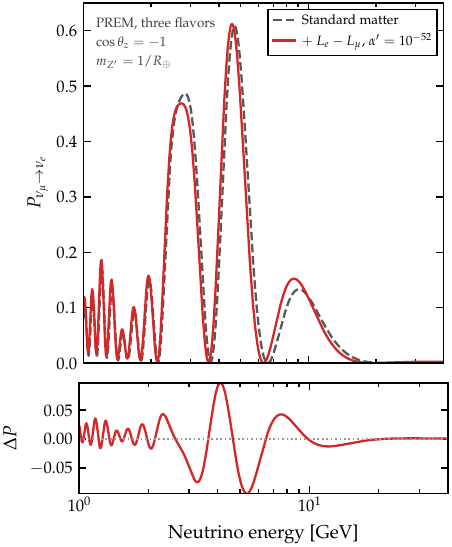}
 \caption{\label{fig:lri_earth}An interaction no code anticipates, at the cost of one extra term.  $P_{\nu_\mu \to \nu_e}$ for a neutrino crossing the Earth along a diameter ($\cos\theta_z = -1$), through the PREM profile, with and without a long-range $L_e - L_\mu$ interaction whose mediator has range $1/m_{Z^\prime} = R_\oplus$.  \emph{Top}: the probability.  \emph{Bottom}: the difference between them.  The Yukawa potential is sourced by the electrons of the whole Earth and evaluated at the radius of each slab; it is added to the matter Hamiltonian as $V_{e\mu}\,{\rm diag}(1,-1,0)$, which is the entirety of the modification.  Both curves are exact for their own Hamiltonian --- no expansion in $V_{e\mu}$ is made, and none is needed, since the construction does not care what the Hermitian matrix contains.  The coupling $\alpha^\prime = 10^{-52}$ is chosen for visibility.  Produced by notebook 10; notebook 20 develops and validates the potential.}
\end{figure}
%%%%%%%%%%%%%%%%%%%%%%%%%%%%%%%%%%%%%%%%%%%%%%%%%%%%%

The usefulness of {\tt NuOscProbExact} stems in part from its ability to compute oscillation probabilities for any arbitrary time- or position-independent Hamiltonian.   Listing \ref{lst:3nu_arbitrary} shows a code template to compute three-neutrino oscillation probabilities using a user-supplied Hamiltonian that is added to the Hamiltonian for oscillations in matter.

The template is written around a concrete instance: a long-range neutrino-electron interaction made possible by a new, light neutral gauge boson, $Z^\prime$, resulting from gauging the anomaly-free combination $L_e - L_\mu$\ \cite{He:1990pn, Foot:1994vd}, where $L_e$ and $L_\mu$ are the electron and muon lepton numbers.  We explore its effect for neutrinos propagating through the Earth. The charge under it, $Q \equiv L_e - L_\mu$, is $+1$ for the electron and $\nu_e$, $-1$ for the muon and $\nu_\mu$, and zero for every other Standard-Model field, quarks included.  Therefore, a new matter potential term, sourced by electrons, enters the Hamiltonian, \ie,
\begin{equation}
 \label{equ:h3_lri}
 \mathbb{H}_3^{\rm LRI}(E)
 =
 \mathbb{H}_3^{\rm vac}(E)
 +
 \mathbb{A}_3
 +
 V_{e\mu}(r) \, {\rm diag}\left(1,-1,0\right) \;,
\end{equation}
where the first two terms are those of Section \ref{sec:prob_3nu_matter} and the third is the modification, which varies along the path of the neutrino.

What makes it a useful example is that it is not merely a new number in an existing formula.  Under $L_e - L_\mu$, an electron sources a Yukawa potential centered on its position.  Thus, the potential experienced by a neutrino at position $\mathbf{r}$ is a Yukawa integral over the electrons that surround it,
\begin{equation}
 \label{equ:yukawa}
 V_{e\mu}(\mathbf{r})
 =
 \frac{g^{\prime 2}}{4\pi}
 \int d^3r^\prime \, n_e(\mathbf{r}^\prime) \,
 \frac{e^{-m_{Z^\prime} \lvert \mathbf{r} - \mathbf{r}^\prime \rvert}}
      {\lvert \mathbf{r} - \mathbf{r}^\prime \rvert} \;,
\end{equation}
with $g^\prime$ the new gauge coupling, so that $\alpha^\prime \equiv g^{\prime 2}/4\pi$.  Where $V_{\rm CC}$ reads the density at the neutrino's own position, \equ{yukawa} is \emph{non-local}: at mediator masses light enough to matter, $m_{Z^\prime} \lesssim 1/R_\oplus$, its value at any radius is an integral over the whole electron distribution, so it depends on how many electrons there are and only weakly on where they sit\ \cite{Wise:2018rnb}.  Its variation along a trajectory is therefore a matter of geometry rather than of the local density, and survives even where that density is constant. In Listing \ref{lst:3nu_arbitrary}, for simplicity, the source is a uniform-density ball, and the potential still falls by half between the center of the Earth and the surface.  (Figure~\ref{fig:lri_earth} does employ the PREM.)

Matter enters as $V_{\rm CC}\,{\rm diag}(1,0,0)$ and the new term as $V_{e\mu}\,{\rm diag}(1,-1,0)$, and no value of $V_{\rm CC}$ turns one into the other.  Nor can the potential be reintroduced as an effective density profile: $V_{e\mu}(r)$ is not a function of the density at $r$, it cannot be reintroduced as an effective profile either.  No specialized expression for it exists in any of the codes of Section \ref{sec:comparison} either.

Taken as it stands, \equ{yukawa} would be a three-dimensional integral for every slab.  For a spherically symmetric $n_e$ it is not.  The angular average of the Yukawa kernel over a shell of radius $r^\prime$ separates it into an interior and an exterior piece,
\begin{equation}
 \label{equ:yukawa_radial}
 \begin{split}
  V_{e\mu}(r)
  =
  g^{\prime 2} \bigg[
  &\frac{e^{-m_{Z^\prime} r}}{r}
  \int_0^r \! dr^\prime \, r^{\prime 2} \, n_e(r^\prime) \,
  {\rm shc}\!\left(m_{Z^\prime} r^\prime\right) \\
  &+ {\rm shc}\!\left(m_{Z^\prime} r\right)
  \int_r^{R_\oplus} \! dr^\prime \, r^\prime \, n_e(r^\prime) \,
  e^{-m_{Z^\prime} r^\prime} \bigg] \;,
 \end{split}
\end{equation}
with ${\rm shc}(x) \equiv \sinh(x)/x$, which is $1$ at the origin.  The radius $r$ now appears only in the limits of the two integrals and in the factors outside them, never in the integrands.  Both can therefore be accumulated once across the profile---the interior piece outward from the center, the exterior piece inward from the surface---and $V_{e\mu}$ read off at any radius afterward.  Since it depends on radius alone, that single pass serves every slab of every chord, so the potential is built once per body rather than once per slab.  Nothing diverges as $m_{Z^\prime} \to 0$ either, where the bracket becomes the enclosed-charge form of an ordinary $1/r$ potential.  Only the electrons of the body being crossed are counted here; adding the rest of the Universe, which takes over for a mediator lighter still, is a different calculation.  The scenario and its phenomenology are developed in \Refs\ \cite{Joshipura:2003jh, Bandyopadhyay:2006uh, Bustamante:2018mzu, Agarwalla:2024ylc}, and the probability that comes out of \equ{h3_lri} is exact.

Figure~\ref{fig:lri_earth} shows the result for a neutrino crossing the full diameter of the Earth: the same PREM slab decomposition evaluated with and without the extra term, the two lines of Listing \ref{lst:3nu_arbitrary} being the whole of the difference between the calculations.  The potential at the center of the Earth is about $1\%$ of $V_{\rm CC}$ there, and the appearance probability is displaced by up to $0.10$, with the largest effect near $4$~GeV, where the standard matter resonance sits.  (The figure sources $V_{e\mu}$ from PREM, which piles electrons into the core; the uniform ball of the listing gives the same shape at about four-fifths the size.)  Notebook \#20 of {\tt NuOscProbExact} builds the potential, fixes its normalization, and checks it against its short- and long-range limits.

%%%%%%%%%%%%%%%%%%%%%%%%%%%%%%%%%%%%%%%%%%%%%%%%%%%%%
%%%%%%%%%%%%%%%%%%%%%%%%%%%%%%%%%%%%%%%%%%%%%%%%%%%%%

\section{Performance and precision}
\label{sec:performance}

Every timing quoted in this paper---in this section, in \figu{plane}, in \figu{prem_plane} and in \figu{prem_plane_3plus1}---was measured on one laptop: an Intel Core i5-1334U, ten cores and twelve threads at up to $4.6$~GHz, with $16$~GB of RAM, running Ubuntu $24.04.4$ LTS on kernel $7.0.0$-$28$-generic, with {\tt Python}~$3.12.7$, {\tt NumPy}~$1.26.4$, {\tt Numba}~$0.60.0$ and {\tt gcc}~$13.3$.  The external codes were compiled on the same machine with the optimization flags their own documentation recommends.  Absolute times therefore mean little on their own and would be perhaps a factor of two different on a workstation; the ratios between codes, which is what these figures are for, are far more stable than that.

%%%%%%%%%%%%%%%%%%%%%%%%%%%%%%%%%%%%%%%%%%%%%%%%%%%%%
\begin{figure}[t!]
 \centering
 \includegraphics[width=\columnwidth]{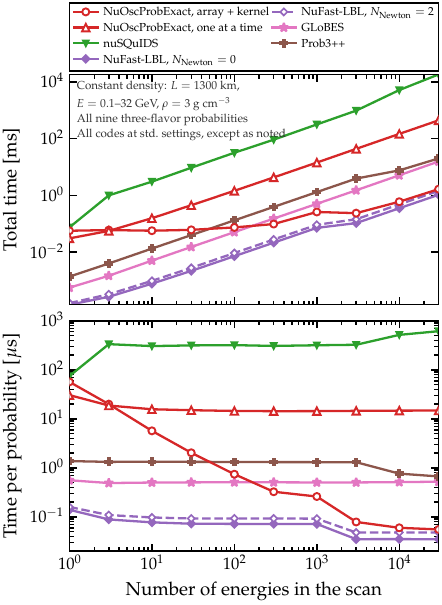}
 \caption{\label{fig:performance}Cost of one three-flavor probability, as a function of how many are asked for at once, for an energy scan at $L = 1300$~km in matter of constant density.  The two curves for {\tt NuOscProbExact} are the same calculation routed through the compiled kernel (solid) and evaluated one point at a time (dashed); the intermediate route, a single call on a stack of Hamiltonians without the kernel, lies between them and is omitted to keep the legend short.  Shown alongside are {\tt nuSQuIDS}\ \cite{Delgado:2014kpa, NuSQuIDS}, through its Python interface, and {\tt NuFast-LBL}\ \cite{Denton:2024pzc}, {\tt GLoBES}\ \cite{GLoBES} and {\tt Prob3++}\ \cite{Prob3pp}, each timed inside C or C++; {\tt NuFast-LBL} is drawn at both of its Newton settings.  {\tt nuCraft} has no constant-density mode and appears in \figu{prem_plane} instead.  Section\ \ref{sec:performance} gives the machine, how each code was driven, and what each setting buys in accuracy.}
\end{figure}
%%%%%%%%%%%%%%%%%%%%%%%%%%%%%%%%%%%%%%%%%%%%%%%%%%%%%

%%%%%%%%%%%%%%%%%%%%%%%%%%%%%%%%%%%%%%%%%%%%%%%%%%%%%

\subsection{The cost of a probability}

Figure~\ref{fig:performance} shows what one three-flavor probability costs, against the number asked for at once.  The closed forms implemented in {\tt NuOscProbExact} are cheap, but how cheap depends almost entirely on how they are called.

Called one point at a time, the cost per probability is flat, at about 14~$\mu$s, and is dominated by Python rather than by arithmetic.  Passing a stack of Hamiltonians instead---which every core routine accepts, together with an array of baselines, broadcast against each other---amortizes that overhead over the whole stack, and by a few thousand points is 30--40 times faster per probability.  At a million points, evaluated in blocks as below, that is the difference between some 14~s and about 0.3~s.  Routing the same stack through the compiled kernel gains a further speed-up factor that grows with the stack: about 4 at a few thousand points, and 8.5 at thirty thousand.  The kernel is used only where it has been measured to win: at three and four flavors, that is any size of stack at all; at two flavors, only stacks of some tens of thousands, below which the {\tt NumPy} path is faster than the cost of dispatching to the kernel.

This performance axis is the one {\tt NuOscProbExact} has and the four external codes in \figu{performance} do not, for a reason about their interfaces rather than their arithmetic.  {\tt GLoBES}, {\tt Prob3++}, and {\tt NuFast-LBL} take one energy per call, and {\tt nuSQuIDS}  exposes a multiple-energy mode, but its cost per probability does not fall as the request grows, costing about twice as much per probability at tens of thousands of energies as at a handful.  Two of the curves in \figu{performance} do fall at the largest stacks, both by about a factor of two against their own plateaus: {\tt NuFast-LBL} from $72$ to $35$~ns, and {\tt Prob3++} from $1310$ to $671$~ns.  Since neither batches, that is not amortization; {\tt GLoBES}, timed the same way on the same machine, shows no such fall, so it is not an artifact of the timing either, and we have not established what it is.  Regardless, at a factor of two against the two orders of magnitude batching buys, it alters no conclusion that rests on this figure.  What these curves are not is a comparison: each code is timed at a setting of its own, and holding speed against accuracy is left to Section \ref{sec:comparison}.

Finally, diagonalizing costs more than the closed form even when it is done well.  Handed the same stack of three-flavor matter Hamiltonians and vectorized through the batched {\tt eigh} of {\tt NumPy}, a route that diagonalizes, evolves, and rotates back to the flavor basis takes roughly an order of magnitude longer per probability than the expansion does.  At four flavors, the gap narrows to about three times, because the quartic and its roots are real work while a $4 \times 4$ matrix costs a library eigensolver little more than a $3 \times 3$ one.  This is the comparison the construction of Section \ref{sec:general_n} is against, since diagonalization is the other way to reach an exact probability; every other code in Section \ref{sec:comparison} reaches something less than exact.

%%%%%%%%%%%%%%%%%%%%%%%%%%%%%%%%%%%%%%%%%%%%%%%%%%%%%

\subsection{Memory and large scans}
\label{sec:memory}

Memory is the constraint that replaces time once the stacks are large.  A batched call holds three things at once: the stack of Hamiltonians, the result, and its own working space.  At three flavors in {\tt NuOscProbExact}, these are $144$, $72$, and about $80$ bytes per energy, so a scan of $10^6$ points needs some $300$~MB and one of $10^7$ some $3$~GB, a fifth of the memory on the machine above and a poor use of it.  At four flavors, the Hamiltonian and the output both grow as $n^2$, sixteen entries in place of nine, so a stack of the same length costs about $1.8$ times as much.  

The remedy is not to abandon batching but to batch in chunks.  Along an Earth chord, the code already does this.  The {\tt earth} entry points cut a long array of energies into pieces, sized from the last-level cache of the processor (L3 in the test machine)---clamped between $4$ and $64$~MB, and $16$~MB where the cache cannot be read---so an oscillogram never builds the whole stack however many energies it is handed.  That size is chosen for the cache rather than for the memory, because the batched kernel is memory-bound rather than compute-bound: the cost per probability was measured at $7.9~\mu$s against an $8$~MB working set and $16.4~\mu$s against a $540$~MB one, a factor of two paid for traffic alone.  Capping the memory usage follows from that choice but is not the reason for it.  Wrapping a call to {\tt earth} in a loop of one's own replaces a cache-matched block size with an arbitrary one, and pays for it in speed.

Everywhere else the chunking is the user's to do, since {\tt oscprob$n$nu} and {\tt slabs} evaluate whatever stack they are given.  Little is lost by doing it: the cost per probability in \figu{performance} is within a factor of two of its best by a few thousand points, and within about a tenth of it by ten thousand on either route, so a long scan in blocks of, say, $10^5$ with an accumulated result costs no more than one enormous call.  One safeguard does fire on its own, guarding against refinement rather than length.  Given a tolerance in place of a slab count, the routines double the count until the estimated error meets it.  A tolerance finer than the discretization can reach would go on doubling until memory ran out, so they stop at $1024$ sub-slabs per segment and raise, naming the error actually reached.  The ceiling moves with {\tt n\_max}.

Three habits follow from all of this, to be recommended to the user.  First, pass arrays rather than looping, since that is worth more than any other single choice.  Second, build the energy-independent part of the vacuum Hamiltonian once and reuse it---the matter, NSI, and Lorentz-violating builders all take it as their first argument, at two, three, and four flavors---rather than rebuilding it inside a scan.

%%%%%%%%%%%%%%%%%%%%%%%%%%%%%%%%%%%%%%%%%%%%%%%%%%%%%
\begin{lstlisting}[language=Python, caption={Choosing the slab count once and reusing it.  A relative tolerance makes the {\tt earth} routines refine until the estimated error meets it, doubling the count as they go; {\tt return\_n\_slabs} hands back the count they settled on, which every later call can be given directly.  The two calls here return identical arrays, the second repeating the last evaluation the first performed.}, label={lst:tolerance}, float=t]
import numpy as np

import earth
import hamiltonians3nu
from globaldefs import *

energies = np.logspace(0., 2., 200)*1.e9  # 1-100 GeV
costhz = -0.9                             # Arrival direction

h_vac = hamiltonians3nu.\
    hamiltonian_3nu_vacuum_energy_independent(
        S12_NO_BF, S23_NO_BF, S13_NO_BF,
        DCP_NO_BF, D21_NO_BF, D31_NO_BF)

# Once, on the scan actually wanted: ask for the
# accuracy and keep the slab count it took
p, n_slabs = earth.probabilities_3nu_earth(
    h_vac, energies, costhz,
    rtol=1.e-3, return_n_slabs=True)

# Thereafter: pass that count, skip the refinement
p = earth.probabilities_3nu_earth(
    h_vac, energies, costhz,
    n_slabs_per_segment=n_slabs)

# atol defaults to zero, so a bare rtol must hold at the
# oscillation minima too, where P is small.  The test is
# |error| <= atol + rtol*|P|, so an absolute floor keeps
# small probabilities from setting the count: adding one
# of 1e-4 here cuts it fourfold
p, n_slabs = earth.probabilities_3nu_earth(
    h_vac, energies, costhz,
    rtol=1.e-3, atol=1.e-4, return_n_slabs=True)
\end{lstlisting}
%%%%%%%%%%%%%%%%%%%%%%%%%%%%%%%%%%%%%%%%%%%%%%%%%%%%%

Listing \ref{lst:tolerance} shows the third habit: choosing the slab count for a chord through Earth without guessing it.  The cost is linear in that count, and past the point where the probability settles, raising it buys nothing.  Passing a relative tolerance in place of a count makes the routines find one by iteration, doubling until the estimated error meets it.  Asked with {\tt return\_n\_slabs=True}, a first call to the probability function returns the count beside the probabilities; every later call is given the count in place of the tolerance.  The refinement costs about $2.7$ times a direct call at that count, so paying it once and reusing the answer beats paying it on every call.

One warning goes with the reuse.  The tolerance has to hold at every energy in the array, so the hardest one sets the count.  For instance, from $1$ to $100$~GeV at $\cos\theta_z = -0.9$, a relative tolerance of $10^{-3}$ needs $128$ sub-slabs per PREM segment, whereas $5$~GeV alone needs $32$.  Fixing the count on a single energy and reusing it across a scan therefore under-resolves the rest, and nothing in the call says so.  The user should fix it instead with the array that will actually be scanned, as Listing \ref{lst:tolerance} does, and the count that comes back is safe to reuse across it.

%%%%%%%%%%%%%%%%%%%%%%%%%%%%%%%%%%%%%%%%%%%%%%%%%%%%%

\subsection{Precision and accuracy}
\label{sec:precision}

Precision and accuracy are separate questions and have separate limits.  Precision is internal: whether the same quantity, computed in different ways, comes back with the same digits.  The scalar, vectorized, and compiled paths of {\tt NuOscProbExact} agree with one another to round-off, which the test suite checks directly.  For a given Hamiltonian and baseline, there is one answer and no setting that moves it: the tolerance of Section \ref{sec:memory} chooses how finely a varying profile is cut, not how a probability is computed from it.

Accuracy is measured against something outside the library, and it takes two values, because two different things limit it.  At constant density nothing is approximated: the closed forms are exact, the only error left is floating-point, and the {\tt NuOscProbExact} entry in \figu{plane} sits at $10^{-15}$ against a $50$-digit reference.  That is round-off rather than a property of the method.  Through the Earth, the profile has to be cut into slabs, and what limits the answer is the cut and not the algebra: the {\tt NuOscProbExact} curve in \figu{prem_plane} falls to $4 \cdot 10^{-8}$ as the slabs are refined, and would keep falling.  That is why a chord through the Earth has a precision setting and constant density has none: a dial exists only where something is being approximated, and at constant density nothing is.

The eigenvalue spectrum alone touches both, and then only in a corner.  Near a level crossing, where a resonance brings two eigenvalues together, taking the roots of a characteristic polynomial is ill-conditioned, and the closed form used by {\tt NuOscProbExact} loses digits that a backward-stable eigensolver keeps (Section \ref{sec:limitations}).  Measured against a $50$-digit reference, the loss stays at round-off until two eigenvalues come within about $10^{-5}$ of each other relative to the spread of the spectrum, and at its worst, near $10^{-8}$, it still leaves the probability good to $10^{-7}$. \textbf{This is orders of magnitude finer than any measurement resolves.}  The scenarios of Section \ref{sec:examples} do not come near even that: about $10^{-2}$ at three flavors in matter, $10^{-3}$ at $3+1$.  Only a user who contrives a near-degeneracy and wants the last digits for their own sake need prefer diagonalization.

%%%%%%%%%%%%%%%%%%%%%%%%%%%%%%%%%%%%%%%%%%%%%%%%%%%%%
%%%%%%%%%%%%%%%%%%%%%%%%%%%%%%%%%%%%%%%%%%%%%%%%%%%%%

\section{Comparison to other codes}
\label{sec:comparison}

%%%%%%%%%%%%%%%%%%%%%%%%%%%%%%%%%%%%%%%%%%%%%%%%%%%%%
\begin{figure}[t!]
 \centering
 \includegraphics[width=\columnwidth]{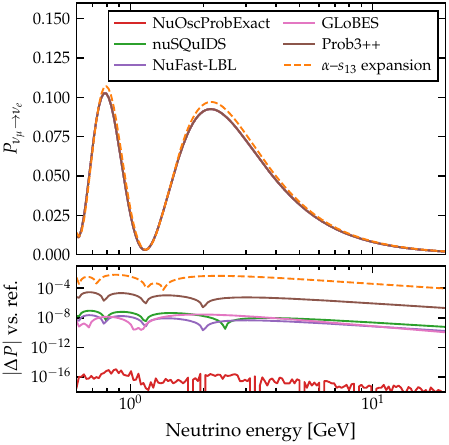}
 \caption{\label{fig:exact_vs_approx}Six ways of computing the same quantity, and the accuracy of each: $P_{\nu_\mu \to \nu_e}$ at $L = 1300$~km through constant matter of $3$~g~cm$^{-3}$.  \emph{Top}: the probability itself, on which all but the expansion coincide to the width of the line.  \emph{Bottom}: the residual of each against an exact reference --- the same $50$-digit {\tt mpmath} matrix exponential of the same Hamiltonian used in \figu{plane}, so that no code here is its own referee, {\tt NuOscProbExact} included.  {\tt NuOscProbExact} sits at about $1 \cdot 10^{-15}$, which is the round-off of double precision rather than a measurement of anything else.  {\tt GLoBES}\ \cite{GLoBES}, {\tt nuSQuIDS}\ \cite{Delgado:2014kpa, NuSQuIDS}, and {\tt Prob3++}\ \cite{Prob3pp} are at their standard settings; {\tt NuFast-LBL}\ \cite{Denton:2024pzc} is shown with $N_{\rm Newton} = 2$ rather than its default of $0$.  The second-order expansion in $\alpha$ and $s_{13}$ is the outlier at about $7 \cdot 10^{-3}$: what it shows is the truncation of the series, which no setting improves.  The matter density each external code is given is rescaled so that all six share the same potential.  The length conventions are not matched here: each code keeps its own value of $\hbar c$, the constant that turns a baseline in kilometers into eV$^{-1}$.  Section\ \ref{sec:differences} accounts for each residual in turn, the unmatched length among them.}
\end{figure}
%%%%%%%%%%%%%%%%%%%%%%%%%%%%%%%%%%%%%%%%%%%%%%%%%%%%%

Several mature codes compute oscillation probabilities.  They overlap less than their common subject suggests.  {\tt GLoBES}\ \cite{GLoBES} is an experiment-simulation framework, in which probabilities are one layer inside a package whose purpose is fluxes, cross-sections, detector response, and the statistics built on them.  {\tt Prob3++}\ \cite{Prob3pp} evaluates the three-flavor probability in matter through the analytic treatment of \Ref\ \cite{Barger:1980tf}, and is long established inside atmospheric-neutrino analyses.  {\tt nuCraft}\ \cite{nuCraft} integrates the evolution numerically along a trajectory through Earth, with sterile states allowed.  {\tt NuFast-LBL}\ \cite{Denton:2024pzc} solves the three-flavor constant-density case specifically, faster than anything else available; {\tt NuFast-Earth}\ \cite{Denton:2025oxf} carries the same three-flavor specialization through the Earth.  {\tt nuSQuIDS}\ \cite{Delgado:2014kpa, NuSQuIDS} is the most general: it propagates a density matrix rather than a state, which admits non-coherent interactions, attenuation, and open-system terms that no closed form can reach, at the cost of solving a differential equation.

What {\tt NuOscProbExact} offers against them is the combination: flexibility, speed, and accuracy at once.  Flexibility and speed can each be found elsewhere, though not in one code.  None of them offers exactness to round-off; reaching it outside this code takes diagonalizing the Hamiltonian numerically, at the cost in speed this construction was built to avoid.  The rest of this section measures each of the three in turn.

\smallskip

\textit{Flexibility.---}{\tt NuOscProbExact} takes an \emph{arbitrary} Hermitian Hamiltonian at two, three, or four flavors and returns the exact probability, with neither approximation nor numerical integration.  Nothing about the scenario needs to be anticipated: a sterile state, a Lorentz-violating background, non-standard interactions, and any mixture of the three are one and the same call, as is a Hamiltonian nobody has yet written down.  Matter of piecewise constant density, and the profile of Earth, follow by composing exact solutions rather than by any new expression.  Section \ref{sec:capabilities} sets that against what the other codes accept.

\smallskip

\textit{Speed.---}Through the Earth, which all but one of the codes in Table~\ref{tab:codes} are built to handle, {\tt NuOscProbExact} reaches a given accuracy sooner than any of the other five, at every accuracy except the very coarsest they share.  The margin widens as the accuracy tightens.  At constant density the ordering changes: {\tt NuFast-LBL}, a closed form specialized to exactly three flavors, is faster, while everything else on that plane remains slower than this code.  Section \ref{sec:plane} puts numbers to that, and Section \ref{sec:performance} to the batching some codes rest on.

\smallskip

\textit{Accuracy.---}At constant density, the answer given by {\tt NuOscProbExact} is exact to round-off: four orders of magnitude below the best any other code reaches there, and eight below {\tt GLoBES}.  Through the Earth, where every code, this one included, must discretize the continuously varying profile, its error sits an order of magnitude below anything else on that plane.  Section \ref{sec:precision} gives both values and says what limits each.

%%%%%%%%%%%%%%%%%%%%%%%%%%%%%%%%%%%%%%%%%%%%%%%%%%%%%

\subsection{Where the differences come from}
\label{sec:differences}

Figure \ref{fig:exact_vs_approx} shows the accuracy claim under test on one problem: six ways of computing the same probability, each measured against a reference obtained by exponentiating the same Hamiltonian in $50$-digit arithmetic.  {\tt NuOscProbExact} sits at about $10^{-15}$, the round-off of double precision rather than a property of the construction.  Each of the other five residuals is made of parts with names---a setting, a convention, a floor that belongs to a method---and we name them below, first at the constant density of the figure, then along a chord through the Earth, where the origins differ and so does the set of codes that can be compared.

We start with the matter potential, which the codes do not define identically.  {\tt NuOscProbExact} takes the electron density as $Y_e \rho / \bar{m}$, with $\bar{m}$ the mean free-nucleon mass, while the other codes refer it to the atomic mass unit.  Each carries its own rounding of that constant, so each needs its own factor---$0.79\%$ for {\tt nuSQuIDS} and {\tt GLoBES}, $0.75\%$ for {\tt NuFast-LBL} and {\tt Prob3++}---all of them the nuclear mass defect of Section \ref{sec:prob_3nu_matter}.  Left unmatched, the mismatch would move $P_{\nu_\mu \to \nu_e}$ by up to $4 \cdot 10^{-4}$, above every residual it would otherwise contaminate, so each external density is rescaled before anything is compared.  The length conventions are left alone on purpose: the codes do not all share the value of $\hbar c$ that converts a baseline in kilometers into eV$^{-1}$, so two of the residuals below carry a convention on top of an approximation.

Against {\tt NuFast-LBL} at its default setting the residual is about $10^{-5}$. That number measures the setting rather than the code. {\tt NuFast-LBL} obtains the eigenvalues of the matter Hamiltonian from a fast approximation and then polishes them with as many Newton steps as the user asks for; its default, $N_{\rm Newton} = 0$, asks for none. Two steps, the setting \figu{exact_vs_approx} draws, bring the residual to about $10^{-8}$. What is left at that point is the length convention. {\tt NuFast-LBL}'s $\hbar c$ differs from ours by about $4 \cdot 10^{-8}$ in relative terms, which the probability reads as a small difference in phase; evaluated at the length their constant implies, this code agrees with {\tt NuFast-LBL} to about $6 \cdot 10^{-12}$. Nothing takes it lower: a third Newton step leaves the residual unchanged, so that last $6 \cdot 10^{-12}$ belongs to {\tt NuFast-LBL}'s own refinement rather than to any disagreement between the two codes. The whole difference is one truncation the user controls and one convention neither code is wrong about.

Against {\tt nuSQuIDS}, the residual is about $10^{-7}$.  A length convention of the same kind accounts for most of it: adopting their kilometer leaves a residual of about $10^{-8}$.  What is left after that is the numerical integration itself.  In vacuum, the two codes agree to about $10^{-15}$, so nothing in the algebra separates them.  The disagreement appears only in matter, where {\tt nuSQuIDS} has a differential equation to solve.  Nor is it a setting: below a solver tolerance of $10^{-10}$, tightening further no longer moves the residual.  (We have not explored tightening further.)  The same integrator that admits the interactions and open-system terms no closed form expresses also sets the finest agreement {\tt nuSQuIDS} reaches in matter.

{\tt GLoBES} and {\tt Prob3++} offer nothing to vary.  The one accuracy dial each provides subdivides a varying matter profile; a constant density gives it nothing to subdivide.  With the potential matched and no setting to move, the residual of each is the intrinsic accuracy of the fixed algorithm it runs: {\tt GLoBES} sits at about $3 \cdot 10^{-8}$ and tracks the probability itself at about one part in $10^{7}$; {\tt Prob3++} sits at about $3 \cdot 10^{-5}$, about one part in $10^{4}$.  The second-order analytic expansion sits at about $7 \cdot 10^{-3}$.  Its residual is not numerical: the terms beyond second order are absent from the formula itself, so no setting tightens it.  The expansion gives up those terms for an expression short enough to read; its residual is the size of what it gave up.

So far we have compared accuracy at constant density.  Through the Earth, the density varies continuously along the trajectory.  The set of codes that can be compared changes with the problem: {\tt NuFast-LBL} and the expansion are limited to constant density, while {\tt NuFast-Earth} and {\tt nuCraft} have no constant-density mode and appear only here.

The six codes that can cross the Earth treat the varying profile in one of two ways.  {\tt NuOscProbExact}, {\tt NuFast-Earth}, {\tt GLoBES}, and {\tt Prob3++} cut the path into pieces of constant density and solve each piece exactly.  Between one boundary and the next, nothing is approximated.  {\tt nuSQuIDS} and {\tt nuCraft} instead integrate the equations of motion numerically along the trajectory.  The treatment of the profile also fixes what refinement can reach.  The four codes that cut have nothing standing between them and an exact answer but the fineness of the cut, so their error falls for as long as the pieces shrink.  The two codes that integrate stop at a floor that no solver setting passes: for {\tt nuSQuIDS}, the same value survives five different steppers, every maximum step size, and every tolerance below $10^{-9}$.  What holds them there is the density jumps of the PREM rather than the integration.  Take the profile away and the floor goes with it---in vacuum along this same chord {\tt nuSQuIDS} reproduces this code to about $10^{-15}$---while placing the jumps exactly, by aligning the grid it is handed to every one of them, improves the floor from about $3 \cdot 10^{-6}$ to about $9 \cdot 10^{-7}$ without lifting it.  \ref{app:prem_plane} measures both families of codes.

Within the family that cuts, the codes differ in what they cut.  {\tt NuFast-Earth}, {\tt GLoBES}, and {\tt Prob3++} divide each of the PREM's major layers into concentric shells and leave the chord to cross the shells where geometry dictates.  {\tt NuOscProbExact} cuts the chord itself: first at the PREM's density jumps, which it places exactly, then each stretch between jumps into slabs, reading the density at each slab's midpoint rather than at its edge.  Taken at the edge, the assumed density is wrong across the whole slab by an amount that grows linearly with distance from that edge; taken at the midpoint, the linear part of the error cancels between the slab's two halves and only the curvature of the profile survives.  Therefore, the error falls as $1/N_{\rm slabs}^2$ rather than as $1/N_{\rm slabs}$.

%%%%%%%%%%%%%%%%%%%%%%%%%%%%%%%%%%%%%%%%%%%%%%%%%%%%%

\subsection{Speed against accuracy}
\label{sec:plane}

%%%%%%%%%%%%%%%%%%%%%%%%%%%%%%%%%%%%%%%%%%%%%%%%%%%%%
\begin{figure*}[t!]
 \centering
 \begin{minipage}[t]{0.48\textwidth}
  \centering
  \includegraphics[width=\linewidth]{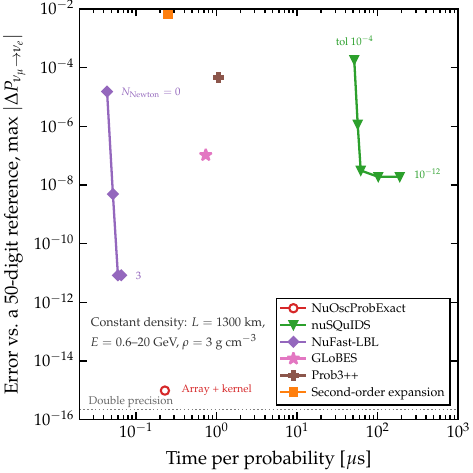}
  \caption{\label{fig:plane}What accuracy costs, in each of six ways of computing the same quantity: $P_{\nu_\mu \to \nu_e}$ at $L = 1300$~km through constant matter of $3$~g~cm$^{-3}$.  The error is the largest over $60$ energies between $0.6$ and $20$~GeV, measured against a $50$-digit reference obtained by exponentiating the same Hamiltonian in arbitrary precision, so that no code is its own referee.  Codes with a precision setting trace a curve --- the solver tolerance of {\tt nuSQuIDS}\ \cite{Delgado:2014kpa, NuSQuIDS}, the Newton refinement of {\tt NuFast-LBL}\ \cite{Denton:2024pzc} --- while {\tt GLoBES}\ \cite{GLoBES}, {\tt Prob3++}\ \cite{Prob3pp}, and the analytic expansion expose none.  {\tt NuOscProbExact} is a single point, drawn for the batched call through the compiled kernel: at constant density its answer is exact, so it has no accuracy to trade for time. Evaluating the same probabilities one at a time moves the point right along the time axis without moving it up, which is what \figu{performance} measures.  Unlike in \figu{exact_vs_approx}, every convention is matched here: all six are given the same $V_{\rm CC}$ and the same baseline in eV$^{-1}$, as the notebook that produces the figure records in full.  Lower and further left is better.}
 \end{minipage}\hfill
 \begin{minipage}[t]{0.48\textwidth}
  \centering
  \includegraphics[width=\linewidth]{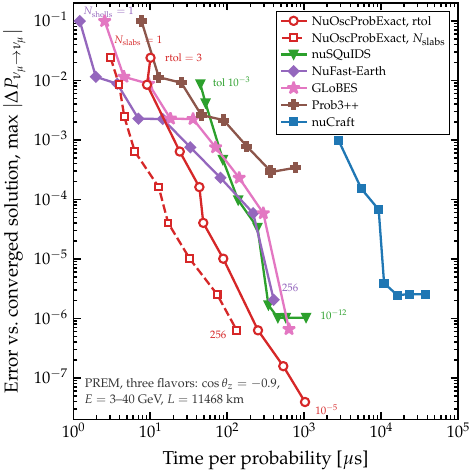}
  \caption{\label{fig:prem_plane}The speed-accuracy plane of \figu{plane}, through the Earth: $P_{\nu_\mu \to \nu_\mu}$ along the PREM chord at $\cos\theta_z = -0.9$, a baseline of $11468$~km, largest error over twelve energies between $3$ and $40$~GeV.  Unlike in \figu{plane} the reference is a \emph{converged} PREM solution and not an exact one, because the density varies continuously and every code approximates the profile; each code therefore sits where its treatment of the profile puts it rather than where the accuracy of its probability formula would.  {\tt NuOscProbExact} is consequently a curve, traced on both dials it exposes --- a relative tolerance and an explicit count of slabs per chord segment --- while {\tt NuFast-Earth}\ \cite{Denton:2025oxf}, {\tt GLoBES}, and {\tt Prob3++} are dialed by a count of shells, the concentric subdivisions of PREM's major layers, and {\tt nuSQuIDS} and {\tt nuCraft} by a solver tolerance.  \ref{app:prem_plane} lists every convention that had to be matched before the five external codes could be compared against this one at all.  Lower and further left is better.}
 \end{minipage}
\end{figure*}
%%%%%%%%%%%%%%%%%%%%%%%%%%%%%%%%%%%%%%%%%%%%%%%%%%%%%

Figure \ref{fig:plane} shows what accuracy costs each code, rather than what each reaches at one setting.  Two of the six expose a setting that trades accuracy for time, so a comparison made at one choice of that setting would say as much about the choice as about the code.  To give a full picture, the figure traces each setting through its range instead.

{\tt NuFast-LBL} spends its setting on Newton steps that polish its eigenvalues.  Its default is the fastest point on the plane, accurate to about $10^{-5}$; two steps take it to some $8 \cdot 10^{-12}$ for about a third more time.  Six orders of magnitude accuracy for a third more time is the best trade on either plane.  {\tt nuSQuIDS} spends its setting on a solver tolerance; its curve bends.  Twenty percent more time carries it from about $2 \cdot 10^{-4}$ to $3 \cdot 10^{-8}$, while tripling the time after that gains less than a factor of two, because by then it is against the integrator floor of Section \ref{sec:differences}.  The tolerance costs almost nothing where it works and gains almost nothing past the floor.

No setting moves the other four codes along the accuracy axis.  {\tt GLoBES} takes three times as long as {\tt NuOscProbExact} and lands at about $10^{-7}$; {\tt Prob3++} takes nearly five times as long, at about $4 \cdot 10^{-5}$; the second-order expansion costs about what this code costs and sits twelve orders further out.  

{\tt NuOscProbExact} has no accuracy setting because at constant density there is nothing to converge: its answer is exact, so it draws a single point near the double-precision round-off line.  (The one thing that moves it, along the time axis, is how many probabilities are asked for at once, which Section \ref{sec:performance} measures.)  The lower-left corner in \figu{plane} is shared between a closed form specialized to three flavors---{\tt NuFast-LBL}---which is fastest, and {\tt NuOscProbExact}, which is exact.  On the standard three-flavor problem in constant matter the trade between the two favors {\tt NuFast-LBL}: timed here over a two-thousand-point scan it leads {\tt NuOscProbExact} by a factor of five, a lead that the larger stacks of \figu{performance} narrow to some twenty percent.

Figure \ref{fig:prem_plane} shows the same trade through the Earth, where it runs the other way.  {\tt NuOscProbExact} now holds the lower-left corner: its tolerance dial carries it to about $4 \cdot 10^{-8}$ for some $1000$~$\mu$s, more than ten times finer than anything else on the plane reaches at any cost.  Short of that it is also the quickest route to a given accuracy, reaching about $6 \cdot 10^{-7}$ in some $130$~$\mu$s where {\tt GLoBES} needs about $640$~$\mu$s for $7 \cdot 10^{-7}$ and {\tt NuFast-Earth} about $400$~$\mu$s for only $2 \cdot 10^{-6}$.  The margin grows as the accuracy tightens: a factor of five at $10^{-3}$, over {\tt NuFast-Earth}, and a factor of eight at $10^{-4}$, over {\tt nuSQuIDS}.

The dial that draws the {\tt NuOscProbExact} curve exists only on this plane, because only here is anything approximated.  At constant density the answer is exact and there is nothing to converge; through the Earth the continuously varying profile must be cut into slabs, so the fineness of the cut, the one approximation in the calculation, becomes a quantity to trade against time (Section \ref{sec:precision}).  The batching of Section \ref{sec:performance} is also in force here from the first energy: along an Earth chord the code builds one Hamiltonian per slab, so even a single energy already presents the compiled kernel with many Hamiltonians at once, where at constant density the same effect needs thousands of energies.  \ref{app:prem_plane} sets out the conventions this comparison rests on, and where each code's curve stops.

In summary, between them, Figs.~\ref{fig:plane} and \ref{fig:prem_plane} say what {\tt NuOscProbExact} costs and what it delivers.  At constant density, it concedes speed to one specialized code ({\tt NuFast-LBL})---the factor of five, narrowing to twenty percent, measured above --and concedes accuracy to nothing.  Through the Earth, past the coarsest settings, it concedes neither.  Neither position depends on what the Hamiltonian contains, since nothing in {\tt NuOscProbExact} is specialized to the standard three-flavor case measured here; for the scenarios that the other entries cannot take at all, the question of which is faster does not arise.  The accuracy has the same reading on both planes: at constant density an exact probability adds nothing to an error budget beyond round-off, while through the Earth it adds at most the error the user's tolerance names, stated in advance rather than estimated afterward.  Both statements hold for any Hermitian Hamiltonian, including the ones for which no tailored expansion has ever been derived.  \ref{app:prem_plane} cements this by showing an example of accuracy~vs.~time for a 3+1 scenario across the Earth.

%%%%%%%%%%%%%%%%%%%%%%%%%%%%%%%%%%%%%%%%%%%%%%%%%%%%%
{\def\arraystretch{1.2}\setlength{\tabcolsep}{4pt}
\begin{table*}[t!]
\centering
\begin{tabular}{l c c c >{\raggedright\arraybackslash}p{1.75cm} c c c c >{\raggedright\arraybackslash}p{1.85cm}}
 \hline\hline
 & & & & & Batch & Precision & \multicolumn{2}{c}{Accuracy} & \\
 \cline{8-9}
 Code & Version & Lang. & Flavors & Matter & over $E$ & setting & \fig~\ref{fig:plane} & \fig~\ref{fig:prem_plane} & Built for \\
 \hline
 {\tt NuOscProbExact}
   & $1.13.1$ & {\tt Python} & $2,3,4$ & Arbitrary $\mathbb{H}$ & yes
   & $N_{\rm slabs}$, rtol & $1 \cdot 10^{-15}$ & $4 \cdot 10^{-8}$
   & Arbitrary $\mathbb{H}$ \\
 {\tt nuSQuIDS}\ \cite{Delgado:2014kpa, NuSQuIDS}
   & $1.13.3$ & {\tt C++} & $\leq 6$ & Any, plus interactions & yes & Solver tol.
   & $2 \cdot 10^{-8}$ & $1 \cdot 10^{-6}$ & Open systems \\
 {\tt nuCraft}\ \cite{nuCraft}
   & r22 & {\tt Python} & $3+n$ & Standard, sterile & no\rlap{$^{\ddagger}$}
   & Solver tol. & --- & $2.5 \cdot 10^{-6}$ & Atmospheric \\
 {\tt NuFast-LBL}\ \cite{Denton:2024pzc}
   & 2026-08 & {\tt C++} & $3$ & Standard & no & Newton steps
   & $8 \cdot 10^{-12}$ & --- & Long-baseline fits \\
 {\tt NuFast-Earth}\ \cite{Denton:2025oxf}
   & 2026-08 & {\tt C++} & $3$ & Standard & yes & $N_{\rm shells}$, Newton
   & --- & $2 \cdot 10^{-6}$ & Atmospheric, solar \\
 {\tt GLoBES}\ \cite{GLoBES}
   & $3.2.18$ & {\tt C} & $3$\rlap{$^{\dagger}$} & Standard & no
   & $N_{\rm shells}$ & $1 \cdot 10^{-7}$ & $7 \cdot 10^{-7}$
   & Experiments \\
 {\tt Prob3++}\ \cite{Prob3pp}
   & 2026-08 & {\tt C++} & $3$ & Standard & no & $N_{\rm shells}$
   & $4 \cdot 10^{-5}$ & $3 \cdot 10^{-4}$ & Atmospheric \\
 \hline\hline
\end{tabular}
\caption{\label{tab:codes}Publicly available codes that compute oscillation probabilities, and what each is able to do.  In the table, ``flavors'' is the number of neutrino states the published interface accepts.  A version given as a month is the month that repository was cloned; every entry is the version actually measured.  ``Batch over $E$'' records whether a stack of energies can be passed in one call; every code here that had it was given it.  Section \ref{sec:performance} measures what batching saves.  ``Precision setting'' is the knob that trades accuracy against time; a count of shells is a dial only where there is a profile to divide, so {\tt GLoBES} and {\tt Prob3++} have no setting at constant density.  Section \ref{sec:plane} and \ref{app:prem_plane} give the two accuracy columns, which record the best accuracy each code reached on the one problem of each figure --- properties of the code \emph{and} of the problem rather than general accuracies.  A dash in either accuracy column means the code was not run on that case: {\tt nuCraft} and {\tt NuFast-Earth} have no constant-density mode, and {\tt NuFast-LBL} is constant-density only, so \figu{prem_plane} carries six codes where this table lists seven.  All but {\tt NuFast-LBL} propagate through Earth; only the first three do so with a sterile state.  Section \ref{sec:capabilities} says what each code is for.  $^{\dagger}${\tt GLoBES} hard-wires a $3 \times 3$ probability matrix, but exposes a hook, {\tt glbRegisterProbabilityEngine}, through which external engines supply more flavors; the {\tt snu} extension\ \cite{Kopp:2006wp} uses it to provide $3+n$.  $^{\ddagger}${\tt nuCraft} accepts a list of particles, but loops over it internally, so a batched call costs what a loop of single calls costs.}
\end{table*}
}
%%%%%%%%%%%%%%%%%%%%%%%%%%%%%%%%%%%%%%%%%%%%%%%%%%%%%

%%%%%%%%%%%%%%%%%%%%%%%%%%%%%%%%%%%%%%%%%%%%%%%%%%%%%

\subsection{What the accuracy is worth}
\label{sec:why_exact}

To understand what accuracy is worth, we examine its impact on a physical measurement: that of the three-flavor CP-violation phase, $\delta_{\rm CP}$.  A residual in the probability mimics a shift of $\delta_{\rm CP}$: the shift that would move the probability by the same amount.  At the baseline and density of \figu{plane}, at the energy between $0.6$ and $20$~GeV where the change is largest, a $15^\circ$ shift of $\delta_{\rm CP}$ moves $P_{\nu_\mu \to \nu_e}$ by about $7 \cdot 10^{-3}$; a $0.1^\circ$ shift, by about $5 \cdot 10^{-5}$; a shift of $0.001^\circ$, by about $5 \cdot 10^{-7}$.  On this scale, an error of $10^{-3}$ in the probability is the size of a $\delta_{\rm CP}$ shift of some $2^\circ$; an error of $10^{-8}$, the size of a shift of $2 \cdot 10^{-5}$ degrees.

An approximation error enters a fit as a bias.  The residual of the second-order expansion at this baseline is a smooth, reproducible function of the same parameters the fit varies: it changes sign four times over the whole band from $0.6$ to $20$~GeV, so it cancels over a spectrum of events only by coincidence.  At the first oscillation maximum, near $2$~GeV, it keeps one sign across the entire range of $\delta_{\rm CP}$, running between about $3 \cdot 10^{-3}$ and $5 \cdot 10^{-3}$ while $P_{\nu_\mu \to \nu_e}$ itself varies by about $0.04$.  An error that follows the parameters displaces the best fit rather than widening the interval quoted around it.

The $7 \cdot 10^{-3}$ of \figu{exact_vs_approx} belongs to one configuration: $1300$~km through $3$~g~cm$^{-3}$.  At $7000$~km through $4$~g~cm$^{-3}$ the expansion is out by about $0.1$; at $10000$~km through $4.5$~g~cm$^{-3}$, where the trajectory crosses the matter resonance, by about $0.6$.  Those last two lie outside the accelerator baselines the expansion was derived for, so neither is a fault of the series; the formula simply gives no sign of where its own boundary runs.  An approximate probability therefore carries an obligation: its error must be bounded against something more accurate before a result is built on it, at every new baseline, density, or set of parameters.  An exact probability, as in {\tt NuOscProbExact}, carries none.

%%%%%%%%%%%%%%%%%%%%%%%%%%%%%%%%%%%%%%%%%%%%%%%%%%%%%

\subsection{What each code is for}
\label{sec:capabilities}

Table \ref{tab:codes} shows what each of these codes was built to do and what its interface accepts.  Those two facts decide more code choices than speed or accuracy do: a code that cannot express the problem at hand is not a candidate at any speed.  Speed and accuracy only ever rank the candidates that remain.  The last column of the table records what each code was built for.

The number of flavors is fixed in most of the interfaces.  {\tt Prob3++}, {\tt NuFast-LBL}, and {\tt NuFast-Earth} hard-wire three, and {\tt GLoBES} does the same, although a documented hook lets an external engine supply more.  A sterile state is therefore not a setting in four of the seven codes but a different program.  The three codes that do accept one were built with it in mind: {\tt nuSQuIDS} and {\tt nuCraft} reach it by propagating numerically in a basis of whatever size is asked for; the {\tt NuOscProbExact} construction of Section \ref{sec:four_flavors} reaches it in closed form.

What may be put in the Hamiltonian is as fixed as the flavor count in most of them.  Most of these codes take oscillation parameters---angles, phases, mass splittings---and build the Hamiltonian themselves.  That design is convenient, because the user supplies familiar parameters, and confining, because only the terms its authors built in can ever appear.  {\tt nuSQuIDS} and {\tt NuOscProbExact} take the Hamiltonian itself: a new oscillation scenario is something the user writes down, not a modification of the source.  

%%%%%%%%%%%%%%%%%%%%%%%%%%%%%%%%%%%%%%%%%%%%%%%%%%%%%

\subsection{What to use, and when}
\label{sec:what_to_use}

Which code to reach for follows from the Hamiltonian and from how many probabilities are wanted at once.  For a production three-flavor fit at a long baseline through roughly constant density, which is the case most of the field spends most of its time on, reach for {\tt NuFast-LBL}: it is faster than {\tt NuOscProbExact} by twenty percent on large stacks and by a factor of five on small ones, and, given two Newton steps, accurate to some $8 \cdot 10^{-12}$.  This is orders of magnitude below the roughly $5 \cdot 10^{-7}$ by which a shift of $\delta_{\rm CP}$ by a thousandth of a degree moves $P_{\nu_\mu \to \nu_e}$ on this baseline, at the energy most sensitive to it (Section \ref{sec:why_exact}).

For a spectrum deliberately tuned to a near-degeneracy, which Section \ref{sec:precision} shows the standard scenarios come nowhere near, prefer a backward-stable eigensolver.  For an open system, or for interactions, no closed form applies and {\tt nuSQuIDS} is the only option.  And for a single coarse probability through the Earth, {\tt NuFast-Earth} returns one in a microsecond at its coarsest shell count; the advantage this code shows in \figu{prem_plane} needs a stack of energies to appear.

Everything that remains calls for {\tt NuOscProbExact}: the arbitrary Hamiltonian, for which no tailored expression exists and none is likely to be derived---and, separately, the case of wanting a great many probabilities at once.

%%%%%%%%%%%%%%%%%%%%%%%%%%%%%%%%%%%%%%%%%%%%%%%%%%%%%
%%%%%%%%%%%%%%%%%%%%%%%%%%%%%%%%%%%%%%%%%%%%%%%%%%%%%

\section{Conclusions}
\label{sec:conclusions}

We have provided {\tt NuOscProbExact}\ \cite{NuOscProbExact}, which computes exact neutrino oscillation probabilities at two, three, and four flavors for arbitrary time- or position-independent Hamiltonians, by the construction of Section \ref{sec:general_n}, so that the Hamiltonian is never diagonalized.  Neither that identity nor the expressions it yields are ours; what we provide is their implementation at two, three, and four flavors, the composition of exact solutions across slabs and along any chord through Earth, and the fast, accurate evaluation of whole stacks of Hamiltonians.

Set beside existing oscillation codes (Section \ref{sec:comparison}), {\tt NuOscProbExact} makes good on three claims: flexibility, speed, and accuracy.  The call accepts the Hamiltonian itself, any Hermitian matrix, rather than a list of oscillation parameters, so a non-standard interaction, a Lorentz-violating term, an added sterile state, all three at once, or a completely novel oscillation scenario first is the same computation as the standard case.  Among the six codes that cross the Earth that we examined, {\tt NuOscProbExact} is the quickest route to every accuracy except the coarsest they share (\figu{prem_plane}); a scan requested as one call comes out two orders of magnitude cheaper per probability than the same scan run point by point (\figu{performance}).  And at constant density its answer differs from a $50$-digit reference by round-off alone, four decades finer than the closest any other code comes (\figu{plane}); no accuracy setting exists there because no approximation is made for one to refine.

{\tt NuOscProbExact} does not treat open systems: a neutrino that decays to invisible states or decoheres leaks out of the flavor basis, its evolution moves to the space of density matrices, and no enlargement of that basis brings it back (Section \ref{sec:limitations}).  It is not the quickest route to the standard three-flavor probability at constant density: {\tt NuFast-LBL}\ \cite{Denton:2024pzc} is quicker there, on a small stack of energies by a factor of five and on a large one by some twenty percent.  And near an eigenvalue level crossing, where two eigenvalues nearly meet, it is not the most accurate route: a backward-stable eigensolver is.  The loss is bounded: it begins only once the gap between the eigenvalues falls below about $10^{-5}$ of the spectral spread, nearer to degeneracy than any scenario in this paper goes; even at its worst it leaves the probability good to about $10^{-7}$ (Section \ref{sec:precision}).

None of those limitations touches the case the code was built for: an arbitrary Hamiltonian, scanned across a broad parameter space, for which nothing specialized exists and nothing specialized is coming.  {\tt NuOscProbExact} is available from PyPI, with {\tt pip install nuoscprobexact}, and from its repository\ \cite{NuOscProbExact}.  We provide it so that such a scenario can be explored before anyone has worked out how to approximate it.

%%%%%%%%%%%%%%%%%%%%%%%%%%%%%%%%%%%%%%%%%%%%%%%%%%%%%
%%%%%%%%%%%%%%%%%%%%%%%%%%%%%%%%%%%%%%%%%%%%%%%%%%%%%

\section*{Acknowledgments}

We thank Carlos Arg\"uelles, Peter Denton, Francisco De Zela, Joachim Kopp, Stephen Parke, Serguey Petcov, Jordi Salvad\'o, and Irene Tamborra for helpful discussion and suggestions.  We thank especially Shirley Li and Tommy Ohlsson for that and for carefully reading the manuscript, and the latter for pointing out the references of the Ohlsson-Snellman method.  MB is supported by the Villum Fonden projects no.~13164 and 29388.  Early phases of this work were supported by a High Energy Physics Latin American European Network (HELEN) grant and by a grant from the Direcci\'on Acad\'emica de Investigaci\'on of the Pontificia Universidad Cat\'olica del Per\'u. 

%%%%%%%%%%%%%%%%%%%%%%%%%%%%%%%%%%%%%%%%%%%%%%%%%%%%%
%%%%%%%%%%%%%%%%%%%%%%%%%%%%%%%%%%%%%%%%%%%%%%%%%%%%%

% \newpage

% \bibliographystyle{elsarticle-num}
% \bibliography{refs.bib}

%%%%%%%%%%%%%%%%%%%%%%%%%%%%%%%%%%%%%%%%%%%%%%%%%%%%%
%%%%%%%%%%%%%%%%%%%%%%%%%%%%%%%%%%%%%%%%%%%%%%%%%%%%%

\appendix

\renewcommand{\thefigure}{\Alph{section}.\arabic{figure}}
\renewcommand{\thetable}{\Alph{section}.\arabic{table}}
\newcommand{\appresetfloats}{\setcounter{figure}{0}\setcounter{table}{0}}

%%%%%%%%%%%%%%%%%%%%%%%%%%%%%%%%%%%%%%%%%%%%%%%%%%%%%
%%%%%%%%%%%%%%%%%%%%%%%%%%%%%%%%%%%%%%%%%%%%%%%%%%%%%

% \newpage

\section{SU($n$) matrices}
\label{app:matrices}
\appresetfloats

For completeness, and to avoid any ambiguity in the method presented in the main text, we show here explicitly the SU(2) and SU(3) matrices, and the construction of the SU(4) matrices.  

%%%%%%%%%%%%%%%%%%%%%%%%%%%%%%%%%%%%%%%%%%%%%%%%%%%%%

\subsection{Two neutrino flavors, SU(2)}

The three Pauli matrices, $\sigma_k = \sigma^k$, are:
\begin{equation}
 \sigma_1
 =
 \left(\begin{array}{cc}
  0 & 1 \\
  1 & 0 \\
 \end{array}\right)
 \;, \;\;
 \sigma_2
 =
 \left(\begin{array}{cc}
  0 & -i \\
  i & 0 \\
 \end{array}\right)
 \;, \;\;
 \sigma_3
 =
 \left(\begin{array}{cc}
  1 & 0 \\
  0 & -1 \\
 \end{array}\right) \;.
\end{equation}

%%%%%%%%%%%%%%%%%%%%%%%%%%%%%%%%%%%%%%%%%%%%%%%%%%%%%

\subsection{Three neutrino flavors, SU(3)}

The eight Gell-Mann matrices, $\lambda_k = \lambda^k$, are:
\begin{eqnarray}
 &&
 \lambda_1
 =
 \left(\begin{array}{ccc}
  0 & 1 & 0 \\
  1 & 0 & 0 \\
  0 & 0 & 0 \\
 \end{array}\right)
 \;, \;\;
 \lambda_2
 =
 \left(\begin{array}{ccc}
  0 & -i & 0 \\
  i &  0 & 0 \\
  0 &  0 & 0 \\
 \end{array}\right)
 \;, \nonumber \\
 &&
 \lambda_3
 =
 \left(\begin{array}{ccc}
  1 &  0 & 0 \\
  0 & -1 & 0 \\
  0 &  0 & 0 \\
 \end{array}\right)
 \;, \;\;
 \lambda_4
 =
 \left(\begin{array}{ccc}
  0 & 0 & 1 \\
  0 & 0 & 0 \\
  1 & 0 & 0 \\
 \end{array}\right)
 \;,  \\
 &&
 \lambda_5
 =
 \left(\begin{array}{ccc}
  0 & 0 & -i \\
  0 & 0 &  0 \\
  i & 0 &  0 \\
 \end{array}\right)
 \;, \;\;
 \lambda_6
 =
 \left(\begin{array}{ccc}
  0 & 0 & 0 \\
  0 & 0 & 1 \\
  0 & 1 & 0 \\
 \end{array}\right)
 \;, \nonumber \\
 &&
 \lambda_7
 =
 \left(\begin{array}{ccc}
  0 & 0 &  0 \\
  0 & 0 & -i \\
  0 & i &  0 \\
 \end{array}\right)
 \;, \;\;
 \lambda_8
 =
 \frac{1}{\sqrt{3}}
 \left(\begin{array}{ccc}
  1 & 0 &  0 \\
  0 & 1 &  0 \\
  0 & 0 & -2 \\
 \end{array}\right)
 \;. \nonumber
\end{eqnarray}

%%%%%%%%%%%%%%%%%%%%%%%%%%%%%%%%%%%%%%%%%%%%%%%%%%%%%

\subsection{Four neutrino flavors, SU(4)}
\label{app:matrices_4nu}

For $n=2$ and $n=3$ the numbering of the generators is fixed by convention.  For $n=4$, none exists, so we state ours; it is the one used in Table \ref{tab:prob_4nu} and in {\tt NuOscProbExact}.  Writing $E_{\alpha\beta}$ for the $4 \times 4$ matrix with a single $1$ in row $\alpha$ and column $\beta$, the fifteen SU(4) generators are, in order,
\begin{equation}
 \lambda^{2p-1} = E_{\alpha\beta} + E_{\beta\alpha} \;, \quad
 \lambda^{2p} = -i\left(E_{\alpha\beta} - E_{\beta\alpha}\right) \;,
\end{equation}
for $p = 1, \ldots, 6$ running over the six flavor pairs $(\alpha,\beta)$ in lexicographic order---$(e,\mu)$, $(e,\tau)$, $(e,s)$, $(\mu,\tau)$, $(\mu,s)$, $(\tau,s)$---followed by the three Cartan generators,
\begin{equation}
 \lambda^{12+l}
 =
 \sqrt{\frac{2}{l\left(l+1\right)}}\,
 {\rm diag}\big(\underbrace{1, \ldots, 1}_{l}, -l, 0, \ldots\big) \;,
 \quad
 l = 1, 2, 3 \;.
\end{equation}
So, for instance, $\lambda^1 = E_{e\mu} + E_{\mu e}$ has a $1$ in the $(e,\mu)$ and $(\mu,e)$ slots, $\lambda^2 = -i(E_{e\mu} - E_{\mu e})$ is its antisymmetric partner, and $\lambda^{13} = {\rm diag}(1,-1,0,0)$ is the first Cartan generator.  They satisfy ${\rm Tr}(\lambda^a\lambda^b) = 2\delta^{ab}$, as do the Pauli and Gell-Mann matrices.

Three properties motivate this choice.  The eight Gell-Mann matrices survive it: deleting the six generators of the pairs that involve the fourth state leaves them exactly, on the upper-left $3 \times 3$ block, so the SU(3) subalgebra is embedded as it stands.  The two generators of a given pair are adjacent, so the T-conjugate structure of Table~\ref{tab:prob_4nu} reads off consecutive indices.  And the generator that singles out the sterile state, $\lambda^{15}$, comes last, so that a fourth state appends to the three-flavor labeling rather than renumbering it.

What the embedding does not preserve is the order.  Gell-Mann interleaves the two Cartan generators; this scheme collects all three of its own at the end, so the surviving eight appear as $\lambda^1, \lambda^2, \lambda^4, \lambda^5, \lambda^6, \lambda^7, \lambda^3, \lambda^8$ in Gell-Mann labels.  Sections \ref{sec:two_flavors} and \ref{sec:three_flavors}, with Tables \ref{tab:h2_coefficients} to \ref{tab:prob_3nu}, use the Gell-Mann numbering; Section \ref{sec:four_flavors} and Table~\ref{tab:prob_4nu} use the one above.  That is why Table~\ref{tab:prob_4nu} does not reduce entry by entry to Table~\ref{tab:prob_3nu}.

%%%%%%%%%%%%%%%%%%%%%%%%%%%%%%%%%%%%%%%%%%%%%%%%%%%%%
%%%%%%%%%%%%%%%%%%%%%%%%%%%%%%%%%%%%%%%%%%%%%%%%%%%%%

\section{Structure of the code}
\label{app:structure}
\appresetfloats

%%%%%%%%%%%%%%%%%%%%%%%%%%%%%%%%%%%%%%%%%%%%%%%%%%%%%
\begin{figure*}[t!]
 \centering
 \includegraphics[width=0.98\textwidth]{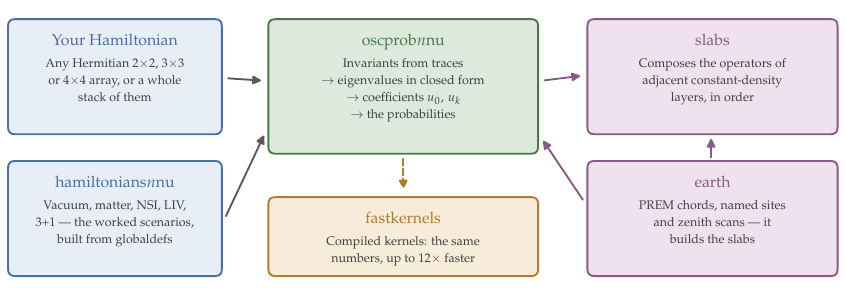}
 \caption{\label{fig:architecture}How the modules of {\tt NuOscProbExact} fit together.  A Hamiltonian enters from the left---either one the user supplies, of any Hermitian $2\times2$, $3\times3$, or $4\times4$ form, or one built by the {\tt hamiltonians} modules for a standard scenario---and the {\tt oscprob} modules turn it into probabilities through the construction of Sections \ref{sec:general_n}--\ref{sec:four_flavors}.  The compiled kernels change nothing but the speed: behind each of them is a vectorized {\tt NumPy} path computing the same quantity, and the test suite holds the two to round-off.  The right-hand pair is for matter that is not uniform: {\tt slabs} composes the solutions of adjacent constant-density layers, and {\tt earth} generates those layers from the PREM profile along a chord.}
\end{figure*}
%%%%%%%%%%%%%%%%%%%%%%%%%%%%%%%%%%%%%%%%%%%%%%%%%%%%%

Figure~\ref{fig:architecture} shows the structure of {\tt NuOscProbExact}: ten modules and some $13\,700$ lines, more than half of them docstrings.  Nothing inherits from anything, and no result depends on a previous call.  What survives a call is a cache and two switches.  The cache is in {\tt earth}, which keeps the slab widths and densities it worked out for a chord, keyed on the arrival direction and the number of slabs per segment, so that a repeated scan does not repeat the geometry.  The switches are in {\tt fastkernels}: {\tt USE\_NUMBA} selects the compiled kernel or the {\tt NumPy} path, and {\tt USE\_PALINDROME} selects whether to compose every slab operator in turn or to compute each distinct one once, which a chord permits because its sequence of slabs reads the same from either end.  Neither switch moves the probability that comes back; both move only what it costs.

%%%%%%%%%%%%%%%%%%%%%%%%%%%%%%%%%%%%%%%%%%%%%%%%%%%%%

\subsection{The modules}

Table \ref{tab:modules} says what each module is for, and \ref{app:tests} says how each is tested.  The division is along one line: a module either builds a Hamiltonian or consumes one.  Everything in the first group is a convenience---the scenarios of Section \ref{sec:examples} are common enough to be worth shipping, but nothing in the second group knows or cares where its input came from, which is what makes an arbitrary Hamiltonian a first-class case rather than an escape hatch.  (Two sign conventions are in use internally: {\tt oscprob3nu} works with the latent roots of Section \ref{sec:evolution_operator}, whereas {\tt oscprob4nu} uses the eigenvalues of $\tilde{\mathbb{H}}$ and the opposite sign in the exponent.  The two are equivalent, both reproduce \equ{sylvester} to round-off; neither is visible outside its module.)

%%%%%%%%%%%%%%%%%%%%%%%%%%%%%%%%%%%%%%%%%%%%%%%%%%%%%

\subsection{Entry points}

The names follow a pattern, so the call for a new case can usually be guessed.  Probabilities come from {\tt probabilities\_$n$nu} and the evolution operator from {\tt evolution\_operator\_$n$nu}, both in {\tt oscprob$n$nu}, and both take a Hamiltonian and a baseline, or stacks of either.  Hamiltonians for the standard scenarios come from {\tt hamiltonian\_$n$nu\_}\emph{scenario} in {\tt hamiltonians$n$nu}, with \emph{scenario} one of {\tt vacuum\_energy\_independent}, {\tt matter}, {\tt nsi}, or {\tt liv}; what they return is an ordinary array, so one built by hand serves just as well.  A trajectory of constant-density layers goes to {\tt probabilities\_$n$nu\_slabs}, and a chord through Earth to {\tt probabilities\_$n$nu\_earth}, which cuts the layers from PREM itself.  Those few names cover every case in this paper.  The full interface---some ninety public functions, including the named observing sites, the tolerance-driven refinement---is documented on the accompanying documentation\ \cite{NuOscProbExactDocs}.

%%%%%%%%%%%%%%%%%%%%%%%%%%%%%%%%%%%%%%%%%%%%%%%%%%%%%

\subsection{Adding a scenario}

Extending the code means writing a Hamiltonian, and nothing else.  The {\tt oscprob} modules accept any Hermitian $2\times2$, $3\times3$, or $4\times4$ array and do not ask where it came from, so a new term is added to whatever an existing builder returns and passed on unchanged.  Listing \ref{lst:3nu_arbitrary} is that in full: one array of potentials, one matrix of charges, one addition.  The same holds along a trajectory---build one Hamiltonian per layer and hand the stack to {\tt slabs}---which is how Figs.~\ref{fig:density_arrangement} and \ref{fig:lri_earth} are computed.  Nothing is registered, subclassed, or recompiled, and the compiled kernels pick up the new Hamiltonian with every other.

What cannot be added this way is anything that is not a Hermitian Hamiltonian.  Absorption, decay to invisible states, and open-system terms are not missing features but a different object, evolving the density matrix rather than the state.  Section \ref{sec:limitations} gives the cost of that; {\tt nuSQuIDS}\ \cite{Delgado:2014kpa, NuSQuIDS} is the tool for it.

%%%%%%%%%%%%%%%%%%%%%%%%%%%%%%%%%%%%%%%%%%%%%%%%%%%%%
{\def\arraystretch{1.2}
\begin{table*}[t!]
  \begin{tabular}{ll}
   Module & Responsibility \\
   \hline
   {\tt oscprob2nu} & Two-flavor probabilities and evolution operator \\
   {\tt oscprob3nu} & Three-flavor, via Table \ref{tab:prob_3nu} \\
   {\tt oscprob4nu} & Four-flavor, via Table \ref{tab:prob_4nu} \\
   {\tt hamiltonians2nu} & Two-flavor Hamiltonians for the standard scenarios \\
   {\tt hamiltonians3nu} & Three-flavor, including the PMNS matrix of \equ{pmns} \\
   {\tt hamiltonians4nu} & Four-flavor, for 3+1 scenarios \\
   {\tt slabs} & Composition across constant-density layers, one chord or a batch \\
   {\tt earth} & PREM profile, chord geometry, named sites, batched energy and zenith scans \\
   {\tt fastkernels} & Compiled kernels for the batched paths \\
   {\tt globaldefs} & Constants, unit conversions, NuFit 4.0 defaults \\
  \end{tabular}
  \caption{\label{tab:modules}The modules of {\tt NuOscProbExact} v1.13.1.  The three {\tt oscprob} modules are the library proper; the rest either prepare their input or arrange their output.  {\tt NumPy} and {\tt Numba} are the only dependencies, and only {\tt fastkernels} touches the second; everything else runs on {\tt NumPy} alone.}
\end{table*}
}
%%%%%%%%%%%%%%%%%%%%%%%%%%%%%%%%%%%%%%%%%%%%%%%%%%%%%

%%%%%%%%%%%%%%%%%%%%%%%%%%%%%%%%%%%%%%%%%%%%%%%%%%%%%
%%%%%%%%%%%%%%%%%%%%%%%%%%%%%%%%%%%%%%%%%%%%%%%%%%%%%

\section{The test suite}
\label{app:tests}
\appresetfloats

Table \ref{tab:tests} lists what the 882 tests of {\tt NuOscProbExact} cover; \ref{app:structure} shows the modules they exercise.  They reach every line and every branch of the library: coverage is $100\%$ in v1.13.1, held above a floor of $98\%$ that fails the build if it drops.

The tests run on every push and every pull request from the GitHub repository, so a contributor need not invoke them; but they are also the quickest way to confirm a local installation.  The tests are not packaged with the library, but from a clone:
\begin{verbatim}
pip install -e ".[test]"
pytest
\end{verbatim}
which takes a few seconds.  Adding {\tt -{}-cov} measures coverage against the floor above.

The test suite is not only a regression net.  Roughly half of it checks the physics and the algebra against something external to the implementation---closed-form eigenvalues, an independently written code, direct matrix exponentiation, the standard two-flavor formula, and the defining properties of the SU(3) $d$ tensor---and the rest checks the properties that a probability must have whatever the Hamiltonian: unitarity, the behavior of reversed channels, and agreement between the scalar, vectorized, and compiled paths, which are three separate pieces of code that must return the same numbers.

{\tt NuOscProbExact} enforces continuous integration: every code pull request on GitHub runs the test suite, across the five supported versions of Python, together with a linter.  Static type checking is not part of that; instead the suite itself checks that every routine carries annotations and that they agree with its docstring, since Sphinx renders both and two different answers to the same question are worse than one absent.  The Jupyter tutorial notebooks are executed in the same workflow, so an example that stops working cannot survive a merge.  This ensures continuity and correctness in future contributions.

%%%%%%%%%%%%%%%%%%%%%%%%%%%%%%%%%%%%%%%%%%%%%%%%%%%%%
{\def\arraystretch{1.15}
\begin{table}[t!]
  \begin{tabular}{lr}
   What is tested & Tests \\
   \hline
   Batching and tolerance selection         & 168 \\
   PREM, geometry, and Earth crossings       & 110 \\
   Numerical edge cases and degeneracies    & 92 \\
   Compiled kernels against {\tt NumPy}      & 69 \\
   Type annotations                         & 54 \\
   Four-flavor construction                 & 45 \\
   Docstrings                               & 40 \\
   Slab composition                         & 37 \\
   Vectorized against scalar evaluation     & 36 \\
   Sample Hamiltonians                      & 33 \\
   Physical scales and unit conversions     & 30 \\
   Agreement with {\tt nuSQuIDS}             & 28 \\
   Standard analytic formulas               & 25 \\
   Matter eigenvalues, Zaglauer--Schwarzer  & 24 \\
   SU(3) algebra and star products          & 22 \\
   Figures reproduced from the documentation & 19 \\
   Probabilities                            & 11 \\
   Evolution operator                       & 11 \\
   Vectorized Hamiltonian builders          & 9 \\
   Repository layout                        & 8 \\
   Version consistency                      & 7 \\
   This paper, against the repository       & 4 \\
   \hline
   Total                                    & 882 \\
  \end{tabular}
  \caption{\label{tab:tests}The test suite of {\tt NuOscProbExact} v1.13.1, grouped by what each file covers, in descending order.  Coverage of the library is complete: every line and every branch is executed.  The comparison with {\tt nuSQuIDS}\ \cite{Delgado:2014kpa, NuSQuIDS} runs against frozen reference values rather than against a live installation, so that a release of another code cannot break this one's continuous integration.}
\end{table}
}
%%%%%%%%%%%%%%%%%%%%%%%%%%%%%%%%%%%%%%%%%%%%%%%%%%%%%

%%%%%%%%%%%%%%%%%%%%%%%%%%%%%%%%%%%%%%%%%%%%%%%%%%%%%
%%%%%%%%%%%%%%%%%%%%%%%%%%%%%%%%%%%%%%%%%%%%%%%%%%%%%

\section{The tutorial notebooks}
\label{app:notebooks}
\appresetfloats

Twenty tutorial Jupyter notebooks sit in the {\tt notebooks/} directory of the repository\ \cite{NuOscProbExact}, numbered in reading order; Table \ref{tab:notebooks} says what each one does.  They are the long form of Section \ref{sec:examples}: what is a paragraph and a figure there is, in a notebook, the same calculation with the reasoning around it---why a convention is what it is, what happens at the edges of the parameter range, and what the numbers were checked against.

They are not part of the installed package, so they come with a clone rather than with {\tt pip}; running them from one needs only
\begin{verbatim}
pip install nuoscprobexact[notebooks]
jupyter-lab notebooks/
\end{verbatim}
Reading them needs less than that: each is stored with its output, so a notebook can be read on GitHub without being run, and each ends with a pointer to its neighbors in the reading order and to the API reference.  All twenty are written by one generator script, also in the repository, which executes each of them in turn.  Continuous integration then re-executes them on every change to the repository, under the conditions Section \ref{sec:documentation} gives for failing the build.

Three of them carry more of this paper than the others.  Every figure in this paper is produced by notebook \#10: the probabilities in them are computed as the notebook runs, except in the figures that compare this code against another, where every code's numbers, including this one's, are read from data frozen in the repository.  Notebook \#17 works through the external corroboration summarized in Section \ref{sec:validation}: what {\tt nuSQuIDS} and the Zaglauer--Schwarzer closed form each agree with, to what accuracy, and---the laborious part---which conventions had to be matched before either comparison meant anything.  And notebook \#20 is where the long-range potential of Section \ref{sec:arbitrary_hamiltonian} is built and validated, Listing \ref{lst:3nu_arbitrary} being the core of it.

%%%%%%%%%%%%%%%%%%%%%%%%%%%%%%%%%%%%%%%%%%%%%%%%%%%%%
{\def\arraystretch{1.1}
\begin{table*}[t!]
  {\small
  \begin{tabular}{rlp{9.2cm}}
   & Notebook & What it does \\
   \hline
   \multicolumn{3}{l}{\emph{Getting started}} \\
   1 & Basics & Units, one probability, and why to pass arrays rather than loop \\
   2 & Oscillations in vacuum & The probabilities against baseline and against energy \\
   \multicolumn{3}{l}{\emph{Matter, and new physics}} \\
   3 & Matter, NSI, and Lorentz-invariance violation & Constant-density matter, then non-standard interactions and Lorentz-invariance violation on top of it: three Hamiltonians, one routine \\
   4 & Oscillograms & A two-dimensional map in energy and baseline, from a single call \\
   5 & Bi-probability plots & CP violation as an ellipse, and where it degenerates to a point \\
   \multicolumn{3}{l}{\emph{Matter that varies along the trajectory}} \\
   6 & The Earth: PREM, chords, and slabs & How a varying profile becomes a sequence of exact pieces, and how the error falls with the number of them \\
   7 & Probabilities through the Earth & Zenith-angle scans, an Earth oscillogram, and the baselines of real experiments \\
   8 & Unusual density profiles & Profiles built by hand, and why the arrangement of the matter matters and not only its mean \\
   \multicolumn{3}{l}{\emph{Speed, and the figures of this paper}} \\
   9 & Performance & Looping against broadcasting, and the compiled kernels, measured as the notebook runs \\
   10 & The paper's figures & Every figure in this paper, produced by one run \\
   \multicolumn{3}{l}{\emph{Where the approximations fail}} \\
   11 & Exact versus the textbook approximations & Where the standard formulas depart from the exact result, and by how much \\
   12 & Mass ordering and the octant & Both orderings, and the octant degeneracy of $\theta_{23}$: where it holds, and the channel that breaks it \\
   13 & Antineutrinos, done properly & The two operations of \equ{antineutrino}, and two ways of applying only one \\
   \multicolumn{3}{l}{\emph{Hard cases}} \\
   14 & Solar neutrinos, the MSW resonance, and the limits of slabs & The adiabatic resonance, checked against the analytic result, and where a Magnus-type method takes over \\
   15 & Numerical edge cases & Degeneracies and vanishing denominators, and what is returned in place of NaN \\
   16 & Four neutrinos, and a sterile state & A 3+1 system through SU(4), and why the construction stops at four \\
   17 & Cross-checks with other codes & Agreement with {\tt nuSQuIDS} and with the Zaglauer--Schwarzer eigenvalues, and the conventions matched to reach it \\
   \multicolumn{3}{l}{\emph{The machinery, and extending it}} \\
   18 & The evolution operator, and the SU($n$) coefficients & The operator and the coefficients $h_a$ themselves, for composing evolution or extending the method \\
   19 & Animated scenes & Four parameter sweeps, drawn as filmstrips, with the animation an opt-in switch \\
   20 & An arbitrary Hamiltonian, through three profiles & A long-range $L_e - L_\mu$ force carried through an invented body, an exponential one, and the Earth \\
  \end{tabular}}
  \caption{\label{tab:notebooks}The twenty tutorial notebooks of {\tt NuOscProbExact} v1.13.1, in reading order.  The groupings are a suggested path rather than a division of the code: any notebook may be read on its own.  This table is checked against that repository by the test suite of \ref{app:tests}, which derives the numbering and the titles from the notebooks themselves, so a notebook added, removed, reordered, or retitled fails the build until the table follows it.}
\end{table*}
}
%%%%%%%%%%%%%%%%%%%%%%%%%%%%%%%%%%%%%%%%%%%%%%%%%%%%%

%%%%%%%%%%%%%%%%%%%%%%%%%%%%%%%%%%%%%%%%%%%%%%%%%%%%%
%%%%%%%%%%%%%%%%%%%%%%%%%%%%%%%%%%%%%%%%%%%%%%%%%%%%%

\section{Derivation of \equ{prob_2nu_general}}
\label{app:derivation_prob_2nu_general}
\appresetfloats

The two-flavor probabilities follow from a single matrix product.  Operating on $\nu_\alpha = \left(1~0\right)^{\rm T}$ with the linear combination $h_k \sigma^k$ that appears in the expansion of the evolution operator, \equ{evol_op_2nu}, gives
\begin{equation}
 \label{equ:h_on_nu_2nu}
 \left(\begin{array}{cc}
  h_3      & h_1-ih_2 \\
  h_1+ih_2 & -h_3     \\
 \end{array}\right)
 \left(\begin{array}{c}
  1 \\
  0 \\
 \end{array}\right)
 =
 \left(\begin{array}{c}
  h_3      \\
  h_1+ih_2 \\
 \end{array}\right) \;,
\end{equation}
and each entry carries one of the probabilities, with the coefficients $h_k$ of Table \ref{tab:h2_coefficients}: contracting on the left with $\nu_\alpha^\dagger$ keeps the first, and with $\nu_\beta^\dagger = \left(0~1\right)$ the second.

Take the transition first.  The identity term of \equ{evol_op_2nu} is diagonal and cannot connect $\nu_\alpha$ to $\nu_\beta$, so the whole amplitude comes from the second entry of \equ{h_on_nu_2nu},
\begin{equation}
 \nu_\beta^\dagger \mathbb{U}_2(L) \nu_\alpha
 =
 -i \frac{\sin\left(\lvert h \rvert L\right)}{\lvert h \rvert}
 \left(h_1 + i h_2\right) \;,
\end{equation}
whose squared modulus is \equ{prob_2nu_general}: the prefactor contributes $\sin^2\left(\lvert h \rvert L\right)/\lvert h \rvert^2$, and $\lvert h_1 + i h_2 \rvert^2 = h_1^2 + h_2^2$ because both are real.

The first entry gives the survival probability, and the two then sum to one.  The amplitude is
\begin{equation}
 \nu_\alpha^\dagger \mathbb{U}_2(L) \nu_\alpha
 =
 \cos\left(\left\vert h\right\vert L\right)
 -
 i \frac{h_3}{\lvert h \rvert} \sin\left(\left\vert h\right\vert L\right) \;,
\end{equation}
whose two terms are its real and imaginary parts, $h_3$ being real, so that the squared modulus is the sum of their squares:
\begin{eqnarray}
 P_{\nu_\alpha \to \nu_\alpha}(L)
 \!\!\!&=&\!\!\!
 \cos^2(\lvert h \rvert L)
 +
 \frac{h_3^2}{\lvert h \rvert^2} \sin^2(\lvert h \rvert L) \nonumber \\
 \!\!\!&=&\!\!\!
 1 - \frac{h_1^2 + h_2^2}{\lvert h \rvert^2} \sin^2(\lvert h \rvert L) \;,
\end{eqnarray}
the second step using $h_3^2 = \lvert h \rvert^2 - h_1^2 - h_2^2$.  This is $1 - P_{\nu_\alpha \to \nu_\beta}$, as conservation of probability requires.  Starting from $\nu_\beta$ reverses the sign of the $h_3$ term and leaves the modulus untouched, so a two-flavor system has only these two numbers in it: $P_{\nu_\beta \to \nu_\beta} = P_{\nu_\alpha \to \nu_\alpha}$, and $P_{\nu_\beta \to \nu_\alpha} = P_{\nu_\alpha \to \nu_\beta}$, the latter already plain in \equ{prob_2nu_general}, whose right-hand side carries no flavor label at all.

Both moduli divide by $\lvert h \rvert^2$, which vanishes at one point only; Section \ref{sec:limitations} says what the code returns there.

%%%%%%%%%%%%%%%%%%%%%%%%%%%%%%%%%%%%%%%%%%%%%%%%%%%%%
%%%%%%%%%%%%%%%%%%%%%%%%%%%%%%%%%%%%%%%%%%%%%%%%%%%%%

\section{Two-flavor oscillations in vacuum}
\label{app:2nu_vacuum}
\appresetfloats

As a simple example and cross-check of the OS method, we compute $P_{\nu_e \to \nu_\mu}$ for two-flavor oscillations in vacuum.  These are driven by the mass-squared difference between two mass eigenstates, $\nu_1$ and $\nu_2$, with masses $m_1$ and $m_2$, out of which the flavor states $\nu_e$ and $\nu_\mu$ are constructed.  Any other pair of flavors does as well---$\nu_\mu$ and $\nu_\tau$, say---since in vacuum nothing distinguishes one pair from another once the angle and the splitting are fixed; the paragraph that closes this appendix says what fixes them once a term is added.  The spaces of flavor and mass states are connected by a unitary rotation that is parametrized by a mixing angle $\theta$, \ie,
\begin{equation}
 \label{equ:u_rotation_2nu}
 \mathbb{R}_{2,\theta}
 =
 \left(\begin{array}{cc}
  \cos \theta  & \sin \theta \\
  -\sin \theta & \cos \theta 
 \end{array}\right) \;.
\end{equation}
The Hamiltonian in the flavor basis, at neutrino energy $E$, is
\begin{equation}
 \label{equ:h2_vacuum}
 \mathbb{H}_2^{\rm vac}(E)
 =
 \frac{1}{2E}
 \mathbb{R}_{2,\theta}
 \mathbb{M}_2^2
 \mathbb{R}_{2,\theta}^\dagger \;,
\end{equation}
where $\mathbb{M}_2^2~\equiv~$${\rm diag}(-\Delta m^2/2, \Delta m^2/2)$ is the mass-squared matrix, and $\Delta m^2\equiv m_2^2-m_1^2$.  The larger $m_i^2$ value sits on the \emph{second} mass eigenstate, which is the textbook traceless form and the ordering the code uses.

Two conventions meet here.  Section \ref{sec:three_flavors}, and the four-flavor Hamiltonians of \ref{app:sample_2nu_hamiltonians}, put the zero on $\nu_1$ rather than making the mass matrix traceless, as is usual at three and four flavors; the two differ by a multiple of the identity, which Section \ref{sec:general_remarks} discards, so nothing below turns on which is used.  The other convention is which eigenstate carries the larger $m_i^2$, and that one is not free of consequences; the paragraph that closes this appendix says where they fall.

Using Table \ref{tab:h2_coefficients}, we identify the squares that \equ{prob_2nu_general} calls for,
\begin{equation}
 h_1^2 = \left(\frac{\Delta m^2}{4E}\right)^2 \sin^2\left(2\theta\right) \;, \;\;
 h_3^2 = \left(\frac{\Delta m^2}{4E}\right)^2 \cos^2\left(2\theta\right) \;,
\end{equation}
with $h_2 = 0$, so that $\lvert h \rvert^2 = \left(\Delta m^2/4E\right)^2$ and $h_1^2 / \lvert h \rvert^2  = \sin^2\left(2\theta\right)$.  From \equ{prob_2nu_general}, the probability is
\begin{equation}
 P_{\nu_e \to \nu_\mu}^{\rm vac} (E, L)
 =
 \sin^2 \left(2\theta\right)
 \sin^2\left(\frac{\Delta m^2}{4E}L\right) \;,
\end{equation}
which is the standard expression for two-neutrino oscillations in vacuum; see, \eg, \Refs\ \cite{Giunti:2007ry, Tanabashi:2018oca}.

Now, the ordering.  Putting the larger $m_i^2$ on the first mass eigenstate instead gives the negative of $\mathbb{M}_2^2$, and hence of $\mathbb{H}_2^{\rm vac}$, which nothing above would notice: the probability depends on the $h_k$ only through $h_1^2 + h_2^2$, $\lvert h \rvert^2$, and $\sin^2\left(\lvert h \rvert L\right)$, each of them even under $h_k \to -h_k$, so the formula just recovered stands either way.  What the sign does affect is every Hamiltonian that \emph{adds} a term to $\mathbb{H}_2^{\rm vac}$: the matter, NSI, and LIV cases of \ref{app:sample_2nu_hamiltonians}.  Negating only the vacuum term reverses its sign relative to the matter potential, and what comes out is the negative of the antineutrino Hamiltonian of \equ{antineutrino}, up to the sign of $h_2$ --- which \equ{prob_2nu_general} does not see either.  The probabilities are therefore exactly those of the corresponding \emph{antineutrino} process, and for the LIV case those of the CPT-conjugate one, with $b_i \to -b_i$: an identity, not a numerical near-agreement, and the code returns the two bit for bit.

\begin{figure}[t!]
 \centering
 \includegraphics[width=\columnwidth]{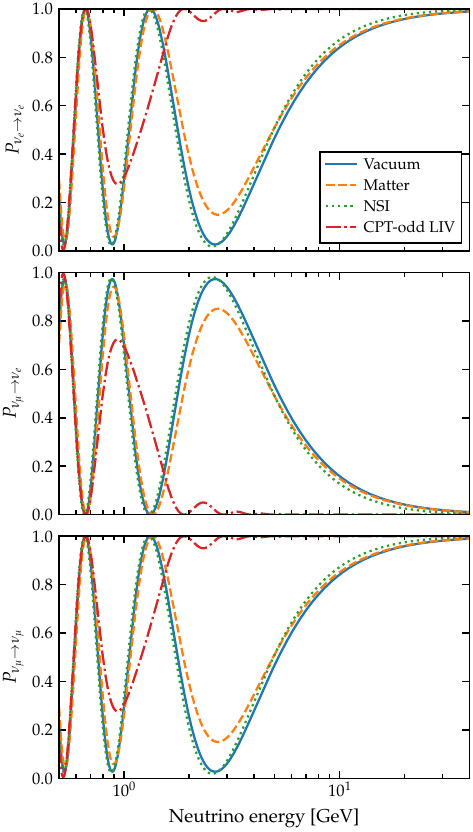}
 \caption{\label{fig:prob_2nu}Two-neutrino oscillation probabilities $P_{\nu_e \to \nu_e}$ ({\it top}), $P_{\nu_\mu \to \nu_e}$ ({\it center}), and $P_{\nu_\mu \to \nu_\mu}$ ({\it bottom}), computed using the method presented here, via {\tt NuOscProbExact}\ \cite{NuOscProbExact}.  This figure is the two-neutrino counterpart of \figu{prob_3nu}.  See the main text and \ref{app:sample_2nu_hamiltonians} for details.}
\end{figure}

Those cases also fix the labels that vacuum leaves free, each in its own way.  The potential $\mathbb{A}_2$ of \ref{app:sample_2nu_hamiltonians} acts on one state and not the other, so with matter and with NSI the first is $\nu_e$ in earnest rather than by convention; the Lorentz-violating case carries no potential at all, and there it is $\mathbb{B}_2$, with $b_1 \neq b_2$, that does the same work.  The matter, NSI, and LIV panels of \figu{prob_2nu} are computed with the ordering above, and would show the antineutrino probabilities under the other one; the vacuum panel does not depend on the choice.  Nothing in Section\ \ref{sec:examples} is affected, since this concerns the two-flavor Hamiltonian alone.

%%%%%%%%%%%%%%%%%%%%%%%%%%%%%%%%%%%%%%%%%%%%%%%%%%%%%
%%%%%%%%%%%%%%%%%%%%%%%%%%%%%%%%%%%%%%%%%%%%%%%%%%%%%

%%%%%%%%%%%%%%%%%%%%%%%%%%%%%%%%%%%%%%%%%%%%%%%%%%%%%
%%%%%%%%%%%%%%%%%%%%%%%%%%%%%%%%%%%%%%%%%%%%%%%%%%%%%

\section{Derivation of \equ{uk_expanded}}
\label{app:derivation_uk_expanded}
\appresetfloats

We start from the definition $u_k = -\frac{i}{2}{\rm Tr}\left(\lambda^k \mathbb{U}_3\right)$ of Section \ref{sec:readings}, and assume throughout the non-degenerate spectrum that \equ{sylvester} and \equ{uk_expanded} both require.  The definition lets $u_k$ be obtained by differentiating ${\rm Tr}\,\mathbb{U}_3$, which needs the latent roots and nothing else, since the projectors are then rank one and ${\rm Tr}\,\mathbb{P}_m = 1$; evaluating ${\rm Tr}\left(\lambda^k \mathbb{U}_3\right)$ directly would need the projectors themselves.  So,
\begin{eqnarray}
 u_k 
 &=& -\frac{1}{2}{\rm Tr}\frac{\partial}{\partial\left(-h_kL\right)}e^{-ih_a\lambda^aL} \nonumber \\
 &=& -\frac{1}{2}\frac{\partial}{\partial\left(-h_kL\right)}{\rm Tr}~\mathbb{U}_3 \nonumber \\
 &=& -\frac{i}{2}\sum_{m=1}^3 e^{iL\psi_m} \frac{\partial\left(L\psi_m\right)}{\partial\left(-h_kL\right)} \;,
\end{eqnarray}
where the first step holds under the trace and not outside it: $\lambda^k$ does not commute with $h_a \lambda^a$, so differentiating the exponential term by term leaves a $\lambda^k$ in every position of every power, and it is the cyclic property of the trace that gathers them into ${\rm Tr}\left(i \lambda^k \mathbb{U}_3\right)$.  The last step uses ${\rm Tr}\,\mathbb{U}_3 = \sum_m e^{iL\psi_m}$ and brings down a factor of $i$ from each exponential.  What is left is the derivative of a latent root, which follows from implicit differentiation of $\chi\left(\psi_m\right) = 0$; the assumption above is what makes that step legitimate, since it is $\chi^\prime\left(\psi_m\right) = 3\psi_m^2 - \lvert h \rvert^2$ that must not vanish, the same quantity that ends up as the denominator of \equ{uk_expanded}.  The baseline is held fixed, so it cancels:
\begin{equation}
 \frac{\partial\left(L\psi_m\right)}{\partial\left(-h_kL\right)}
 =
 \frac{\partial\psi_m}{\partial\left(-h_k\right)}
 =
 \frac{2\left[\left(h \ast h\right)_k - \psi_m h_k\right]}{3\psi_m^2-\lvert h \rvert^2} \;.
\end{equation}
Substituting this into the sum that closes the first display --- the $2$ canceling the $\tfrac{1}{2}$, and the bracket written the other way round to absorb the sign that is left --- gives
\begin{equation}
 u_k
 =
 i\sum_{m=1}^3
 e^{i L \psi_m} \frac{\psi_m h_k - \left(h \ast h\right)_k}{3\psi_m^2 - \left\vert h \right\vert^2} \;,
\end{equation}
which is \equ{uk_expanded} of the main text.

%%%%%%%%%%%%%%%%%%%%%%%%%%%%%%%%%%%%%%%%%%%%%%%%%%%%%
%%%%%%%%%%%%%%%%%%%%%%%%%%%%%%%%%%%%%%%%%%%%%%%%%%%%%

\section{Sample two- and four-flavor Hamiltonians}
\label{app:sample_2nu_hamiltonians}
\appresetfloats

Here we present the two- and four-flavor Hamiltonians included as examples in {\tt NuOscProbExact}.  These are the counterparts of the three-flavor examples presented in Section\ \ref{sec:examples}; we refer to that section for a description of each scenario.  Every Hamiltonian is written in the flavor basis, as in that section.

%%%%%%%%%%%%%%%%%%%%%%%%%%%%%%%%%%%%%%%%%%%%%%%%%%%%%

\subsection{Two-flavor Hamiltonians}

Figure \ref{fig:prob_2nu} shows the probabilities $P_{\nu_e \to \nu_e}$, $P_{\nu_\mu \to \nu_e}$, and $P_{\nu_\mu \to \nu_\mu}$ for the same four scenarios as in \figu{prob_3nu}, computed using {\tt NuOscProbExact}\ \cite{NuOscProbExact}.  Again, we set the baseline to $L = 1300$~km.  The parameters and their values used in each example case are introduced below; they are a selection of the ones used in \figu{prob_3nu}.  The two non-standard cases follow the two-flavor treatments of \Refs\ \cite{GonzalezGarcia:2004wg, GonzalezGarcia:2005xw}; Section \ref{sec:examples} carries the references for the physics itself.

For oscillations in vacuum, we use \equ{h2_vacuum}, \ie,
\begin{equation}
 \mathbb{H}_2^{\rm vac}(E)
 =
 \frac{1}{2E}
 \mathbb{R}_{2,\theta}
 \mathbb{M}_2^2
 \mathbb{R}_{2,\theta}^\dagger \;,
\end{equation}
where the rotation matrix $\mathbb{R}_{2,\theta}$ is given in \equ{u_rotation_2nu}, in terms of the mixing angle $\theta$.  In \figu{prob_2nu}, we set $\theta$ and $\Delta m^2$, respectively, to the values of $\theta_{23}$ and $\Delta m_{31}^2$ used in \figu{prob_3nu}.  The labels $e$ and $\mu$ on a pair carrying atmospheric-sector parameters---and $\theta_{23}$ with $\Delta m_{31}^2$ rather than $\Delta m_{32}^2$, which differ by $3\%$---are names the vacuum panel does not fix, as \ref{app:2nu_vacuum} notes; in the other three it is the state that feels $V_{\rm CC}$ that is $\nu_e$.  These are also the values used by the two-neutrino examples in the repository, and they are the ones the figure requires: with $\Delta m_{21}^2$ the first minimum of $P_{\nu_e \to \nu_e}$ at $L = 1300$~km would fall at $E \simeq 0.08$~GeV, far below the range plotted, and its depth would be set by $\sin^2 2\theta_{12}$; \figu{prob_2nu} instead shows that minimum at $E \simeq 2.6$~GeV with near-maximal depth, which is what $\Delta m_{31}^2$ and $\theta_{23}$ give.

For oscillations in matter, we use
\begin{equation}
 \mathbb{H}_2^{\rm matt}(E) 
 = 
 \mathbb{H}_2^{\rm vac}(E)  + \mathbb{A}_2 \;,
\end{equation}
where $\mathbb{A}_2 \equiv {\rm diag}(V_{\rm CC}, 0)$ and $V_{\rm CC}$ is defined as before.  In \figu{prob_2nu}, we set $\rho = 3$~g~cm$^{-3}$.

For oscillations in matter with non-standard interactions, we use
\begin{equation}
 \mathbb{H}_2^{\rm NSI}(E) 
 = 
 \mathbb{H}_2^{\rm vac}(E) + \mathbb{A}_2 + \mathbb{V}_2 \;,
\end{equation}
where $\mathbb{V}_2 \equiv V_{\rm CC} \epsilon_2$ and the matrix of NSI strength parameters is
\begin{equation}
 \epsilon_2
 =
 \left(
  \begin{array}{cc}
   \epsilon_{ee}         & \epsilon_{e\mu}         \\
   \epsilon_{e\mu}^\ast  & \epsilon_{\mu\mu}       \\
  \end{array}
 \right) \;.
\end{equation}
In \figu{prob_2nu}, we set $\epsilon_{ee} = -\epsilon_{e\mu} = 0.06$ and $\epsilon_{\mu\mu} = 1.2$.

For oscillations in a CPT-odd Lorentz-violating background, we use
\begin{equation}
 \mathbb{H}_2^{\rm LIV}(E) 
 = 
 \mathbb{H}_2^{\rm vac}(E) + \frac{E}{\Lambda} \mathbb{R}_{2,\xi} \mathbb{B}_2 \mathbb{R}_{2,\xi}^\dagger \;,
\end{equation}
where $\mathbb{R}_{2,\xi}$ is a $2 \times 2$ rotation matrix, like \equ{u_rotation_2nu}, but evaluated at a different mixing angle $\xi$.  In \figu{prob_2nu}, we set $\mathbb{R}_{2,\xi} = \mathbb{1}$, $b_1/\Lambda = 10^{-21}$, and $b_2/\Lambda = 2 \cdot 10^{-21}$.  These are the first and the third of the three eigenvalues that {\tt globaldefs} carries for \figu{prob_3nu}, and deliberately not the first two: that set has $b_1 = b_2$, which would make $\mathbb{B}_2$ proportional to the identity and leave the probabilities untouched.  A two-flavor Lorentz-violating term needs two \emph{different} eigenvalues, as a two-flavor mass matrix needs two different masses.

%%%%%%%%%%%%%%%%%%%%%%%%%%%%%%%%%%%%%%%%%%%%%%%%%%%%%

\subsection{Four-flavor Hamiltonians}

The four-flavor examples are 3+1 systems: the three active flavors, plus one sterile state $\nu_s$ that has no standard-model interactions.  Figure~\ref{fig:validation} shows one such case and \figu{sterile_earth} another; in both, $\Delta m_{41}^2 = 1$~eV$^2$ and $\sin^2\theta_{14} = \sin^2\theta_{24} = 0.1$, with $\theta_{34} = 0$, and the three active parameters keep the values used elsewhere in this paper.

For oscillations in vacuum, we use
\begin{equation}
 \mathbb{H}_4^{\rm vac}(E)
 =
 \frac{1}{2E}
 \mathbb{R}_4
 \mathbb{M}_4^2
 \mathbb{R}_4^\dagger \;,
\end{equation}
where $\mathbb{M}_4^2 = {\rm diag}\left(0, \Delta m_{21}^2, \Delta m_{31}^2, \Delta m_{41}^2\right)$ and $\mathbb{R}_4$ is the $4 \times 4$ mixing matrix, built from the six angles $\theta_{12}$, $\theta_{13}$, $\theta_{23}$, $\theta_{14}$, $\theta_{24}$, $\theta_{34}$, and three phases, of which the code takes $\delta_{14}$ and $\delta_{24}$ to be zero by default.  Section \ref{sec:three_flavors} writes the three-flavor matrix out because conventions differ between codes.  No standard ordering exists at $3+1$, any more than one exists for the SU(4) generators of \ref{app:matrices_4nu}, so we state ours,
\begin{equation}
 \mathbb{R}_4
 =
 \mathbb{R}_{34}
 \mathbb{R}_{24}\left(\delta_{24}\right)
 \mathbb{R}_{14}\left(\delta_{14}\right)
 \mathbb{R}_{23}
 \mathbb{R}_{13}\left(\delta_{\rm CP}\right)
 \mathbb{R}_{12} \;,
\end{equation}
with $\mathbb{R}_{ij}$ the rotation by $\theta_{ij}$ in the $i$--$j$ plane: $\cos\theta_{ij}$ in its $\left(i,i\right)$ and $\left(j,j\right)$ entries, $\sin\theta_{ij} e^{-i\delta}$ in its $\left(i,j\right)$ entry, $-\sin\theta_{ij} e^{i\delta}$ in its $\left(j,i\right)$ entry, $1$ on the rest of the diagonal and $0$ everywhere else, and $\delta = 0$ where no phase is shown.  Setting the three new angles to zero leaves $\mathbb{R}_{23} \mathbb{R}_{13}\left(\delta_{\rm CP}\right) \mathbb{R}_{12}$, which is \equ{pmns}, so $\mathbb{H}_3^{\rm vac}$ returns in the active block and $\nu_s$ decouples.

For oscillations in matter, we use
\begin{equation}
 \mathbb{H}_4^{\rm matt}(E)
 =
 \mathbb{H}_4^{\rm vac}(E) + \mathbb{A}_4 \;,
\end{equation}
where
\begin{equation}
 \label{equ:a4}
 \mathbb{A}_4 = {\rm diag}\left(V_{\rm CC}, 0, 0, V_s\right) \;,
 \quad
 V_s = -V_{\rm NC} = \frac{G_{\rm F} n_n}{\sqrt{2}} \;,
\end{equation}
so that $V_s$ is exactly half of $V_{\rm CC}$ wherever there is one neutron per electron, which is what {\tt globaldefs} takes the crust to be.  That last entry is the one feature with no two- or three-flavor counterpart, and it is not a convention: the neutral-current potential is common to the three active flavors and may be removed as a global phase, but the sterile state never had it, and subtracting a term it does not feel leaves $-V_{\rm NC}$ behind.  The sterile state does not feel the charged-current potential either, while the other three do not feel $V_s$; it is that difference that drives the resonance of \figu{sterile_earth}.  A 3+1 system is therefore sensitive to the neutron density as well as the electron density, which no active-only scenario is.  In \figu{sterile_earth} the density is the PREM profile\ \cite{Dziewonski:1981xy} along the chord, so both $V_{\rm CC}$ and $V_{\rm NC}$ vary from slab to slab.

For oscillations in matter with non-standard interactions, we use
\begin{equation}
 \mathbb{H}_4^{\rm NSI}(E)
 =
 \mathbb{H}_4^{\rm vac}(E) + \mathbb{A}_4 + \mathbb{V}_4 \;,
 \quad
 \mathbb{V}_4 \equiv V_{\rm CC}
 \left(
  \begin{array}{cc}
   \epsilon_3 & 0 \\
   0          & 0 \\
  \end{array}
 \right) \;,
\end{equation}
where $\epsilon_3$ is the $3 \times 3$ matrix of Section\ \ref{sec:examples}.  The sterile row and column vanish because a state with no standard-model interactions has no non-standard ones either.  The sterile entry keeps the $-V_{\rm NC}$ of $\mathbb{A}_4$.

For oscillations in an isotropic CPT-even Lorentz-violating background, we use
\begin{equation}
 \mathbb{H}_4^{\rm LIV}(E)
 =
 \mathbb{H}_4^{\rm vac}(E) + \frac{E}{\Lambda} \mathbb{R}_{4,\xi} \mathbb{B}_4 \mathbb{R}_{4,\xi}^\dagger \;,
\end{equation}
with $\mathbb{B}_4 = {\rm diag}\left(b_1, b_2, b_3, b_4\right)$ and $\mathbb{R}_{4,\xi}$ ordered as $\mathbb{R}_4$ above but carrying its own six angles $\xi_{ij}$ and one phase.  Here, unlike in $\mathbb{A}_4$ and $\mathbb{V}_4$, nothing privileges the fourth state: $b_4$ is an eigenvalue like the others, so the sterile neutrino may couple to the Lorentz-violating background even though it is blind to matter.  Setting the three new angles to zero recovers the three-flavor term in the active block whatever $b_4$ is---with $\mathbb{R}_{4,\xi}$ block diagonal, $b_4$ acts on $\nu_s$ alone, and a term on a decoupled state is a phase.  We have checked that the active-flavor probabilities are unchanged to $10^{-13}$ as $b_4$ is varied over four orders of magnitude and both signs.

Neither of these last two Hamiltonians is drawn on in this paper.  Every four-flavor figure in it rests on $\mathbb{H}_4^{\rm matt}$ instead: the four-flavor case of \figu{validation}, at constant density, and \figu{sterile_earth} and \figu{prem_plane_3plus1}, slab by slab through PREM, the first of those two with the antineutrino exchange of \equ{antineutrino}.  The suite of \ref{app:tests} exercises the NSI and the Lorentz-violating Hamiltonians, and notebook \#16 of \ref{app:notebooks} works through the NSI one.

%%%%%%%%%%%%%%%%%%%%%%%%%%%%%%%%%%%%%%%%%%%%%%%%%%%%%
%%%%%%%%%%%%%%%%%%%%%%%%%%%%%%%%%%%%%%%%%%%%%%%%%%%%%

\section{Speed against accuracy through the Earth}
\label{app:prem_plane}
\appresetfloats

\begin{figure}[t!]
 \centering
 \includegraphics[width=\columnwidth]{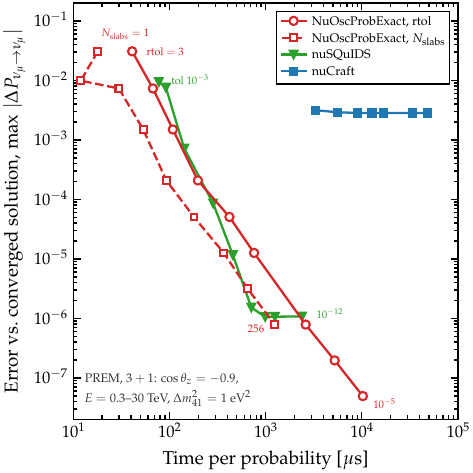}
 \caption{\label{fig:prem_plane_3plus1}Same as \figu{prem_plane}, but with a sterile state: $P_{\nu_\mu \to \nu_\mu}$ along the same chord, at the mixing parameters of \figu{sterile_earth}, $\Delta m_{41}^2 = 1$~eV$^2$ and $\sin^2\theta_{14} = \sin^2\theta_{24} = 0.1$, over twelve energies between $0.3$ and $30$~TeV.  These are neutrinos, where \figu{sterile_earth} shows antineutrinos, on the same PREM profile; what is measured here is the cost of resolving that profile, not any feature of the probability.  The reference is built the same way as in \figu{prem_plane} and checked the same way, to about $10^{-9}$.  {\tt NuOscProbExact} is again v1.13.1 with the compiled kernels, called on the whole energy stack at once.  {\tt nuCraft} is drawn as released, flat at about $3 \cdot 10^{-3}$, a floor that comes from the rounding of two constants in its own source rather than from its solver.}
\end{figure}

In the main text, Sections \ref{sec:differences} and \ref{sec:plane} read \figu{prem_plane}; this appendix records what had to be true before it could be drawn and where each code's curve stops.  At constant density, every code is handed the same Hamiltonian, so Figure~\ref{fig:plane} measures how exactly each code evaluates the probability from it.  Through the Earth (\figu{prem_plane}), there is no single Hamiltonian to hand over: the density changes continuously along the chord and the Hamiltonian with it.  Every code here, {\tt NuOscProbExact} among them, must first build its own piecewise approximation of the profile, whether by cutting it into pieces of constant density or by sampling it as it integrates.  Therefore, \figu{prem_plane} (and \figu{prem_plane_3plus1}) measures how finely a code resolves the profile rather than the exactness of its probability formula.  Because every answer through the Earth carries an approximation of the profile, no exact answer is available to measure the codes against.

Hence, the reference against which we compare the performance of different oscillation codes is a \emph{converged} PREM solution rather than an exact one.  Cutting each segment of the chord into $N_{\rm slabs}$ slabs and composing them, as \equ{composition} does, leaves an error that falls as $1/N_{\rm slabs}^2$ (Section \ref{sec:differences}); refining the count and fitting that power returns 2.  Doubling the count therefore divides the error by four, so the probabilities at $256$ and at $128$ slabs combine as $\left[4 P(256) - P(128)\right]/3$ to cancel the leading error rather than shrink it.  Both are computed at $30$ decimal digits rather than the 16 of double precision, which leaves the reference limited by neither round-off nor the slab count.

That combination is built by cutting the chord the same way the code being judged cuts it, so an error in the cutting would sit in both and cancel from the comparison instead of showing up in it.  It is therefore checked against a calculation that shares none of that machinery: an adaptive integration of the flavor state along the chord, $d\nu/dx = -i \mathbb{H}(x) \nu$, through the continuous profile, restarted at every density jump so that no step straddles one.  The two have nothing in common but PREM itself.  They agree to about $10^{-11}$ at three flavors, some three decades below the smallest error plotted in \figu{prem_plane}, so the reference contributes nothing to any point there.  At $3+1$ they agree to about $10^{-9}$, only a factor of about $40$ below the smallest error in \figu{prem_plane_3plus1}: the last point of the sterile panel is the one place in either figure where the reference comes within sight of what it measures.

None of the conventions behind \figu{prem_plane} is optional.  Each external code is dialed over the range its own interface offers: $N_{\rm shells}$ from $1$ to $256$ for {\tt NuFast-Earth}, {\tt GLoBES}, and {\tt Prob3++}, and solver tolerances from $10^{-3}$ to $10^{-12}$ for {\tt nuSQuIDS} and from $10^{-2}$ to $10^{-10}$ for {\tt nuCraft}.  Every code is given the same oscillation parameters --- the NuFit 4.0 values of Section\ \ref{sec:examples}, and at $3+1$ the $\Delta m^2_{41}$ and mixings of \figu{sterile_earth} --- each expressed in the convention that code expects, which for {\tt Prob3++} means $\Delta m^2_{32}$ rather than $\Delta m^2_{31}$.  All five are handed the PREM, as implemented in {\tt NuOscProbExact} at $Y_e = 0.5$, the same $V_{\rm CC}$, and the same chord in eV$^{-1}$, so that what is compared is the propagation and not the Earth model.  Each is called through whatever batched interface it has--- {\tt nuSQuIDS} through its multiple-energy constructor, {\tt NuFast-Earth} through {\tt Set\_Spectra}, and {\tt nuCraft} through the particle list its {\tt CalcWeights} takes, though that list is looped over internally---while {\tt GLoBES} and {\tt Prob3++} expose no such entry point and are looped, which is their interface and not a handicap imposed here.  {\tt NuOscProbExact} is called with one batched request for all twelve energies.

%%%%%%%%%%%%%%%%%%%%%%%%%%%%%%%%%%%%%%%%%%%%%%%%%%%%%
%%%%%%%%%%%%%%%%%%%%%%%%%%%%%%%%%%%%%%%%%%%%%%%%%%%%%

In \figu{prem_plane}, {\tt NuOscProbExact} is drawn as two curves rather than as the single point of \figu{plane}, one for each dial it exposes: the number of slabs per chord segment, and the relative tolerance of Section \ref{sec:memory}.  The tolerance sets approximately the quantity the vertical axis measures; a slab count yield an accuracy consequence that is unknown in advance.  The convenience costs about a factor of two in time---the gap between the two red curves---because the search evaluates at $1, 2, 4, \ldots$ slabs and keeps only the last.  The coarsest settings measure overhead rather than the slab product: at one slab per segment the chord is seventeen slabs long, too short to amortize the kernel launch.

{\tt Prob3++} flattens first, at about $3 \cdot 10^{-4}$.  That floor is its own arithmetic rather than its treatment of the profile: on this same chord in vacuum, with no profile to resolve, it returns some $6 \cdot 10^{-4}$.  {\tt nuSQuIDS} flattens at about $1 \cdot 10^{-6}$ and {\tt nuCraft} at $2.5 \cdot 10^{-6}$, the floors that Section \ref{sec:differences} ties to PREM's density jumps after exhausting the solver settings; the {\tt nuSQuIDS} curve drawn here is the run whose grid puts a point on every one of those jumps, the better of the two runs measured there.  The tie holds for {\tt nuCraft} too: handed a constant profile through the same hook that installs PREM, so that it has no jumps to cross, it agrees with this code to about $3 \cdot 10^{-11}$.  {\tt Prob3++} aside, the codes that cut the profile into pieces of constant density show no comparable floor: this code reaches about $4 \cdot 10^{-8}$, while {\tt GLoBES} and {\tt NuFast-Earth} reach about $7 \cdot 10^{-7}$ and about $2 \cdot 10^{-6}$ where their dial runs out at $256$ shells, {\tt NuFast-Earth} still gaining a factor of about $30$ per doubling between its last two settings.  These six values are the ones \tabl{codes} records in its \figu{prem_plane} accuracy column.

Figure~\ref{fig:prem_plane_3plus1} shows the same plane with a sterile state.  {\tt NuFast-Earth}, {\tt GLoBES}, and {\tt Prob3++} hard-wire three flavors and drop out.  {\tt nuCraft} runs, drawn as released: flat at about $3 \cdot 10^{-3}$ whatever it is asked for.  That floor is not its solver.  Its sterile and charged-current matter entries come from two independently rounded constants whose ratio is $0.5016$ where the isoscalar value is exactly $1/2$.  No setting removes the floor: rescaling the density scales both entries together.  Forcing the ratio to $1/2$ by hand drops the same calculation to about $3 \cdot 10^{-7}$, which is what its solver is worth.  The released curve is the one drawn because it is what a user of {\tt nuCraft} gets.  {\tt nuSQuIDS} again floors near $10^{-6}$.  The slab product does not care how large the Hamiltonian is: the same dial buys the same second-order convergence at four flavors as at three.

\newpage

\end{document}